\documentclass{article}

\usepackage[utf8]{inputenc}
\usepackage[english]{babel}
\usepackage[a4paper, margin=1in]{geometry}
\usepackage{graphicx}
\usepackage{amsmath}
\usepackage{amsfonts}
\usepackage{physics}
\usepackage{multicol}
\usepackage{float}
\usepackage{bbold}
\usepackage{bm}
\usepackage{xcolor}
\usepackage{subcaption}
\usepackage{tikz}
\usetikzlibrary{external}
\usepackage{pgfplots}
\pgfplotsset{compat=1.16}
\pgfplotsset{scaled x ticks=false}
\definecolor{cb-blue}{RGB}{0,119,187}
\definecolor{cb-cyan}{RGB}{51,187,238}
\definecolor{cb-teal}{RGB}{0,153,136}
\definecolor{cb-orange}{RGB}{238,119,51}
\definecolor{cb-red}{RGB}{204,51,17}
\definecolor{cb-magenta}{RGB}{238,51,119}
\definecolor{cb-grey}{RGB}{187,187,187}
\usepackage{algorithm}
\usepackage{algpseudocode}

\usepackage{doi}
\usepackage[numbers, sort&compress]{natbib}
\usepackage{hyperref}
\usepackage{cleveref}
\crefformat{equation}{(#2#1#3)} 

\newtheorem{theorem}{Theorem}[section]
\newtheorem{lemma}{Lemma}[theorem]

\begin{document}

\renewcommand{\abstractname}{\vspace{-\baselineskip}}

\title{Non-Pauli Errors in the Three-Dimensional Surface Code}
\author{Thomas R. Scruby$^{1,2,3}$\thanks{thomas.scruby@oist.jp}, Michael Vasmer$^{4,5}$\thanks{mvasmer@perimeterinstitute.ca}, Dan E. Browne$^3$\thanks{d.browne@ucl.ac.uk}}
\date{%
    \small{\textit{
    $^1$Okinawa Institute of Science and Technology, Okinawa, 904-0495, Japan\\%
    $^2$National Institute of Informatics, Tokyo, 101-8430, Japan\\%
    $^3$Dept. of Physics and Astronomy, University College London, London, WC1E 6BT, UK\\%
    $^4$Perimeter Institute for Theoretical Physics, Waterloo, ON N2L 2Y5, Canada\\%
    $^5$Institute for Quantum Computing, University of Waterloo, Waterloo, ON N2L 3G1, Canada
    }}
    }

\maketitle

\begin{abstract}
    A powerful feature of stabiliser error correcting codes is the fact that stabiliser measurement projects arbitrary errors to Pauli errors, greatly simplifying the physical error correction process as well as classical simulations of code performance. However, logical non-Clifford operations can map Pauli errors to non-Pauli (Clifford) errors, and while subsequent stabiliser measurements will project the Clifford errors back to Pauli errors the resulting distributions will possess additional correlations that depend on both the nature of the logical operation and the structure of the code. Previous work has studied these effects when applying a transversal $T$ gate to the three-dimensional colour code and shown the existence of a non-local ``linking charge'' phenomenon between membranes of intersecting errors. In this work we generalise these results to the case of a $CCZ$ gate in the three-dimensional surface code and find that many aspects of the problem are much more easily understood in this setting. In particular, the emergence of linking charge is a local effect rather than a non-local one. We use the relative simplicity of Clifford errors in this setting to simulate their effect on the performance of a single-shot magic state preparation process --- the first such simulation to account for the full effect of these errors --- and find that their effect on the threshold is largely determined by probability of $X$ errors occurring immediately prior to the application of the gate, after the most recent stabiliser measurement.
\end{abstract}

\vspace{1cm}

\begin{multicols}{2}

\section{Introduction}
\label{section:intro}
Traditional wisdom in the field of quantum error correction says that, although there are infinitely many possible quantum errors, measurements made as part of the error correction process will project this spectrum of errors to a discrete set and it is sufficient to consider only this set when examining the performance of the code~\cite{shor1995,ekert1996error,nielsen_quantum_2011}. In particular, for stabiliser error correcting codes this set is just the $n$-qubit Pauli group (up to phases). This fact greatly simplifies the study of these codes, and in particular enables efficient numerical investigation of code performance~\cite{gottesman1998a,wang2003,qecsim,gidney2021stim}. 

In recent years further study has shown that things are not quite so simple. For example, \textit{coherent errors} (small rotations on all qubits of a code) will indeed be projected to distributions of Pauli errors, but this projection can also create a phase conditional on the presence/absence of a logical operator and this will cause a rotation in the logical space~\cite{wallman_estimating_2015,kueng_comparing_2016,greenbaum_modeling_2017,bravyi_correcting_2018,beale2018a,iverson2020}. Another kind of non-Pauli error (the subject of this work) is produced when Pauli errors in a code are mapped to Clifford errors by the application of a non-Clifford transversal gate~\cite{gottesman1999a}. Subsequent stabiliser measurements will map these errors back to Paulis again but not to the same error we started with, instead producing errors with complex non-local correlations~\cite{yoshida_topological_2015,bombin_transversal_2018,beverland_cost_2021}. The effects of Clifford errors and approaches to correcting them have also been studied in more general settings~\cite{chamberland_fault-tolerant_2018, chamberland_hard_2017}, and error membranes very similar to those considered in this work previously appeared in~\cite{zhu_topological_2021}.

\begin{figure*}[t]
    \centering
    \begin{subfigure}{0.3\textwidth}
        \includegraphics[width=\textwidth]{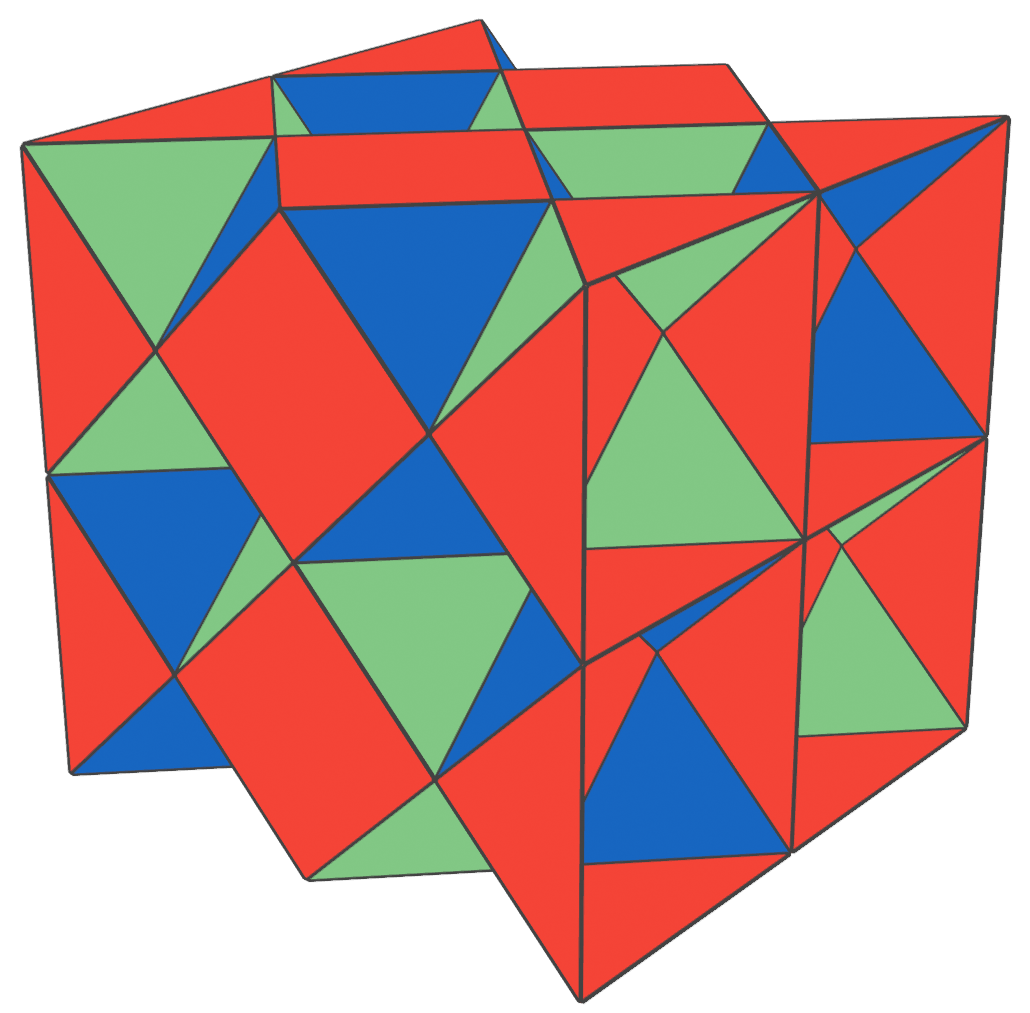}
        \subcaption{}
        \label{subfig:rectifiedA}
    \end{subfigure}
    ~~
    \begin{subfigure}{0.3\textwidth}
        \includegraphics[width=\textwidth]{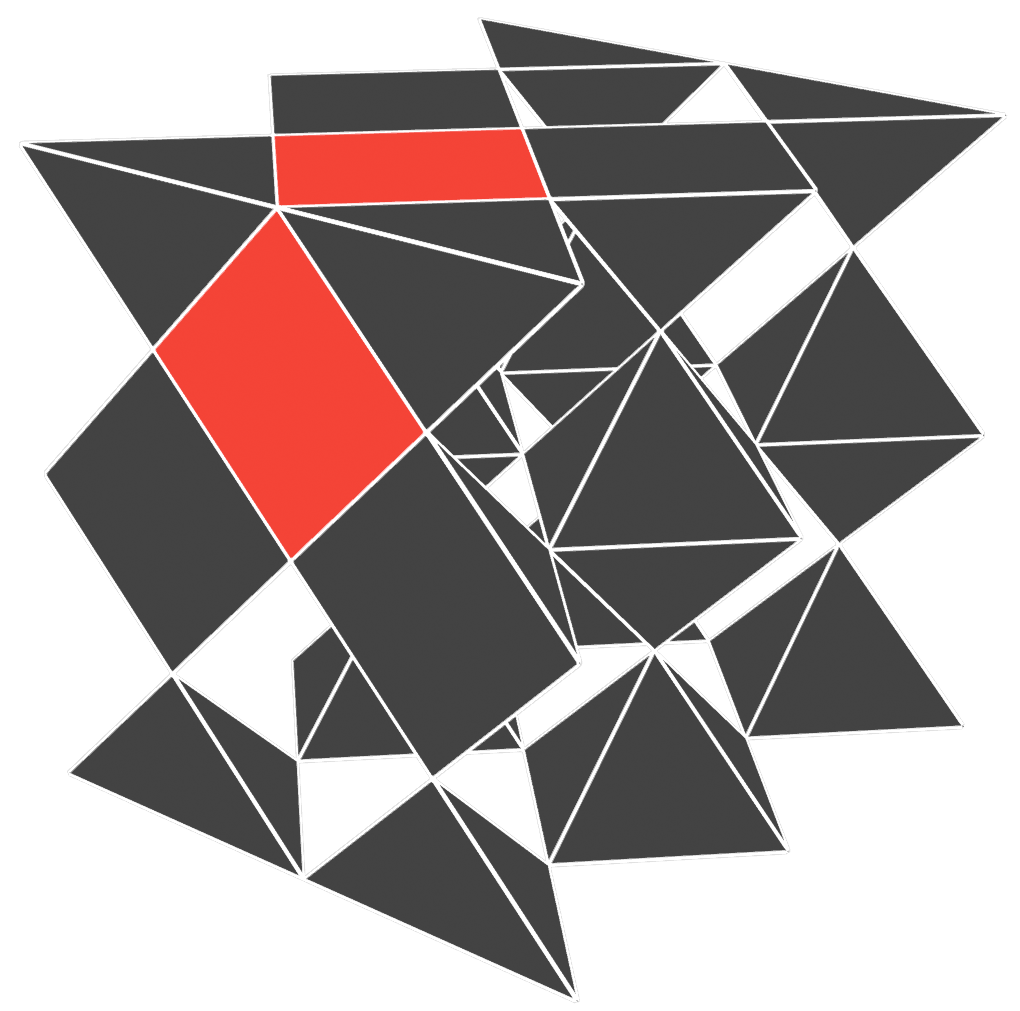}
        \subcaption{}
        \label{subfig:rectifiedB}
    \end{subfigure}
    ~~
    \begin{subfigure}{0.3\textwidth}
        \includegraphics[width=\textwidth]{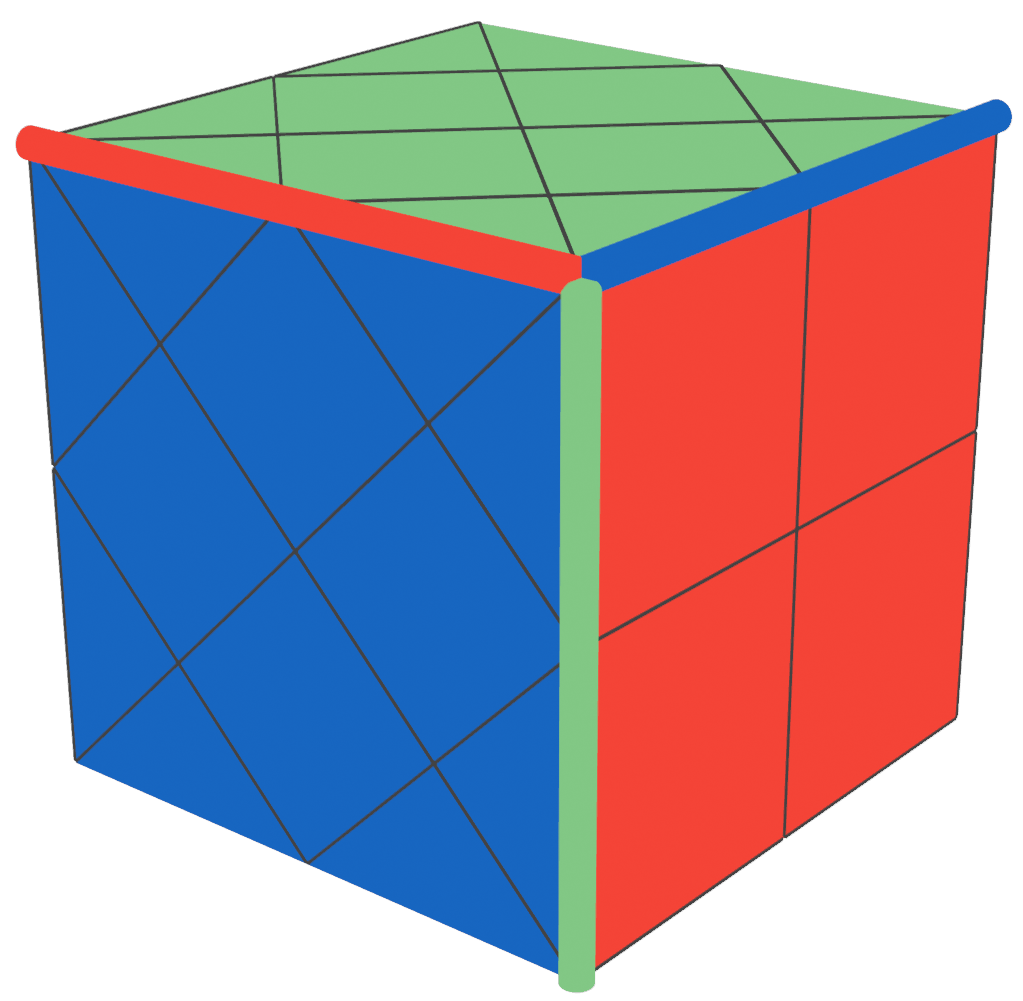}
        \subcaption{}
        \label{subfig:rectifiedC}
    \end{subfigure}
    \caption{(Colour) (a) A rectified lattice supporting three surface codes. (b) One of the three surface codes. $X$ stabilisers are on octahedral cells and $Z$ stabilisers are on square faces. (c) The logical operator structure of the three codes. Logical $X$s are sheets and logical $Z$s are strings. The intersection of logical $X$s from any pair of codes is the support of a logical $Z$ in the third}
    \label{fig:rectified}
\end{figure*}

Previous studies of Clifford errors due to transversal non-Clifford gates have mostly been restricted to the setting of the colour code~\cite{bombin_topological_2007,bombin2006,kubica2015}, where (in the three-dimensional (3D) variant) transversal application of $T$ and $T^\dag$ can map distributions of $X$ errors to distributions of $S$ and $S^\dag$ (in addition to the original Pauli error). In \cref{section:background} we review these results and in \cref{section:sc} we generalise them to the setting of the 3D surface code, which admits a transversal $CCZ$~\cite{vasmer_three-dimensional_2019} gate that can map $X$ errors to $CZ$ errors. We demonstrate similar behaviour to what was observed in the colour code, and additionally (in \cref{appendix:cc}) we show that our proof technique reproduces previous results when applied to the colour code, providing a fresh perspective on the problem of Clifford errors in this setting. 

Following this analytical study of these Clifford errors we numerically investigate their impact on code performance in \cref{section:numerics}. We simulate a magic state preparation scheme that uses a transversal $CCZ$ gate between three 3D surface codes followed by a dimension jump~\cite{bombin_dimensional_2016} to prepare the magic state $CCZ\ket{{+}{+}{+}}$ in constant time. To assess the impact of Clifford errors on code performance we also simulate the codestate initialisation and dimension jump steps in the absence of the $CCZ$ gate and compare the performance of the two cases. We find that $X$ errors occurring immediately before the $CCZ$ gate (after the most recent stabiliser measurement) can have a large impact on the threshold, while the impact of earlier errors is fairly minimal. 

Finally, we discuss the implications of these results and also their generalisations to other codes in \cref{section:discussion}. 

\section{Background}
\label{section:background}
In this section we provide a review of the 3D surface (\cref{subsection:bg_sc}) and colour (\cref{subsection:bg_cc}) codes and review previous results regarding Clifford errors in the latter (\cref{subsection:bg_errors}).

\subsection{Surface Codes}
\label{subsection:bg_sc}

A 2D surface code can be defined by placing qubits on the edges of a 2D lattice and then assigning a $Z$ stabiliser generator to each face and an $X$ stabiliser generator to each vertex~\cite{kitaev_fault-tolerant_2003}. A 3D surface code can be defined in an identical fashion using a 3D lattice rather than a 2D one~\cite{dennis_topological_2002}. There also exists a ``rotated'' description of the 2D surface code which is more qubit efficient, and assigns qubits to vertices and $X$ and $Z$ stabilisers to a bicolouring of faces~\cite{horsman_surface_2012}. A similar description exists for the 3D surface code, where qubits are placed on the vertices of a ``rectified'' lattice~\cite{vasmer_three-dimensional_2019}, obtained from the original lattice by placing a vertex at the centre of each edge, joining these vertices with edges if they are part of the same face in the original lattice, and then deleting all the vertices and edges of the original lattice. 

One such rectified lattice (obtained from a simple cubic lattice) is shown in \cref{subfig:rectifiedA}, and an appropriate colouring of this lattice allows us to simultaneously define three distinct 3D surface codes. This is done by placing three qubits at each vertex (one for each code) and then assigning $Z$ stabilisers of code $i$ to faces with colour $\kappa_i$ and $X$ stabilisers to cells that have no $\kappa_i$-coloured faces~\cite{vasmer_three-dimensional_2019}. \Cref{subfig:rectifiedB} shows one such code, where $Z$ stabilisers are on square faces and $X$ stabilisers are on octahedral cells. This code is equivalent to the 3D surface code obtained by assigning $Z$ ($X$) stabilisers to faces (vertices) in the original simple cubic lattice. The other codes will have $Z$ stabilisers on triangular faces and $X$ stabilisers on cuboctahedral cells. In what follows we refer to these two types of 3D surface code as octahedral and cuboctahedral codes respectively. 

\begin{figure*}
    \centering
    \begin{subfigure}{.35\textwidth}
        \includegraphics[width=\textwidth]{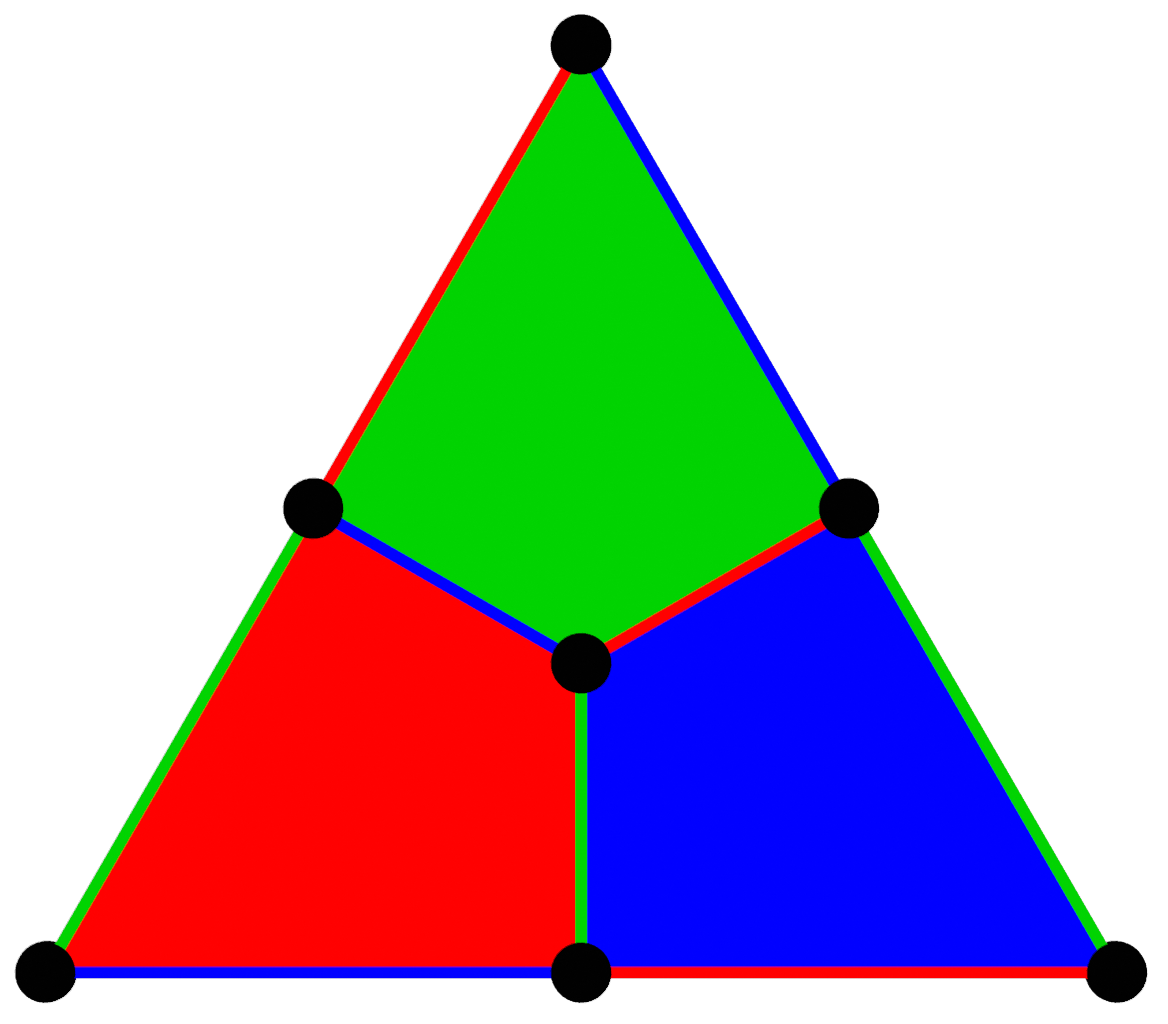}
        \subcaption{}
        \label{subfig:cc1A}
    \end{subfigure}
    ~~~~
    \begin{subfigure}{.35\textwidth}
        \includegraphics[width=\textwidth]{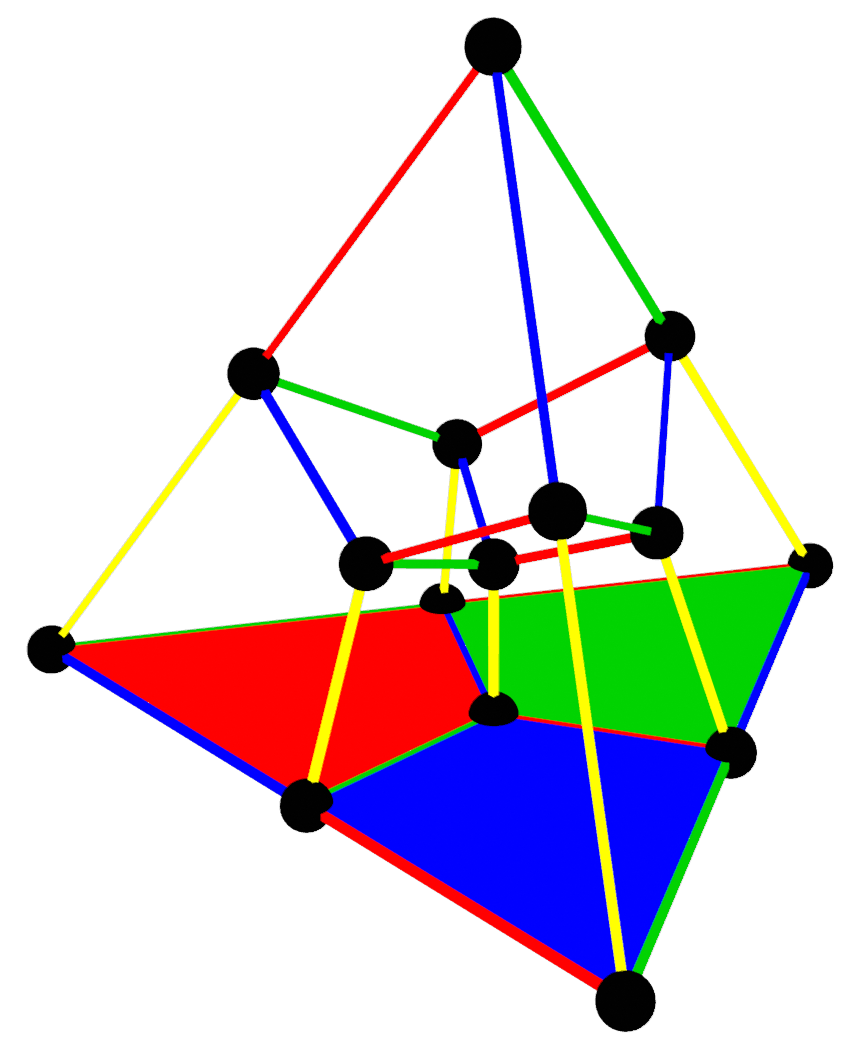}
        \subcaption{}
        \label{subfig:cc1B}
    \end{subfigure}
    \caption{(Colour) (a) A 7-qubit 2D colour code. (b) A 15-qubit 3D colour code. A Z stabiliser is
supported on each face and an X stabiliser is supported on each cell. The bottom
boundary of this code matches the 2D colour code shown in (a).}
    \label{fig:cc1}
\end{figure*}

The logical $X$ operator of each code is a membrane supported on a boundary of the lattice, while the logical $Z$ is a string supported on the intersection of two boundaries (up to composition with stabilisers). The code distance is therefore the lattice size $L$ (for a lattice with dimensions $L \times L \times L$). An appropriate choice of boundary stabilisers ensures that the logical $X$ operators for the three codes are all supported on different boundaries, and that the logical $Z$ of code $k$ is supported on the intersection of the logical operators of code $i$ and $j$, as shown in \cref{subfig:rectifiedC}. These three surface codes admit a transversal implementation of $CCZ$~\cite{kubica2015a,vasmer_three-dimensional_2019}.

\subsection{Colour Codes}
\label{subsection:bg_cc}

The 2D colour code is defined using a trivalent lattice with faces that are 3-colourable and have even numbers of vertices~\cite{bombin_topological_2007}. A qubit is associated with each vertex of this lattice and an X and a Z stabiliser generator are associated with each face. The fact that faces have even numbers of vertices ensures that X and Z stabilisers from the same face commute, while the trivalency/3-colourability of the lattice ensures that neighbouring faces always meet at an edge, so X and Z stabilisers from different faces also commute. An example of a distance-3 2D colour code is shown in \cref{subfig:cc1A}. Logical $X$ and $Z$ operators are supported on any boundary of this code.

The 3D colour code is defined using a four-valent lattice with four-colourable cells and qubits on vertices. If an edge connects two $\kappa$ coloured cells then we can colour that edge $\kappa$, meaning that $\kappa$ colour cells will be formed from three colours of edge (the three colours not equal to $\kappa$). Each face of the lattice supports a $Z$ stabiliser and each cell supports an $X$ stabiliser. An example of a 15-qubit 3D colour code is shown in \cref{subfig:cc1B}. An implementation of logical $X$ in this code is supported on any boundary of the tetrahedron while an implementation of logical $Z$ is supported on any edge of the tetrahedron. The colour code admits a transversal $T$ gate, which involves the application of $T$ and $T^\dag$ to a bicolouring of qubits in the lattice~\cite{bombin_gauge_2015}.

\subsection{Clifford Errors in the Colour Code}
\label{subsection:bg_errors}

\begin{figure*}
    \centering
    \begin{subfigure}{.5\textwidth}
        \includegraphics[width=\textwidth]{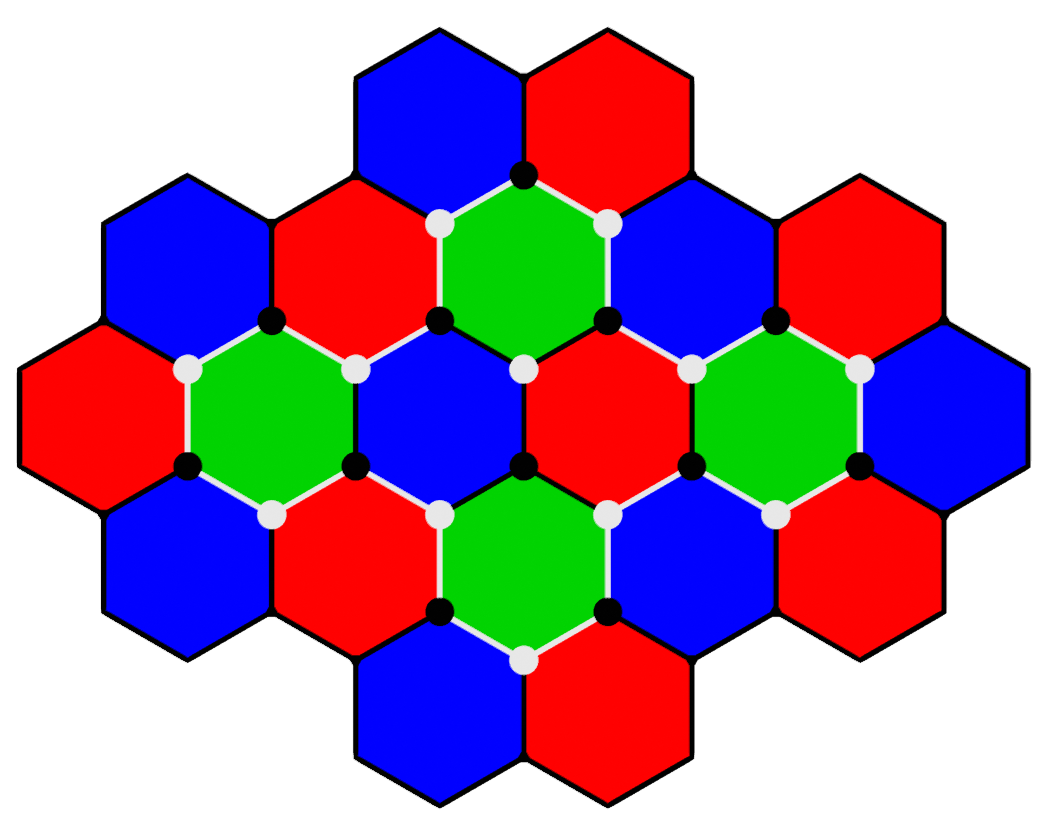}
        \subcaption{}
        \label{subfig:s_error_2d_a}
    \end{subfigure}
    ~
    \begin{subfigure}{.14\textwidth}
        \includegraphics[width=\textwidth]{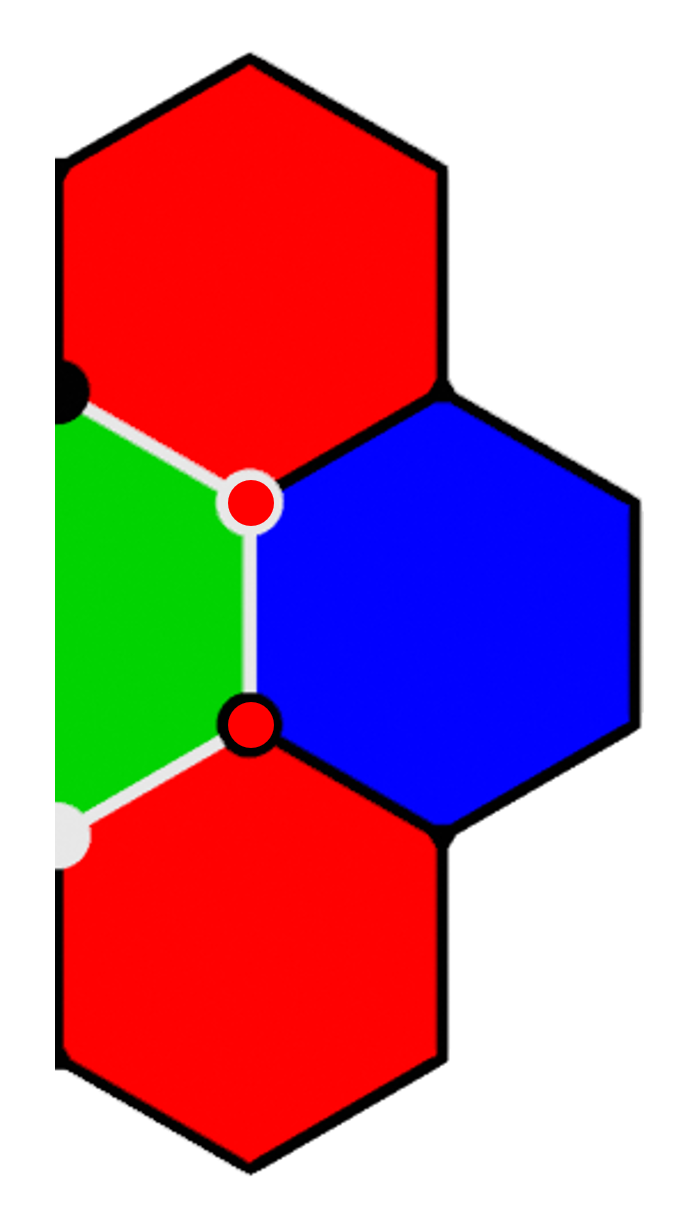}
        \subcaption{}
        \label{subfig:s_error_2d_b}
    \end{subfigure}
    \caption[(Colour) A region of $S$ errors in the 2D colour code]{A region of $S$ errors in the 2D colour code. $S$ is applied to all qubits marked with a white circle and $S^\dag$ is applied to all qubits marked with a black circle. The qubits in this region are those on the vertices of the four green plaquettes inside the loop of white edges while the plaquettes that border this region are either red or blue. (b) Part of the boundary region of (a). A $ZZ$ error acting on the qubits marked with red circles anticommutes with the $X$ stabilisers on the two red plaquettes.}
    \label{fig:s_error_2d}
\end{figure*}

The problem of Clifford errors in the 2D colour code was examined by Yoshida in \cite{yoshida_topological_2015}, although they are not referred to as such and instead are considered in the context of excitations in a symmetry-protected topological phase. Specifically, Yoshida examined the effect of applying a pattern of alternating $S$ and $S^\dag$ to all qubits within a particular region $\mathcal{R}$ defined by a subset of plaquettes of a particular colour as in \cref{fig:s_error_2d}. The 2D colour code possesses a transveral $S$ gate which can be implemented via such an application of $S$ and $S^\dag$ to all qubits in the code and so this error can be thought of as a partial or incomplete logical operator. We can define the boundary of $\mathcal{R}$, $\partial \mathcal{R}$, to be the set of all plaquettes/stabiliser generators partially supported on $\mathcal{R}$. An important observation in this and all subsequent cases is that because logical operators must preserve the codespace (and therefore are not detectable by stabilisers of the code) this region of Clifford errors should only be detected by stabilisers in $\partial \mathcal{R}$. If it could be detected by a stabiliser not in $\partial \mathcal{R}$ then that stabiliser should also detect the logical $S$ gate as these operators are not locally distinguishable except on the boundary of $\mathcal{R}$.

The formally derived result of \cite{yoshida_topological_2015} agrees with this intuition. It says that if we apply the Clifford error shown in \cref{fig:s_error_2d}, for example, and then measure the stabilisers of the code (specifically we only need to measure the $X$ stabilisers as $Z$ and $S$ commute) we will project the $S$ and $S^\dag$ errors to a distribution of $Z$ errors that anticommute only with the red and blue stabilisers in $\partial \mathcal{R}$, with each such stabiliser returning a $-1$ measurement outcome with probability $p=0.5$. Error strings in the 2D colour code must either anticommute with stabilisers of all three colours or anticommute with a pair of stabilisers of the same colour. The former is not possible in this case as there are only two colours of plaquette in $\partial \mathcal{R}$ and so the error distribution must be a collection of $Z$ strings supported on qubits of $\mathcal{R}$ that run between same-coloured plaquettes in $\partial \mathcal{R}$. Because each such string anticommutes with a pair of plaquettes the total number of $-1$ outcomes from plaquettes of each colour should be even. For example, \cref{subfig:s_error_2d_b} shows a two-qubit $Z$ error that anticommutes with a pair of red plaquettes in $\partial \mathcal{R}$.

A corresponding analysis of $S$ errors in the 3D colour code can be found in \cite{bombin_transversal_2018}. This case is arguably of greater practical relevance because, as mentioned previously, the 3D colour code admits a transversal $T$ gate implemented by an application of $T$ and $T^\dag$ to a bicolouring of the vertices (qubits) of the lattice. This gate will create regions of $S$ and $S^\dag$ errors wherever we have regions of $X$ errors as $TXT^\dag = e^{-i\pi/4}SX$ and $T^\dag X T = e^{i\pi/4}S^\dag X$. 

\begin{figure*}
    \hspace{.8cm}
    \begin{subfigure}{.4\textwidth}
        \centering
        \begin{subfigure}{.7\textwidth}
            \includegraphics[width=\textwidth]{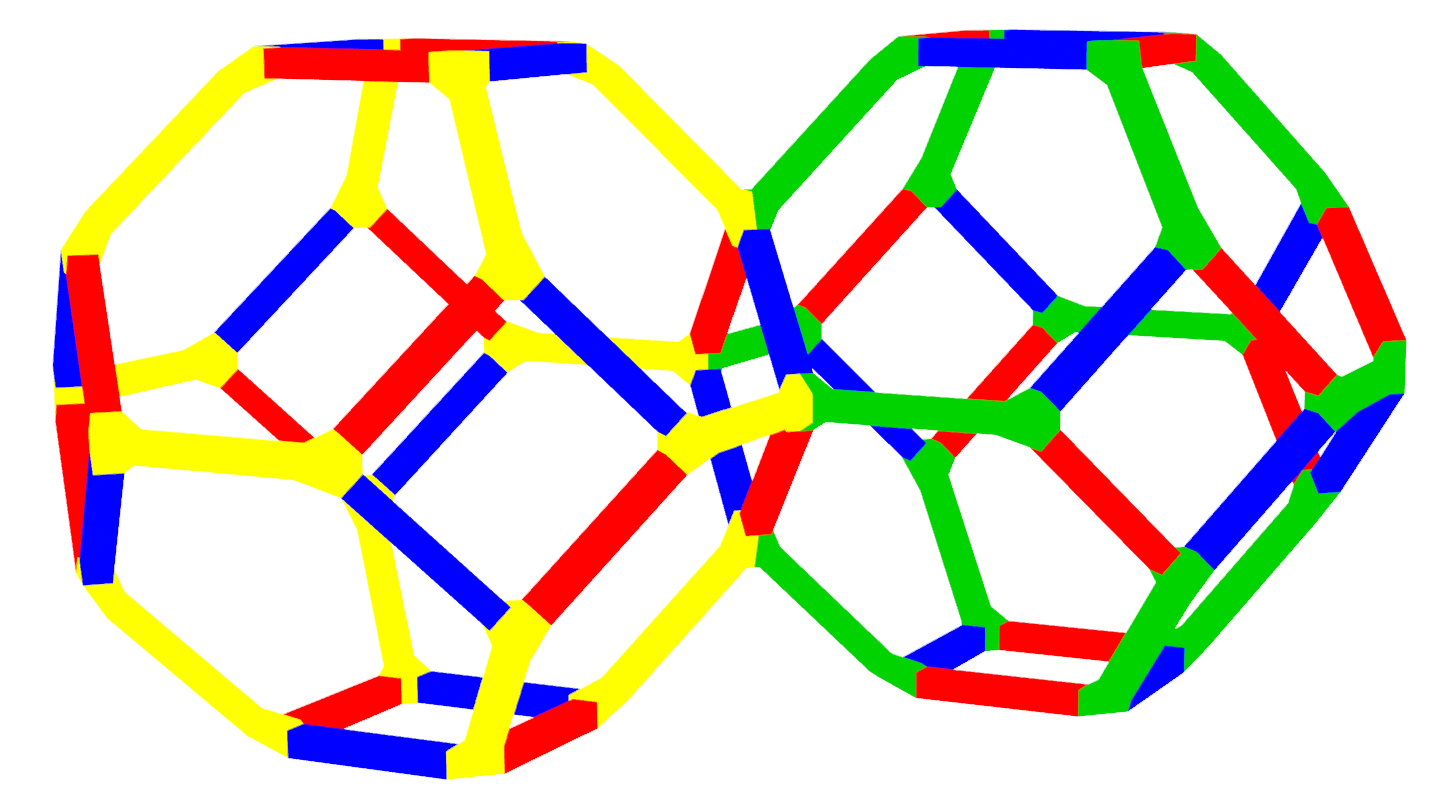}
        \end{subfigure}
        \\
        \vspace{.2cm}
        \begin{subfigure}{\textwidth}
            \includegraphics[width=\textwidth]{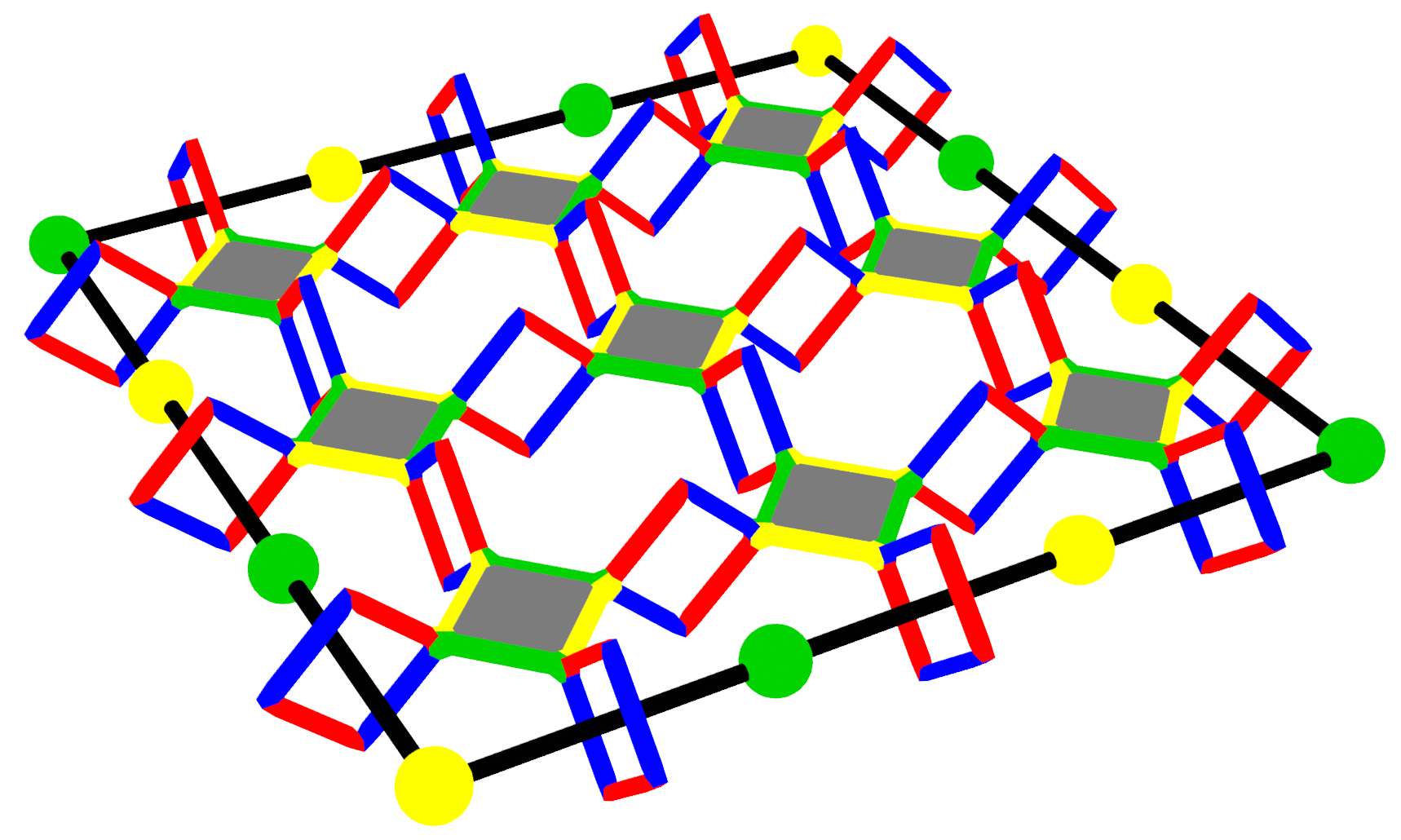}
        \end{subfigure}
        \subcaption{}
        \label{subfig:s_error_3d_a}
    \end{subfigure}
    \hspace{-1cm}
    \begin{subfigure}{.6\textwidth}
        \centering
        \includegraphics[width=.8\textwidth]{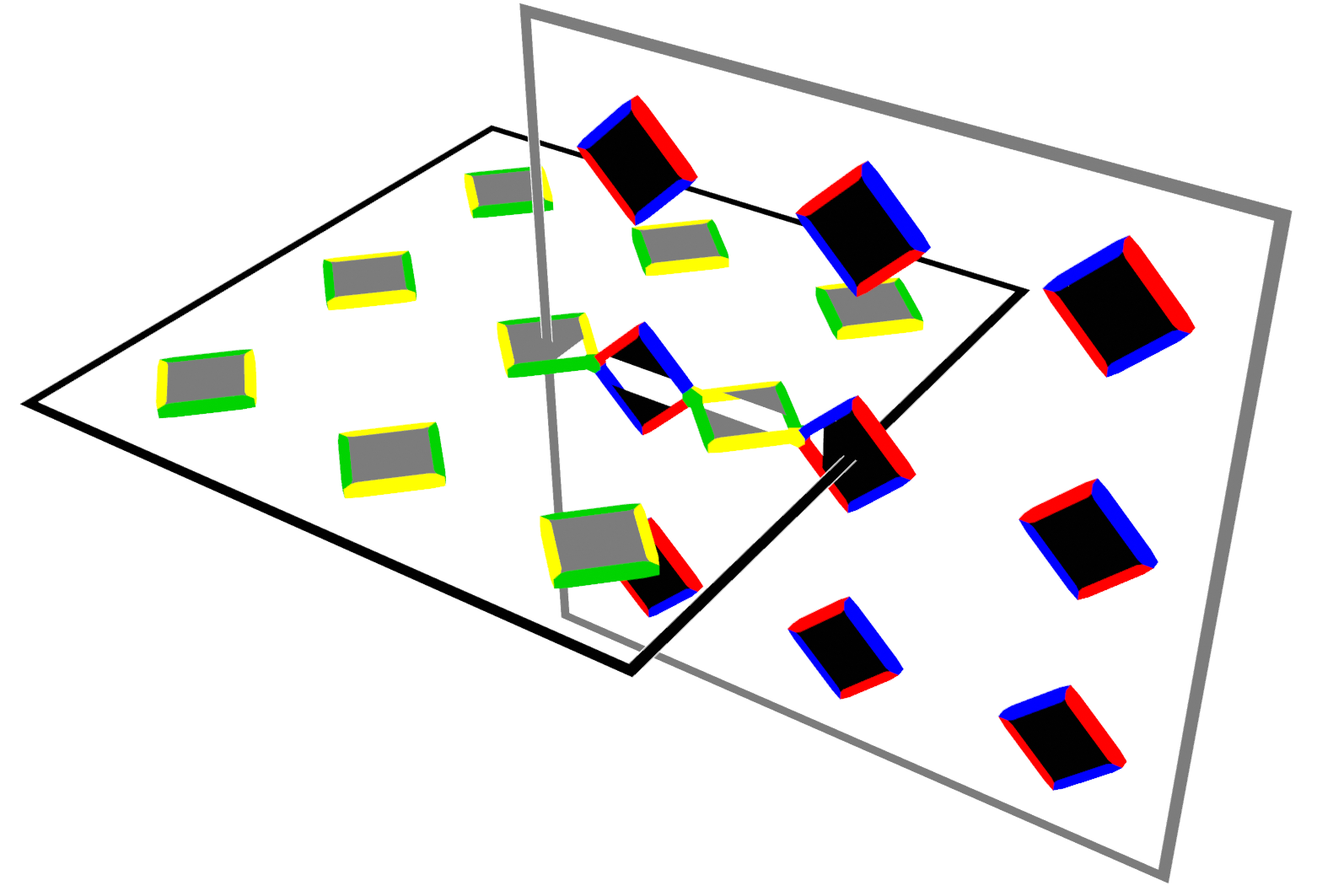}
        \subcaption{}
        \label{subfig:s_error_3d_b}
    \end{subfigure}
    \caption{(Colour) (a:above) G (left) and Y (right) cells of a 3D colour code. These cells meet at an RB face. (a:below) A membrane of $X$ errors in the 3D colour code. The error is supported on qubits on the vertices of YG faces (grey) and detected by $Z$ stabilisers on RB faces on the membrane's boundary. The resulting syndrome is shown by the black loop, which passes through the centres of the violated $Z$ stabilisers. The two colours of dots on this syndrome mark the places where it passes through the centre of a G or Y cell. In order to improve visual clarity full cells are not shown. (b) Two intersecting membranes of $X$ errors with linked syndromes in the 3D colour code. One is defined on YG faces and the other on RB faces. This error is the product of the two individual membranes so errors on the intersection cancel. One face of the YG membrane supports a $Z$ stabiliser that detects the RB membrane and vice versa.}
    \label{fig:s_error_3d}
\end{figure*}

We can initially consider an $X$ error membrane defined on the vertices of a set of faces of colour $\kappa_1\kappa_2$ (by which we mean they are formed from edges of colour $\kappa_1$ and $\kappa_2$). This error will be detected by $Z$ stabiliser generators on faces of colour $\kappa_3\kappa_4$ at the boundary of the membrane. An example is shown in \cref{subfig:s_error_3d_a}. When we apply the $\overline{T}$ gate described previously we will create $S$ errors wherever we apply $T$ and $S^\dag$ errors wherever we apply $T^\dag$. 

Like the 2D colour code, the 3D colour code admits a transversal $S$ gate. This gate can be implemented by a membrane of $S$ and $S^\dag$ (using the same colouring as the transversal $T$) with its edges at the boundaries of the code\footnote{This membrane of qubits also supports a logical $X$ operator. We can see that this configuration of $S$ and $S^\dag$ should implement transversal $S$ because it will be created by applying transversal $T$ to a code in the logical $\ket*{1}$ state, and in the logical space we should have $\overline{T}\overline{X}\overline{T}^\dag = e^{-i\pi/4}\overline{S}\overline{X}$}. As with the 2D case, an $S$ error membrane such as the one in \cref{subfig:s_error_3d_a} should only be detected by stabilisers at its boundary (i.e. by $X$ stabilisers on cells that the syndrome loop passes through) because it is only locally distinguishable from the logical $S$ operator in this region.

The results of \cite{bombin_transversal_2018} agree with \cite{yoshida_topological_2015} for this case, i.e. we expect measurement outcomes of $+1$ from all stabilisers except for $X$ stabilisers on the membrane boundary, which we expect to return random outcomes but with an even parity of $-1$s for each colour. However, more complex errors are possible in the 3D colour code and these are where the results diverge from the 2D case. 

Consider a pair of intersecting membranes with linked syndromes as in \cref{subfig:s_error_3d_b}. This error is the product of two $X$ error membranes of the form discussed above, and so one might expect that application of transversal $T$ and measurement of the $X$ stabilisers on the boundaries of these membranes would once again give random outcomes with an even parity of violated stabilisers of each colour. However, what is shown in \cite{bombin_transversal_2018} is that we actually observe an odd number of violated stabilisers of each colour. This is consistent with a distribution of $Z$ errors as described previously plus an additional $Z$ error string running between the two membrane boundaries (such an error string anticommutes with a G and Y cell on one boundary and an R and a B cell on the other boundary). This is termed a ``linking charge'' of the two membranes, since in the topological phase perspective of the 3D colour code the charge distributions on the boundaries of the individual membranes are no longer independent. Previously each boundary was charge-neutral overall, whereas now the distribution for the pair of membranes is charge neutral but the distributions on individual membrane boundaries are not. 

In the next section we provide a proof that these same phenomena occur in the 3D surface code, and explain their origin in depth. In \cref{appendix:cc} we show that our proof technique recovers the results of \cite{bombin_transversal_2018} when applied to the 3D colour code and also examine the cases of $CS$ and $CCZ$ applied between multiple colour codes, so readers wishing to understand the colour code case in more detail can consult that appendix (or alternatively \cite{brown_kesselring_unreleased}) for a different perspective on the problem. 

\section{Clifford Errors in the 3D Surface Code}
\label{section:sc}
\subsection{Single Error Membrane in Cleanable Code Regions}

\begin{figure}[H]
    \centering
    \includegraphics[width=.45\textwidth]{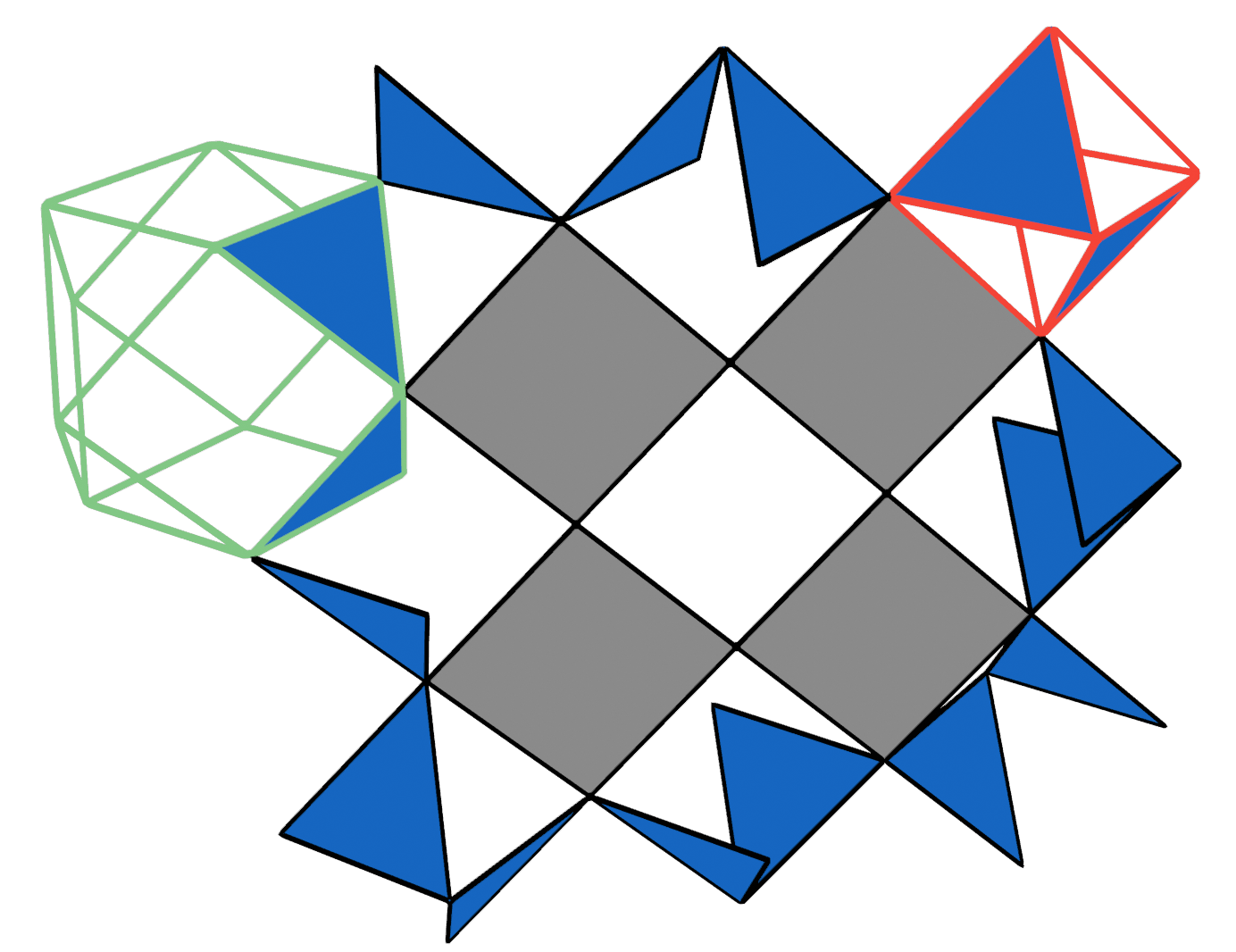}
    \caption{(Colour) An $X$ error membrane in one of the cuboctahedral 3D surface codes that can be defined using the rectified lattice. The error is supported on qubits placed on vertices of the grey faces and is detected by $Z$ stabilisers on the blue faces. These faces are part of cells that support $X$ stabilisers in the other two codes (two examples shown, octahedral for one code and cuboctahedral for the other).}
    \label{fig:cz_error}
\end{figure}

We now consider three 3D surface codes defined on a rectified lattice as in the previous section, and therefore admitting a transversal $CCZ$. We use notation where $X_{\alpha}^c$ means $X$ operators on qubits from code $c \in \{1,2,3\}$ at vertices in the set $\alpha$. We start with a single a membranelike operator $X_\alpha^1$ detected by $Z$ stabilisers of code 1, which are faces of cells in codes 2 and 3 as in \cref{fig:cz_error}. We assume that this membrane exists in a cleanable region~\citep{bravyi2009} of the code, i.e.\ for any given logical Pauli operator of the code we can find an implementation of this operator that has trivial intersection with the membrane. Using the commutation relation $(CCZ)(X \otimes I \otimes I) = (X \otimes CZ)(CCZ)$ we see that applying transversal $CCZ$ in the presence of this error has the effect 

\begin{equation}
    \overline{CCZ} X_\alpha^1 \ket*{\overline{\psi}} = X_\alpha^1 CZ_\alpha^{23} \ket*{\overline{\psi'}}
\end{equation}

\noindent where $\ket*{\overline{\psi}}$ and $\ket*{\overline{\psi'}}$ are states in the codespace of the three codes. $CZ_\alpha^{23}$ is a Clifford error analagous to the $S$ error membrane we observed in the 3D colour code. As with that error, this $CZ$ error becomes a logical operator if applied to the full 

support of a logical $X$ operator (as $\overline{CCZ}$ should preserve the codespace if $X_\alpha^1$ was a logical $X$ operator rather than an error). We therefore expect that, once again, this error should only be detected by stabilisers on the boundaries of the error membrane. 

In order to consider the effect of $CZ_\alpha^{23}$ on the codestate $\ket*{\overline{\psi'}}$ we can consider its effect individually on basis states. The state $\ket*{\overline{000}}$ can be written as 

\begin{equation}
    \label{eq:css_codestate}
    \ket*{\overline{000}} = \frac{1}{\sqrt{n}}\sum_{ijk} X_{\beta_i}^1X_{\beta_j}^2X_{\beta_k}^3\ket*{\bm{0}}
\end{equation}

\noindent where $X_{\beta}^m$ are $X$ stabilisers of code $m$ and $\ket*{\bm{0}}$ is the all-zeros state of the qubits of all three codes. Other basis states can be written in a similar way by replacing these $X$ stabilisers with products of $X$ stabilisers and a logical $X$ operator for a given code. However, as we are currently considering an error membrane in a cleanable region of the code we can always choose these logical operators such that they have trivial intersection with the membrane and thus the analysis of any basis state is equivalent to the analysis for $\ket*{\overline{000}}$ in this case. We can apply $CZ_\alpha^{23}$ to this state and use the commutation relation $(CZ)(X \otimes I) = (X \otimes Z)(CZ)$ to find

\begin{equation}
    \begin{split}
        CZ_\alpha^{23}\ket*{\overline{000}}  
        &= \frac{1}{\sqrt{n}}\sum_{ijk}CZ_\alpha^{23}X_{\beta_i}^1X_{\beta_j}^2X_{\beta_k}^3\ket*{\bm{0}}\\
        &= \frac{1}{\sqrt{n}}\sum_{ijk} X_{\beta_i}^1X_{\beta_j}^2Z_{\alpha \cap \beta_j}^3X_{\beta_k}^3Z_{\alpha \cap \beta_k}^2 \ket*{\bm{0}}
    \end{split}
\end{equation}

\noindent where we have used that $CZ_\alpha^{23}$ acts trivially on $\ket*{\bm{0}}$. We can then commute the $Z$ terms to the right and absorb them into $\ket*{\bm{0}}$ to obtain

\begin{equation}
    \label{eq:cz_codestate}
    \begin{split}
        \ket*{\phi} &= CZ_\alpha^{23}\ket*{\overline{000}} \\
                    &= \frac{1}{\sqrt{n}}\sum_{ijk} (-1)^{|\alpha \cap \beta_j \cap \beta_k|} X_{\beta_i}^1X_{\beta_j}^2X_{\beta_k}^3\ket*{\bm{0}}.
    \end{split}
\end{equation}

In order to investigate the possible measurement outcomes of a specific stabiliser $X_{\beta_l}^q$ we can split the state into $\ket*{\phi} = a\ket*{\phi_l^+} + b\ket*{\phi_l^-}$ where $X_{\beta_l}^q\ket*{\phi_l^+} = \ket*{\phi_l^+}$ and $X_{\beta_l}^q\ket*{\phi_l^-} = - \ket*{\phi_l^-}$. Note that (neglecting normalisation) $\ket*{\phi_l^+}$ will be a sum of pairs

\begin{equation}
    \label{eq:sc_phi_plus}
    \ket*{\phi_l^+} = \sum_x (X_{\beta_x}^{123} + X_{\beta_l}^q X_{\beta_x}^{123})\ket*{\bm{0}}
\end{equation}

\noindent while $\ket*{\phi_l^-}$ will be a sum 

\begin{equation}
    \label{eq:sc_phi_minus}
    \ket*{\phi_l^-} = \sum_x (X_{\beta_x}^{123} - X_{\beta_l}^q X_{\beta_x}^{123})\ket*{\bm{0}}
\end{equation}

\noindent where $X_{\beta_x}^{123} = X_{\beta_i}^1X_{\beta_j}^2X_{\beta_k}^3$ is a product of stabilisers from all three codes. The case where $q=1$ is trivial because the $CZ$ error is not supported in this code, and the cases where $q=2$ and $q=3$ are identical because of the symmetry of the $CZ$ operator. It is therefore sufficient to consider only the case where $q=2$. Then the pairs in \cref{eq:sc_phi_plus} correspond to cases where $(-1)^{|\alpha \cap \beta_j \cap \beta_k|} = (-1)^{|\alpha \cap (\beta_j + \beta_l) \cap \beta_k|}$ which implies $|\alpha \cap \beta_l \cap \beta_k|$ is even ($\beta_j + \beta_l$ is the pointwise addition of these sets modulo 2). On the other hand, pairs in \cref{eq:sc_phi_minus} correspond to cases where $(-1)^{|\alpha \cap \beta_j \cap \beta_k|} = -(-1)^{|\alpha \cap (\beta_j +\beta_l) \cap \beta_k|}$ meaning $|\alpha \cap \beta_l \cap \beta_k|$ must be odd in this case. We now consider three relevant types of stabiliser: 

\begin{itemize}
    \item Stabilisers not on the membrane boundary: Recall that if $X_\alpha^1$ was a logical $X$ operator of code 1 then $CZ_\alpha^{23}$ should be a logical $CZ$ of codes 2 and 3 and so must preserve the stabiliser groups of these codes. This means that for any cell of code 2 not on the boundary of the $CZ$ membrane, the intersection of this cell with the $CZ$ membrane must be the support of a $Z$ stabiliser of code 3 or $X$ stabilisers in one code would be mapped to $Z$ errors in the other by a transversal application of $CZ$. This means that $\alpha \cap \beta_l$ is the support of a $Z$ stabiliser of code 3 and since $\beta_k$ is the support of an $X$ stabiliser of code 3, $|\alpha \cap \beta_l \cap \beta_k|$ must be even for all such $\beta_l$ and $\beta_k$. Therefore $\ket*{\phi} = \ket*{\phi_l^+}$ and we are in a $+1$ eigenstate of these stabilisers.
    \item Stabilisers generators (cells) on the membrane boundary: The error $X_\alpha^1$ is detected by $Z$ stabilisers supported on the faces of these cells as in \cref{fig:cz_error}. Each such face is the intersection between a pair of generators $X_{\beta_l}^2$ and $X_{\beta_k}^3$ and so $|\alpha \cap \beta_l \cap \beta_k|$ must be odd or a $Z$ stabiliser on this face would not detect the error. There are two such faces for every generator on the membrane boundary (because the syndrome is a loop) and so every $X_{\beta_l}^2$ has two $X_{\beta_k}^3$ neighbours for which $|\alpha \cap \beta_l \cap \beta_k|$ is odd, which we will refer to as the ``boundary neighbours'' of $X_{\beta_l}^2$. The boundary neighbours are disjoint, so if $X_{\beta_k}^3$ is instead the product of the boundary neighbours then $|\alpha \cap \beta_l \cap \beta_k|$ is even. Thus, the terms in $\ket*{\phi_l^+}$ are those for which $X_{\beta_x}^{123}$ contains neither or both of the boundary neighbours of $X_{\beta_l}^2$, whereas the terms in $\ket*{\phi_l^-}$ are those where $X_{\beta_x}^{123}$ contains a single one of these neighbours. The overall superposition contains an equal number of each type of term so $\ket*{\phi} = \frac{1}{\sqrt{2}} (\ket*{\phi_l^+} + \ket*{\phi_l^-})$ and we measure a random $\pm1$ outcome from this stabiliser. 
    \item The product of all stabiliser generators on the membrane boundary: For any generator $X_{\beta_j}^2$ on the membrane boundary the intersection $|\alpha \cap \beta_j \cap \beta_k|$ with a neighbouring generator $X_{\beta_k}^3$ is odd (as discussed above). This means a $Z$ operator $Z_{\alpha \cap \beta_j}^3$ anticommutes with these $X_{\beta_k}^3$, and for each such $X_{\beta_k}^3$ there are two generators $X_{\beta_j}^2$ that have this property, so if $X_{\beta_l}^2$ is the product of all generators from code 2 on the membrane boundary then $Z_{\alpha \cap \beta_l}^3$ is a stabiliser. This means that $|\alpha \cap \beta_l \cap \beta_k|$ is even for all $X_{\beta_k}^3$ and so $\ket*{\phi} = \ket*{\phi_l^+}$ for this choice of $X_{\beta_l}^2$.
\end{itemize}

In summary this gives us an analogous result to what was observed for an isolated membrane in the colour code. Stabilisers not on the boundary always give $+1$. Stabiliser generators on the boundary give $\pm1$ randomly but we must get an even number of $-1$ stabilisers in any given code. 

\subsection{Linked Error Membranes in Cleanable Code Regions}

Consider the case where we have $X$ error membranes in codes 1 and 2 

\begin{equation}
    \overline{CCZ} X_\alpha^1X_\gamma^2 \ket*{\overline{\psi}}
\end{equation}

\noindent such that $\alpha \cap \gamma$ is non-empty. Then commuting the $CCZ$ to the right we have

\begin{equation}
    \label{eq:sc_link}
    X_\alpha^1 CZ_\alpha^{23} X_\gamma^2 CZ_\gamma^{13} \ket*{\overline{\psi'}} = X_\alpha^1 X_\gamma^2 Z_{\alpha \cap \gamma}^3 CZ_\alpha^{23} CZ_\gamma^{13} \ket*{\overline{\psi}}
\end{equation}

So we now have some $CZ$ errors acting on a codestate as before, but we also have a $Z$ string $Z_{\alpha \cap \gamma}^3$ on the intersection of these membranes in code 3. This is the surface code equivalent of the linking charge string described in \citep{bombin_transversal_2018}. Thus we expect that in code 1 we get random outcomes from stabilisers on the boundary of $CZ_\gamma^{13}$ and $+1$ outcomes from all others, with the total number of $-1$s even. In code 2 we expect the same thing on the boundary of $CZ_\alpha^{23}$. In code 3 we expect random outcomes from stabilisers on the boundaries of both membranes, and also expect an odd parity of $-1$ stabilisers on each boundary due to the linking charge string. 

The simplicity of this statement stands in stark contrast to the complexity of the original proof of this phemonenon in the case of the colour code as presented in \citep{bombin_transversal_2018}, to the extent that it may seem unremarkable to readers who are not familiar with that work. Additionally, it is not only the mathematical origin of linking charge that is clearer in the surface code but also its physical significance. The transversal $CCZ$ in this code is acheived because the intersection of logical $X$ operators from any pair of codes is the support of a logical $Z$ operator in the third and so the logical action $\overline{CCZ}_{123} \overline{X}_1 \overline{X}_2 \overline{CCZ}_{123} = \overline{X}_1 \overline{X}_2 \overline{Z}_3 \overline{CZ}_{23} \overline{CZ}_{13}$ is correctly implemented. Linking charge in this case is just another example of this creation of $Z$ strings on $X$ membrane intersections. This is consistent with claims made in \citep{bombin_transversal_2018} that linking charge is important for the correct function of the transversal $T$ gate in the 3D colour code.



\subsection{Error Membranes in Non-Cleanable Regions}

Finally we address the case of error membranes in non-cleanable regions of the code. Consider 

\begin{equation}
    \overline{CCZ} X_\alpha^1 \ket*{\overline{\psi}} = X_\alpha^1 CZ_\alpha^{23} \ket*{\overline{\psi'}}
\end{equation}

\noindent where $CZ_\alpha^{23}$ now contains the support of a logical $Z$ operator in code 2 or 3. We choose code 2 but this choice is not important. This means that any implementation of logical $X$ in code 2 must have nontrivial intersection with $X_\alpha^1$ and this changes our analysis for some codewords. For $\ket*{\overline{010}}$ we have

\begin{equation}
    \begin{split}
        &CZ_\alpha^{23} \ket*{\overline{010}} = \frac{1}{\sqrt{n}} \sum_{ijk} CZ_\alpha^{23} X_{\beta_i}^1\overline{X}_L^2X_{\beta_j}^2X_{\beta_k}^3 \ket*{\bm{0}} \\
                                            &= \frac{1}{\sqrt{n}} \sum_{ijk} X_{\beta_i}^1\overline{X}_L^2 Z_{\alpha \cap L}^3  X_{\beta_j}^2Z_{\alpha \cap \beta_j}^3X_{\beta_k}^3Z_{\alpha \cap \beta_k}^2 \ket*{\bm{0}} \\
                                            &= \frac{Z_{\alpha \cap L}^3}{\sqrt{n}} \sum_{ijk} (-1)^{\alpha \cap \beta_j \cap \beta_k} X_{\beta_i}^1 \overline{X}_L^2 X_{\beta_j}^2 X_{\beta_k}^3 \ket*{\bm{0}}
    \end{split}
\end{equation}

Where $\overline{X}_L^2$ is a logical $X$ implementation for code 2. We therefore create a global $Z$ error on the intersection of the error membrane support $\alpha$ and the logical operator support $L$. Both $\alpha$ and $L$ are membranes so we can always choose $L$ such that $\alpha \cap L$ is a string. This does not affect observed syndromes (because $Z_{\alpha \cap L}^3$ runs from one side of the membrane to the other and so anticommutes with a pair of stabilisers on the membrane boundary) and also does not affect the encoded logical information unless $\alpha \cap L$ is the support of a logical $Z$ operator for code 3. Therefore we only get a logical error from applying $\overline{CCZ}$ to $X_\alpha$ if $\alpha$ contains the support of logical operators of both code 2 and code 3.

\section{Effects of Clifford Errors on Code Performance}
\label{section:numerics}
\begin{table*}
    \centering
    \begin{tabular}{c|ccc}
         & Octahedral & Cuboctahedral 1 & Cuboctahedral 2 \\
        \hline
        \\[-5pt]
        $\overline Z$ error threshold (no $CCZ$) & $1.788(8)\%$ & $0.525(8)\%$ & $0.519(8)\%$ \\[5pt]
        $\overline Z$ error threshold (noise before $CCZ$) & $1.065(9)\%$ & $0.278(9)\%$ & $0.276(9)\%$
        \\[5pt]
        $\overline Z$ error threshold (noise after $CCZ$) & $1.747(14)\%$ & $0.542(8)\%$ & $0.533(7)\%$ \\[5pt]
        $\overline X$ error threshold (no $CCZ$) & $2.693(12)\%$ & $2.864(22)\%$ & $2.879(18)\%$ \\[5pt]
        $\overline X$ error threshold (noise before $CCZ$) & $2.630(14)\%$ & $2.750(26)\%$ & $2.750(31)\%$
        \\[5pt]
        $\overline X$ error threshold (noise after $CCZ$) & $2.694(12)\%$ & $2.874(21)\%$ & $2.886(19)\%$ \\[5pt]
        \end{tabular}
    \caption{Error threshold estimates obtained by fitting the data in \cref{fig:threshold_plots_noise_after,fig:threshold_plots_noise_before} to the ansatz in \cref{eq:fit_ansatz}. The indicated errors were derived via bootstrap resampling.}
    \label{tab:thresholds}
\end{table*}

To investigate the effects of Clifford errors on the performance of an error correcting code we simulate a single-shot magic state preparation procedure that makes use of the surface code $CCZ$. Previous work has simulated the effect of Clifford errors arising from the application of transversal $T$ in the colour code~\cite{beverland_cost_2021} but did not include the contribution due to linking charge (as it cannot be modelled locally in that setting). On the other hand, that work used a circuit noise model whereas we use a simpler phenomenological noise model. All code used for our simulations can be found at~\cite{source_code}.

Our simulated procedure consists of the following steps:

\begin{itemize}
    \item Prepare the three codes in the logical $\ket{\overline{+}}$ state by preparing all physical qubits in $\ket{+}$, measuring all $Z$ stabilisers that correspond to faces of the lattice and applying a correction. This can be done in constant time as the 3D surface code allows for single-shot decoding of $X$ errors. 
    \item Apply transversal $CCZ$ to the three codes to create the magic state $CCZ\ket{{+}{+}{+}}$.
    \item Perform a 3D $\rightarrow$ 2D dimension jump in each code, such that we end with three entangled 2D surface codes in the state $CCZ\ket{{+}{+}{+}}$. This can also be done in constant time at the cost of reduced code performance. 
\end{itemize}

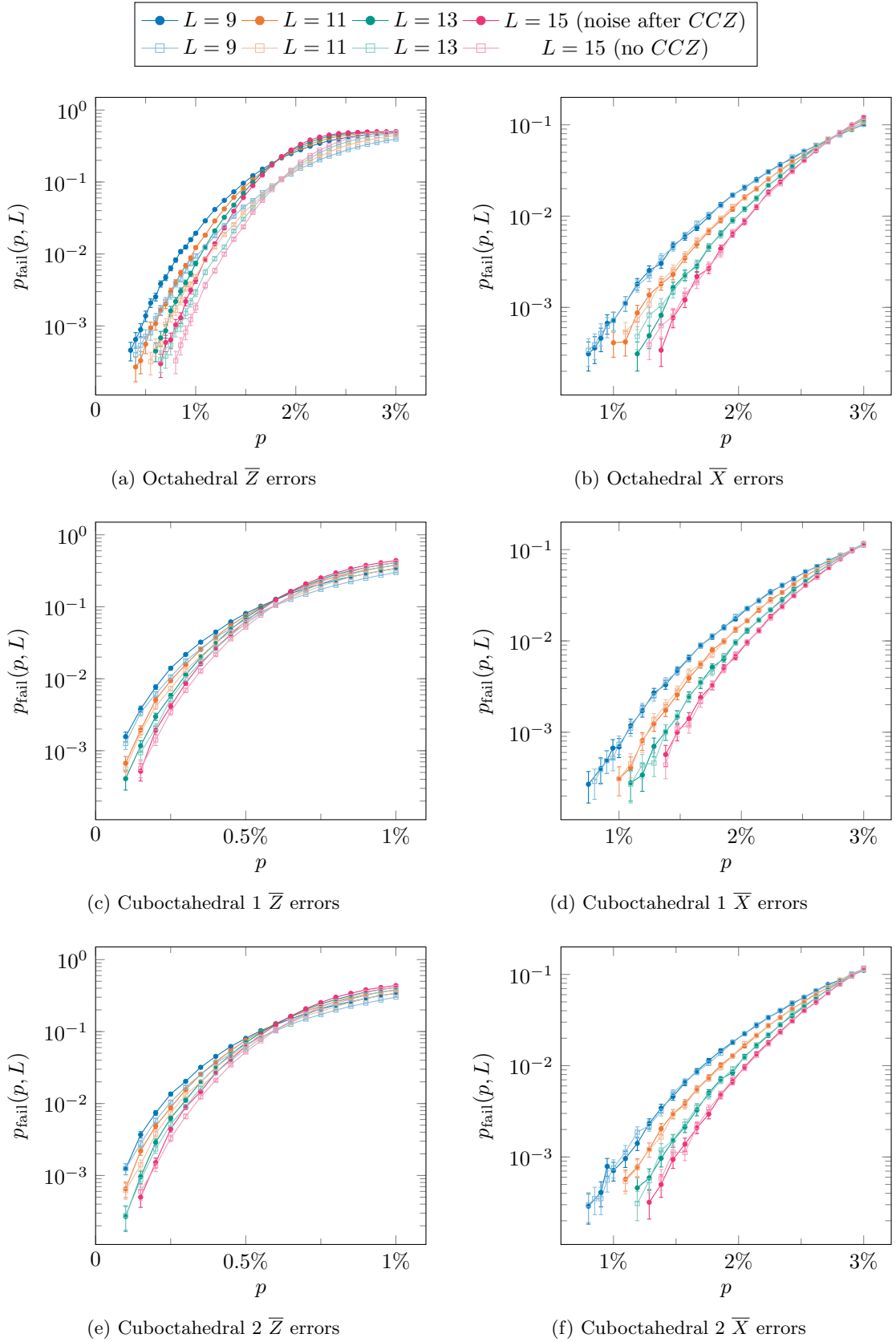
\begin{figure*}
    \centering
    \begin{tikzpicture} 
        \begin{axis}[%
        hide axis,
        xmin=10,
        xmax=50,
        ymin=0,
        ymax=0.4,
        legend columns=4
        ]
        \addlegendimage{cb-blue,mark=*}
        \addlegendentry{$L=9$};
        \addlegendimage{cb-orange,mark=*}
        \addlegendentry{$L=11$};
        \addlegendimage{cb-teal,mark=*}
        \addlegendentry{$L=13$};
        \addlegendimage{cb-magenta,mark=*}
        \addlegendentry{$L=15$ (noise after $CCZ$)};
        \addlegendimage{cb-blue!50!white,mark=square}
        \addlegendentry{$L=9$};
        \addlegendimage{cb-orange!50!white,mark=square}
        \addlegendentry{$L=11$};
        \addlegendimage{cb-teal!50!white,mark=square}
        \addlegendentry{$L=13$};
        \addlegendimage{cb-magenta!50!white,mark=square}
        \addlegendentry{$L=15$ (no $CCZ$)};
        \end{axis}
    \end{tikzpicture}
    \par\bigskip
    \begin{subfigure}[t]{0.45\textwidth}
    \centering
        \begin{tikzpicture}
            \begin{axis}[
                xlabel={$p$},
                ylabel={$p_{\mathrm{fail}}(p,L)$},
                ymode=log,
                width=7cm,
                height=6.5cm,
                ymin=0.00011,
                ymax=1.5,
                xmin=0,
                xtick={0,0.01,0.02,0.03},
                minor xtick={0.005,0.015,0.025},
                xticklabels={$0$,$1\%$,$2\%$,$3\%$},
                legend style={
                    at={(0.95,0.05)},
                    anchor=south east
                }
            ]
            \addplot[
                color=cb-blue,
                mark=*,
                mark size=1,
                error bars/.cd,
                y dir=both,
                y explicit
            ] table [x=p, y=pfail_C_Z, y error=err_C_Z, col sep=comma] {data/data_Z_cubic_L=9_ccz=1.csv};
            \addplot[
                color=cb-orange,
                mark=*,
                mark size=1,
                error bars/.cd,
                y dir=both,
                y explicit
            ] table [x=p, y=pfail_C_Z, y error=err_C_Z, col sep=comma] {data/data_Z_cubic_L=11_ccz=1.csv};
            \addplot[
                color=cb-teal,
                mark=*,
                mark size=1,
                error bars/.cd,
                y dir=both,
                y explicit
            ] table [x=p, y=pfail_C_Z, y error=err_C_Z, col sep=comma] {data/data_Z_cubic_L=13_ccz=1.csv};
            \addplot[
                color=cb-magenta,
                mark=*,
                mark size=1,
                error bars/.cd,
                y dir=both,
                y explicit
            ] table [x=p, y=pfail_C_Z, y error=err_C_Z, col sep=comma] {data/data_Z_cubic_L=15_ccz=1.csv};
            \addplot[
                color=cb-blue!50!white,
                mark=square,
                mark size=1,
                error bars/.cd,
                y dir=both,
                y explicit
            ] table [x=p, y=pfail_C_Z, y error=err_C_Z, col sep=comma] {data/data_Z_cubic_L=9_ccz=0.csv};
            \addplot[
                color=cb-orange!50!white,
                mark=square,
                mark size=1,
                error bars/.cd,
                y dir=both,
                y explicit
            ] table [x=p, y=pfail_C_Z, y error=err_C_Z, col sep=comma] {data/data_Z_cubic_L=11_ccz=0.csv};
            \addplot[
                color=cb-teal!50!white,
                mark=square,
                mark size=1,
                error bars/.cd,
                y dir=both,
                y explicit
            ] table [x=p, y=pfail_C_Z, y error=err_C_Z, col sep=comma] {data/data_Z_cubic_L=13_ccz=0.csv};
            \addplot[
                color=cb-magenta!50!white,
                mark=square,
                mark size=1,
                error bars/.cd,
                y dir=both,
                y explicit
            ] table [x=p, y=pfail_C_Z, y error=err_C_Z, col sep=comma] {data/data_Z_cubic_L=15_ccz=0.csv};
            \end{axis}
        \end{tikzpicture}
        \subcaption{Octahedral $\overline Z$ errors}
        \label{subfig:cubic_pth_z}
    \end{subfigure}
    ~~
    \begin{subfigure}[t]{0.45\textwidth}
    \centering
        \begin{tikzpicture}
            \begin{axis}[
                xlabel={$p$},
                ylabel={$p_{\mathrm{fail}}(p,L)$},
                ymode=log,
                width=7cm,
                height=6.5cm,
                ymin=0.00011,
                ymax=0.2,
                xtick={0.01,0.02,0.03},
                minor xtick={0.015, 0.025},
                xticklabels={$1\%$,$2\%$,$3\%$},
            ]
            \addplot[
                color=cb-blue,
                mark=*,
                mark size=1,
                error bars/.cd,
                y dir=both,
                y explicit
            ] table [x=p, y=pfail_C_X, y error=err_C_X, col sep=comma] {data/data_X_cubic_L=9_ccz=1.csv};
            \addplot[
                color=cb-orange,
                mark=*,
                mark size=1,
                error bars/.cd,
                y dir=both,
                y explicit
            ] table [x=p, y=pfail_C_X, y error=err_C_X, col sep=comma] {data/data_X_cubic_L=11_ccz=1.csv};
            \addplot[
                color=cb-teal,
                mark=*,
                mark size=1,
                error bars/.cd,
                y dir=both,
                y explicit
            ] table [x=p, y=pfail_C_X, y error=err_C_X, col sep=comma] {data/data_X_cubic_L=13_ccz=1.csv};
            \addplot[
                color=cb-magenta,
                mark=*,
                mark size=1,
                error bars/.cd,
                y dir=both,
                y explicit
            ] table [x=p, y=pfail_C_X, y error=err_C_X, col sep=comma] {data/data_X_cubic_L=15_ccz=1.csv};
            \addplot[
                color=cb-blue!50!white,
                mark=square,
                mark size=1,
                error bars/.cd,
                y dir=both,
                y explicit
            ] table [x=p, y=pfail_C_X, y error=err_C_X, col sep=comma] {data/data_X_cubic_L=9_ccz=0.csv};
            \addplot[
                color=cb-orange!50!white,
                mark=square,
                mark size=1,
                error bars/.cd,
                y dir=both,
                y explicit
            ] table [x=p, y=pfail_C_X, y error=err_C_X, col sep=comma] {data/data_X_cubic_L=11_ccz=0.csv};
            \addplot[
                color=cb-teal!50!white,
                mark=square,
                mark size=1,
                error bars/.cd,
                y dir=both,
                y explicit
            ] table [x=p, y=pfail_C_X, y error=err_C_X, col sep=comma] {data/data_X_cubic_L=13_ccz=0.csv};
            \addplot[
                color=cb-magenta!50!white,
                mark=square,
                mark size=1,
                error bars/.cd,
                y dir=both,
                y explicit
            ] table [x=p, y=pfail_C_X, y error=err_C_X, col sep=comma] {data/data_X_cubic_L=15_ccz=0.csv};
            \end{axis}
        \end{tikzpicture}
        \subcaption{Octahedral $\overline X$ errors}
        \label{subfig:cubic_pth_x}
    \end{subfigure}
    \par\bigskip
    \begin{subfigure}[t]{0.45\textwidth}
    \centering
        \begin{tikzpicture}
            \begin{axis}[
                xlabel={$p$},
                ylabel={$p_{\mathrm{fail}}(p,L)$},
                ymode=log,
                width=7cm,
                height=6.5cm,
                ymin=0.00011,
                ymax=1.5,
                xmin=0,
                xtick={0,0.005,0.01},
                minor xtick={0.0025,0.0075},
                xticklabels={$0$,$0.5\%$,$1\%$},
            ]
            \addplot[
                color=cb-blue,
                mark=*,
                mark size=1,
                error bars/.cd,
                y dir=both,
                y explicit
            ] table [x=p, y=pfail_R1_Z, y error=err_R1_Z, col sep=comma] {data/data_Z_rhombic1_L=9_ccz=1.csv};
            \addplot[
                color=cb-orange,
                mark=*,
                mark size=1,
                error bars/.cd,
                y dir=both,
                y explicit
            ] table [x=p, y=pfail_R1_Z, y error=err_R1_Z, col sep=comma] {data/data_Z_rhombic1_L=11_ccz=1.csv};
            \addplot[
                color=cb-teal,
                mark=*,
                mark size=1,
                error bars/.cd,
                y dir=both,
                y explicit
            ] table [x=p, y=pfail_R1_Z, y error=err_R1_Z, col sep=comma] {data/data_Z_rhombic1_L=13_ccz=1.csv};
            \addplot[
                color=cb-magenta,
                mark=*,
                mark size=1,
                error bars/.cd,
                y dir=both,
                y explicit
            ] table [x=p, y=pfail_R1_Z, y error=err_R1_Z, col sep=comma] {data/data_Z_rhombic1_L=15_ccz=1.csv};
            \addplot[
                color=cb-blue!50!white,
                mark=square,
                mark size=1,
                error bars/.cd,
                y dir=both,
                y explicit
            ] table [x=p, y=pfail_R1_Z, y error=err_R1_Z, col sep=comma] {data/data_Z_rhombic1_L=9_ccz=0.csv};
            \addplot[
                color=cb-orange!50!white,
                mark=square,
                mark size=1,
                error bars/.cd,
                y dir=both,
                y explicit
            ] table [x=p, y=pfail_R1_Z, y error=err_R1_Z, col sep=comma] {data/data_Z_rhombic1_L=11_ccz=0.csv};
            \addplot[
                color=cb-teal!50!white,
                mark=square,
                mark size=1,
                error bars/.cd,
                y dir=both,
                y explicit
            ] table [x=p, y=pfail_R1_Z, y error=err_R1_Z, col sep=comma] {data/data_Z_rhombic1_L=13_ccz=0.csv};
            \addplot[
                color=cb-magenta!50!white,
                mark=square,
                mark size=1,
                error bars/.cd,
                y dir=both,
                y explicit
            ] table [x=p, y=pfail_R1_Z, y error=err_R1_Z, col sep=comma] {data/data_Z_rhombic1_L=15_ccz=0.csv};
            \end{axis}
        \end{tikzpicture}
        \subcaption{Cuboctahedral 1 $\overline Z$ errors}
        \label{subfig:rhombic1_pth_z}
    \end{subfigure}
    ~~
    \begin{subfigure}[t]{0.45\textwidth}
    \centering
      \begin{tikzpicture}
            \begin{axis}[
                xlabel={$p$},
                ylabel={$p_{\mathrm{fail}}(p,L)$},
                ymode=log,
                width=7cm,
                height=6.5cm,
                ymin=0.00011,
                ymax=0.2,
                xtick={0.01,0.02,0.03},
                minor xtick={0.015, 0.025},
                xticklabels={$1\%$,$2\%$,$3\%$},
                legend style={
                    at={(0.5,-0.25)},
                    anchor=north
                },
                legend columns=4,
            ]
            \addplot[
                color=cb-blue,
                mark=*,
                mark size=1,
                error bars/.cd,
                y dir=both,
                y explicit
            ] table [x=p, y=pfail_R1_X, y error=err_R1_X, col sep=comma] {data/data_X_rhombic1_L=9_ccz=1.csv};
            \addplot[
                color=cb-orange,
                mark=*,
                mark size=1,
                error bars/.cd,
                y dir=both,
                y explicit
            ] table [x=p, y=pfail_R1_X, y error=err_R1_X, col sep=comma] {data/data_X_rhombic1_L=11_ccz=1.csv};
            \addplot[
                color=cb-teal,
                mark=*,
                mark size=1,
                error bars/.cd,
                y dir=both,
                y explicit
            ] table [x=p, y=pfail_R1_X, y error=err_R1_X, col sep=comma] {data/data_X_rhombic1_L=13_ccz=1.csv};
            \addplot[
                color=cb-magenta,
                mark=*,
                mark size=1,
                error bars/.cd,
                y dir=both,
                y explicit
            ] table [x=p, y=pfail_R1_X, y error=err_R1_X, col sep=comma] {data/data_X_rhombic1_L=15_ccz=1.csv};
            \addplot[
                color=cb-blue!50!white,
                mark=square,
                mark size=1,
                error bars/.cd,
                y dir=both,
                y explicit
            ] table [x=p, y=pfail_R1_X, y error=err_R1_X, col sep=comma] {data/data_X_rhombic1_L=9_ccz=0.csv};
            \addplot[
                color=cb-orange!50!white,
                mark=square,
                mark size=1,
                error bars/.cd,
                y dir=both,
                y explicit
            ] table [x=p, y=pfail_R1_X, y error=err_R1_X, col sep=comma] {data/data_X_rhombic1_L=11_ccz=0.csv};
            \addplot[
                color=cb-teal!50!white,
                mark=square,
                mark size=1,
                error bars/.cd,
                y dir=both,
                y explicit
            ] table [x=p, y=pfail_R1_X, y error=err_R1_X, col sep=comma] {data/data_X_rhombic1_L=13_ccz=0.csv};
            \addplot[
                color=cb-magenta!50!white,
                mark=square,
                mark size=1,
                error bars/.cd,
                y dir=both,
                y explicit
            ] table [x=p, y=pfail_R1_X, y error=err_R1_X, col sep=comma] {data/data_X_rhombic1_L=15_ccz=0.csv};
            \end{axis}
        \end{tikzpicture}
        \subcaption{Cuboctahedral 1 $\overline X$ errors}
        \label{subfig:rhombic1_pth_x}
    \end{subfigure}
    \par\bigskip
    \begin{subfigure}[t]{0.45\textwidth}
    \centering
        \begin{tikzpicture}
            \begin{axis}[
                xlabel={$p$},
                ylabel={$p_{\mathrm{fail}}(p,L)$},
                ymode=log,
                width=7cm,
                height=6.5cm,
                ymin=0.00011,
                ymax=1.5,
                xmin=0,
                xtick={0,0.005,0.01},
                minor xtick={0.0025,0.0075},
                xticklabels={$0$,$0.5\%$,$1\%$},
            ]
            \addplot[
                color=cb-blue,
                mark=*,
                mark size=1,
                error bars/.cd,
                y dir=both,
                y explicit
            ] table [x=p, y=pfail_R2_Z, y error=err_R2_Z, col sep=comma] {data/data_Z_rhombic2_L=9_ccz=1.csv};
            \addplot[
                color=cb-orange,
                mark=*,
                mark size=1,
                error bars/.cd,
                y dir=both,
                y explicit
            ] table [x=p, y=pfail_R2_Z, y error=err_R2_Z, col sep=comma] {data/data_Z_rhombic2_L=11_ccz=1.csv};
            \addplot[
                color=cb-teal,
                mark=*,
                mark size=1,
                error bars/.cd,
                y dir=both,
                y explicit
            ] table [x=p, y=pfail_R2_Z, y error=err_R2_Z, col sep=comma] {data/data_Z_rhombic2_L=13_ccz=1.csv};
            \addplot[
                color=cb-magenta,
                mark=*,
                mark size=1,
                error bars/.cd,
                y dir=both,
                y explicit
            ] table [x=p, y=pfail_R2_Z, y error=err_R2_Z, col sep=comma] {data/data_Z_rhombic2_L=15_ccz=1.csv};
            \addplot[
                color=cb-blue!50!white,
                mark=square,
                mark size=1,
                error bars/.cd,
                y dir=both,
                y explicit
            ] table [x=p, y=pfail_R2_Z, y error=err_R2_Z, col sep=comma] {data/data_Z_rhombic2_L=9_ccz=0.csv};
            \addplot[
                color=cb-orange!50!white,
                mark=square,
                mark size=1,
                error bars/.cd,
                y dir=both,
                y explicit
            ] table [x=p, y=pfail_R2_Z, y error=err_R2_Z, col sep=comma] {data/data_Z_rhombic2_L=11_ccz=0.csv};
            \addplot[
                color=cb-teal!50!white,
                mark=square,
                mark size=1,
                error bars/.cd,
                y dir=both,
                y explicit
            ] table [x=p, y=pfail_R2_Z, y error=err_R2_Z, col sep=comma] {data/data_Z_rhombic2_L=13_ccz=0.csv};
            \addplot[
                color=cb-magenta!50!white,
                mark=square,
                mark size=1,
                error bars/.cd,
                y dir=both,
                y explicit
            ] table [x=p, y=pfail_R2_Z, y error=err_R2_Z, col sep=comma] {data/data_Z_rhombic2_L=15_ccz=0.csv};
            \end{axis}
        \end{tikzpicture}
        \subcaption{Cuboctahedral 2 $\overline Z$ errors}
        \label{subfig:rhombic2_pth_z}
    \end{subfigure}
    ~~
    \begin{subfigure}[t]{0.45\textwidth}
    \centering
        \begin{tikzpicture}
            \begin{axis}[
                xlabel={$p$},
                ylabel={$p_{\mathrm{fail}}(p,L)$},
                ymode=log,
                width=7cm,
                height=6.5cm,
                ymin=0.00011,
                ymax=0.2,
                xtick={0.01,0.02,0.03},
                minor xtick={0.015, 0.025},
                xticklabels={$1\%$,$2\%$,$3\%$},
            ]
            \addplot[
                color=cb-blue,
                mark=*,
                mark size=1,
                error bars/.cd,
                y dir=both,
                y explicit
            ] table [x=p, y=pfail_R2_X, y error=err_R2_X, col sep=comma] {data/data_X_rhombic2_L=9_ccz=1.csv};
            \addplot[
                color=cb-orange,
                mark=*,
                mark size=1,
                error bars/.cd,
                y dir=both,
                y explicit
            ] table [x=p, y=pfail_R2_X, y error=err_R2_X, col sep=comma] {data/data_X_rhombic2_L=11_ccz=1.csv};
            \addplot[
                color=cb-teal,
                mark=*,
                mark size=1,
                error bars/.cd,
                y dir=both,
                y explicit
            ] table [x=p, y=pfail_R2_X, y error=err_R2_X, col sep=comma] {data/data_X_rhombic2_L=13_ccz=1.csv};
            \addplot[
                color=cb-magenta,
                mark=*,
                mark size=1,
                error bars/.cd,
                y dir=both,
                y explicit
            ] table [x=p, y=pfail_R2_X, y error=err_R2_X, col sep=comma] {data/data_X_rhombic2_L=15_ccz=1.csv};
            \addplot[
                color=cb-blue!50!white,
                mark=square,
                mark size=1,
                error bars/.cd,
                y dir=both,
                y explicit
            ] table [x=p, y=pfail_R2_X, y error=err_R2_X, col sep=comma] {data/data_X_rhombic2_L=9_ccz=0.csv};
            \addplot[
                color=cb-orange!50!white,
                mark=square,
                mark size=1,
                error bars/.cd,
                y dir=both,
                y explicit
            ] table [x=p, y=pfail_R2_X, y error=err_R2_X, col sep=comma] {data/data_X_rhombic2_L=11_ccz=0.csv};
            \addplot[
                color=cb-teal!50!white,
                mark=square,
                mark size=1,
                error bars/.cd,
                y dir=both,
                y explicit
            ] table [x=p, y=pfail_R2_X, y error=err_R2_X, col sep=comma] {data/data_X_rhombic2_L=13_ccz=0.csv};
            \addplot[
                color=cb-magenta!50!white,
                mark=square,
                mark size=1,
                error bars/.cd,
                y dir=both,
                y explicit
            ] table [x=p, y=pfail_R2_X, y error=err_R2_X, col sep=comma] {data/data_X_rhombic2_L=15_ccz=0.csv};
            \end{axis}
        \end{tikzpicture}
        \subcaption{Cuboctahedral 2 $\overline X$ errors}
        \label{subfig:rhombic2_pth_x}
    \end{subfigure}
    \caption{(Colour)
        For each of the three codes, we show a plot of the logical error rate $p_{\mathrm{fail}}$ (separately for $\overline X$ and $\overline Z$ errors) as a function of the depolarising/measurement error rate $p$ for various lattice sizes $L$ (with and without the $CCZ$ gate).
        For these simulations, the depolarising noise channel was applied \emph{after} the $CCZ$ gate.
        The error bars show the Agresti-Coull 95\% confidence intervals~\cite{agresti1998,dasgupta2001}.
    }
    \label{fig:threshold_plots_noise_after}
\end{figure*}

\begin{figure*}
    \centering
    \begin{tikzpicture} 
        \begin{axis}[%
        hide axis,
        xmin=10,
        xmax=50,
        ymin=0,
        ymax=0.4,
        legend columns=4
        ]
        \addlegendimage{cb-blue,mark=*}
        \addlegendentry{$L=9$};
        \addlegendimage{cb-orange,mark=*}
        \addlegendentry{$L=11$};
        \addlegendimage{cb-teal,mark=*}
        \addlegendentry{$L=13$};
        \addlegendimage{cb-magenta,mark=*}
        \addlegendentry{$L=15$ (noise before $CCZ$)};
        \addlegendimage{cb-blue!50!white,mark=square}
        \addlegendentry{$L=9$};
        \addlegendimage{cb-orange!50!white,mark=square}
        \addlegendentry{$L=11$};
        \addlegendimage{cb-teal!50!white,mark=square}
        \addlegendentry{$L=13$};
        \addlegendimage{cb-magenta!50!white,mark=square}
        \addlegendentry{$L=15$ (no $CCZ$)};
        \end{axis}
    \end{tikzpicture}
    \par\bigskip
    \begin{subfigure}[t]{0.45\textwidth}
    \centering
        \begin{tikzpicture}
            \begin{axis}[
                xlabel={$p$},
                ylabel={$p_{\mathrm{fail}}(p,L)$},
                ymode=log,
                width=7cm,
                height=6.5cm,
                ymin=0.00011,
                ymax=1.5,
                xmin=0,
                xtick={0,0.01,0.02,0.03},
                minor xtick={0.005,0.015,0.025},
                xticklabels={$0$,$1\%$,$2\%$,$3\%$},
                legend style={
                    at={(0.95,0.05)},
                    anchor=south east
                }
            ]
            \addplot[
                color=cb-blue,
                mark=*,
                mark size=1,
                error bars/.cd,
                y dir=both,
                y explicit
            ] table [x=p, y=pfail_C_Z, y error=err_C_Z, col sep=comma] {data/data_Z_cubic_L=9_ccz=1_nb.csv};
            \addplot[
                color=cb-orange,
                mark=*,
                mark size=1,
                error bars/.cd,
                y dir=both,
                y explicit
            ] table [x=p, y=pfail_C_Z, y error=err_C_Z, col sep=comma] {data/data_Z_cubic_L=11_ccz=1_nb.csv};
            \addplot[
                color=cb-teal,
                mark=*,
                mark size=1,
                error bars/.cd,
                y dir=both,
                y explicit
            ] table [x=p, y=pfail_C_Z, y error=err_C_Z, col sep=comma] {data/data_Z_cubic_L=13_ccz=1_nb.csv};
            \addplot[
                color=cb-magenta,
                mark=*,
                mark size=1,
                error bars/.cd,
                y dir=both,
                y explicit
            ] table [x=p, y=pfail_C_Z, y error=err_C_Z, col sep=comma] {data/data_Z_cubic_L=15_ccz=1_nb.csv};
            \addplot[
                color=cb-blue!50!white,
                mark=square,
                mark size=1,
                error bars/.cd,
                y dir=both,
                y explicit
            ] table [x=p, y=pfail_C_Z, y error=err_C_Z, col sep=comma] {data/data_Z_cubic_L=9_ccz=0.csv};
            \addplot[
                color=cb-orange!50!white,
                mark=square,
                mark size=1,
                error bars/.cd,
                y dir=both,
                y explicit
            ] table [x=p, y=pfail_C_Z, y error=err_C_Z, col sep=comma] {data/data_Z_cubic_L=11_ccz=0.csv};
            \addplot[
                color=cb-teal!50!white,
                mark=square,
                mark size=1,
                error bars/.cd,
                y dir=both,
                y explicit
            ] table [x=p, y=pfail_C_Z, y error=err_C_Z, col sep=comma] {data/data_Z_cubic_L=13_ccz=0.csv};
            \addplot[
                color=cb-magenta!50!white,
                mark=square,
                mark size=1,
                error bars/.cd,
                y dir=both,
                y explicit
            ] table [x=p, y=pfail_C_Z, y error=err_C_Z, col sep=comma] {data/data_Z_cubic_L=15_ccz=0.csv};
            \end{axis}
        \end{tikzpicture}
        \subcaption{Octahedral $\overline Z$ errors}
        \label{subfig:cubic_pth_z_nb}
    \end{subfigure}
    ~~
    \begin{subfigure}[t]{0.45\textwidth}
    \centering
        \begin{tikzpicture}
            \begin{axis}[
                xlabel={$p$},
                ylabel={$p_{\mathrm{fail}}(p,L)$},
                ymode=log,
                width=7cm,
                height=6.5cm,
                ymin=0.00011,
                ymax=0.2,
                xtick={0.01,0.02,0.03},
                minor xtick={0.015, 0.025},
                xticklabels={$1\%$,$2\%$,$3\%$},
            ]
            \addplot[
                color=cb-blue,
                mark=*,
                mark size=1,
                error bars/.cd,
                y dir=both,
                y explicit
            ] table [x=p, y=pfail_C_X, y error=err_C_X, col sep=comma] {data/data_X_cubic_L=9_ccz=1_nb.csv};
            \addplot[
                color=cb-orange,
                mark=*,
                mark size=1,
                error bars/.cd,
                y dir=both,
                y explicit
            ] table [x=p, y=pfail_C_X, y error=err_C_X, col sep=comma] {data/data_X_cubic_L=11_ccz=1_nb.csv};
            \addplot[
                color=cb-teal,
                mark=*,
                mark size=1,
                error bars/.cd,
                y dir=both,
                y explicit
            ] table [x=p, y=pfail_C_X, y error=err_C_X, col sep=comma] {data/data_X_cubic_L=13_ccz=1_nb.csv};
            \addplot[
                color=cb-magenta,
                mark=*,
                mark size=1,
                error bars/.cd,
                y dir=both,
                y explicit
            ] table [x=p, y=pfail_C_X, y error=err_C_X, col sep=comma] {data/data_X_cubic_L=15_ccz=1_nb.csv};
            \addplot[
                color=cb-blue!50!white,
                mark=square,
                mark size=1,
                error bars/.cd,
                y dir=both,
                y explicit
            ] table [x=p, y=pfail_C_X, y error=err_C_X, col sep=comma] {data/data_X_cubic_L=9_ccz=0.csv};
            \addplot[
                color=cb-orange!50!white,
                mark=square,
                mark size=1,
                error bars/.cd,
                y dir=both,
                y explicit
            ] table [x=p, y=pfail_C_X, y error=err_C_X, col sep=comma] {data/data_X_cubic_L=11_ccz=0.csv};
            \addplot[
                color=cb-teal!50!white,
                mark=square,
                mark size=1,
                error bars/.cd,
                y dir=both,
                y explicit
            ] table [x=p, y=pfail_C_X, y error=err_C_X, col sep=comma] {data/data_X_cubic_L=13_ccz=0.csv};
            \addplot[
                color=cb-magenta!50!white,
                mark=square,
                mark size=1,
                error bars/.cd,
                y dir=both,
                y explicit
            ] table [x=p, y=pfail_C_X, y error=err_C_X, col sep=comma] {data/data_X_cubic_L=15_ccz=0.csv};
            \end{axis}
        \end{tikzpicture}
        \subcaption{Octahedral $\overline X$ errors}
        \label{subfig:cubic_pth_x_nb}
    \end{subfigure}
    \par\bigskip
    \begin{subfigure}[t]{0.45\textwidth}
    \centering
        \begin{tikzpicture}
            \begin{axis}[
                xlabel={$p$},
                ylabel={$p_{\mathrm{fail}}(p,L)$},
                ymode=log,
                width=7cm,
                height=6.5cm,
                ymin=0.00011,
                ymax=1.5,
                xmin=0,
                xtick={0,0.005,0.01},
                minor xtick={0.0025,0.0075},
                xticklabels={$0$,$0.5\%$,$1\%$},
            ]
            \addplot[
                color=cb-blue,
                mark=*,
                mark size=1,
                error bars/.cd,
                y dir=both,
                y explicit
            ] table [x=p, y=pfail_R1_Z, y error=err_R1_Z, col sep=comma] {data/data_Z_rhombic1_L=9_ccz=1_nb.csv};
            \addplot[
                color=cb-orange,
                mark=*,
                mark size=1,
                error bars/.cd,
                y dir=both,
                y explicit
            ] table [x=p, y=pfail_R1_Z, y error=err_R1_Z, col sep=comma] {data/data_Z_rhombic1_L=11_ccz=1_nb.csv};
            \addplot[
                color=cb-teal,
                mark=*,
                mark size=1,
                error bars/.cd,
                y dir=both,
                y explicit
            ] table [x=p, y=pfail_R1_Z, y error=err_R1_Z, col sep=comma] {data/data_Z_rhombic1_L=13_ccz=1_nb.csv};
            \addplot[
                color=cb-magenta,
                mark=*,
                mark size=1,
                error bars/.cd,
                y dir=both,
                y explicit
            ] table [x=p, y=pfail_R1_Z, y error=err_R1_Z, col sep=comma] {data/data_Z_rhombic1_L=15_ccz=1_nb.csv};
            \addplot[
                color=cb-blue!50!white,
                mark=square,
                mark size=1,
                error bars/.cd,
                y dir=both,
                y explicit
            ] table [x=p, y=pfail_R1_Z, y error=err_R1_Z, col sep=comma] {data/data_Z_rhombic1_L=9_ccz=0.csv};
            \addplot[
                color=cb-orange!50!white,
                mark=square,
                mark size=1,
                error bars/.cd,
                y dir=both,
                y explicit
            ] table [x=p, y=pfail_R1_Z, y error=err_R1_Z, col sep=comma] {data/data_Z_rhombic1_L=11_ccz=0.csv};
            \addplot[
                color=cb-teal!50!white,
                mark=square,
                mark size=1,
                error bars/.cd,
                y dir=both,
                y explicit
            ] table [x=p, y=pfail_R1_Z, y error=err_R1_Z, col sep=comma] {data/data_Z_rhombic1_L=13_ccz=0.csv};
            \addplot[
                color=cb-magenta!50!white,
                mark=square,
                mark size=1,
                error bars/.cd,
                y dir=both,
                y explicit
            ] table [x=p, y=pfail_R1_Z, y error=err_R1_Z, col sep=comma] {data/data_Z_rhombic1_L=15_ccz=0.csv};
            \end{axis}
        \end{tikzpicture}
        \subcaption{Cuboctahedral 1 $\overline Z$ errors}
        \label{subfig:rhombic1_pth_z_nb}
    \end{subfigure}
    ~~
    \begin{subfigure}[t]{0.45\textwidth}
    \centering
      \begin{tikzpicture}
            \begin{axis}[
                xlabel={$p$},
                ylabel={$p_{\mathrm{fail}}(p,L)$},
                ymode=log,
                width=7cm,
                height=6.5cm,
                ymin=0.00011,
                ymax=0.2,
                xtick={0.01,0.02,0.03},
                minor xtick={0.015, 0.025},
                xticklabels={$1\%$,$2\%$,$3\%$},
                legend style={
                    at={(0.5,-0.25)},
                    anchor=north
                },
                legend columns=4,
            ]
            \addplot[
                color=cb-blue,
                mark=*,
                mark size=1,
                error bars/.cd,
                y dir=both,
                y explicit
            ] table [x=p, y=pfail_R1_X, y error=err_R1_X, col sep=comma] {data/data_X_rhombic1_L=9_ccz=1_nb.csv};
            \addplot[
                color=cb-orange,
                mark=*,
                mark size=1,
                error bars/.cd,
                y dir=both,
                y explicit
            ] table [x=p, y=pfail_R1_X, y error=err_R1_X, col sep=comma] {data/data_X_rhombic1_L=11_ccz=1_nb.csv};
            \addplot[
                color=cb-teal,
                mark=*,
                mark size=1,
                error bars/.cd,
                y dir=both,
                y explicit
            ] table [x=p, y=pfail_R1_X, y error=err_R1_X, col sep=comma] {data/data_X_rhombic1_L=13_ccz=1_nb.csv};
            \addplot[
                color=cb-magenta,
                mark=*,
                mark size=1,
                error bars/.cd,
                y dir=both,
                y explicit
            ] table [x=p, y=pfail_R1_X, y error=err_R1_X, col sep=comma] {data/data_X_rhombic1_L=15_ccz=1_nb.csv};
            \addplot[
                color=cb-blue!50!white,
                mark=square,
                mark size=1,
                error bars/.cd,
                y dir=both,
                y explicit
            ] table [x=p, y=pfail_R1_X, y error=err_R1_X, col sep=comma] {data/data_X_rhombic1_L=9_ccz=0.csv};
            \addplot[
                color=cb-orange!50!white,
                mark=square,
                mark size=1,
                error bars/.cd,
                y dir=both,
                y explicit
            ] table [x=p, y=pfail_R1_X, y error=err_R1_X, col sep=comma] {data/data_X_rhombic1_L=11_ccz=0.csv};
            \addplot[
                color=cb-teal!50!white,
                mark=square,
                mark size=1,
                error bars/.cd,
                y dir=both,
                y explicit
            ] table [x=p, y=pfail_R1_X, y error=err_R1_X, col sep=comma] {data/data_X_rhombic1_L=13_ccz=0.csv};
            \addplot[
                color=cb-magenta!50!white,
                mark=square,
                mark size=1,
                error bars/.cd,
                y dir=both,
                y explicit
            ] table [x=p, y=pfail_R1_X, y error=err_R1_X, col sep=comma] {data/data_X_rhombic1_L=15_ccz=0.csv};
            \end{axis}
        \end{tikzpicture}
        \subcaption{Cuboctahedral 1 $\overline X$ errors}
        \label{subfig:rhombic1_pth_x_nb}
    \end{subfigure}
    \par\bigskip
    \begin{subfigure}[t]{0.45\textwidth}
    \centering
        \begin{tikzpicture}
            \begin{axis}[
                xlabel={$p$},
                ylabel={$p_{\mathrm{fail}}(p,L)$},
                ymode=log,
                width=7cm,
                height=6.5cm,
                ymin=0.00011,
                ymax=1.5,
                xmin=0,
                xtick={0,0.005,0.01},
                minor xtick={0.0025,0.0075},
                xticklabels={$0$,$0.5\%$,$1\%$},
            ]
            \addplot[
                color=cb-blue,
                mark=*,
                mark size=1,
                error bars/.cd,
                y dir=both,
                y explicit
            ] table [x=p, y=pfail_R2_Z, y error=err_R2_Z, col sep=comma] {data/data_Z_rhombic2_L=9_ccz=1_nb.csv};
            \addplot[
                color=cb-orange,
                mark=*,
                mark size=1,
                error bars/.cd,
                y dir=both,
                y explicit
            ] table [x=p, y=pfail_R2_Z, y error=err_R2_Z, col sep=comma] {data/data_Z_rhombic2_L=11_ccz=1_nb.csv};
            \addplot[
                color=cb-teal,
                mark=*,
                mark size=1,
                error bars/.cd,
                y dir=both,
                y explicit
            ] table [x=p, y=pfail_R2_Z, y error=err_R2_Z, col sep=comma] {data/data_Z_rhombic2_L=13_ccz=1_nb.csv};
            \addplot[
                color=cb-magenta,
                mark=*,
                mark size=1,
                error bars/.cd,
                y dir=both,
                y explicit
            ] table [x=p, y=pfail_R2_Z, y error=err_R2_Z, col sep=comma] {data/data_Z_rhombic2_L=15_ccz=1_nb.csv};
            \addplot[
                color=cb-blue!50!white,
                mark=square,
                mark size=1,
                error bars/.cd,
                y dir=both,
                y explicit
            ] table [x=p, y=pfail_R2_Z, y error=err_R2_Z, col sep=comma] {data/data_Z_rhombic2_L=9_ccz=0.csv};
            \addplot[
                color=cb-orange!50!white,
                mark=square,
                mark size=1,
                error bars/.cd,
                y dir=both,
                y explicit
            ] table [x=p, y=pfail_R2_Z, y error=err_R2_Z, col sep=comma] {data/data_Z_rhombic2_L=11_ccz=0.csv};
            \addplot[
                color=cb-teal!50!white,
                mark=square,
                mark size=1,
                error bars/.cd,
                y dir=both,
                y explicit
            ] table [x=p, y=pfail_R2_Z, y error=err_R2_Z, col sep=comma] {data/data_Z_rhombic2_L=13_ccz=0.csv};
            \addplot[
                color=cb-magenta!50!white,
                mark=square,
                mark size=1,
                error bars/.cd,
                y dir=both,
                y explicit
            ] table [x=p, y=pfail_R2_Z, y error=err_R2_Z, col sep=comma] {data/data_Z_rhombic2_L=15_ccz=0.csv};
            \end{axis}
        \end{tikzpicture}
        \subcaption{Cuboctahedral 2 $\overline Z$ errors}
        \label{subfig:rhombic2_pth_z_nb}
    \end{subfigure}
    ~~
    \begin{subfigure}[t]{0.45\textwidth}
    \centering
        \begin{tikzpicture}
            \begin{axis}[
                xlabel={$p$},
                ylabel={$p_{\mathrm{fail}}(p,L)$},
                ymode=log,
                width=7cm,
                height=6.5cm,
                ymin=0.00011,
                ymax=0.2,
                xtick={0.01,0.02,0.03},
                minor xtick={0.015, 0.025},
                xticklabels={$1\%$,$2\%$,$3\%$},
            ]
            \addplot[
                color=cb-blue,
                mark=*,
                mark size=1,
                error bars/.cd,
                y dir=both,
                y explicit
            ] table [x=p, y=pfail_R2_X, y error=err_R2_X, col sep=comma] {data/data_X_rhombic2_L=9_ccz=1_nb.csv};
            \addplot[
                color=cb-orange,
                mark=*,
                mark size=1,
                error bars/.cd,
                y dir=both,
                y explicit
            ] table [x=p, y=pfail_R2_X, y error=err_R2_X, col sep=comma] {data/data_X_rhombic2_L=11_ccz=1_nb.csv};
            \addplot[
                color=cb-teal,
                mark=*,
                mark size=1,
                error bars/.cd,
                y dir=both,
                y explicit
            ] table [x=p, y=pfail_R2_X, y error=err_R2_X, col sep=comma] {data/data_X_rhombic2_L=13_ccz=1_nb.csv};
            \addplot[
                color=cb-magenta,
                mark=*,
                mark size=1,
                error bars/.cd,
                y dir=both,
                y explicit
            ] table [x=p, y=pfail_R2_X, y error=err_R2_X, col sep=comma] {data/data_X_rhombic2_L=15_ccz=1_nb.csv};
            \addplot[
                color=cb-blue!50!white,
                mark=square,
                mark size=1,
                error bars/.cd,
                y dir=both,
                y explicit
            ] table [x=p, y=pfail_R2_X, y error=err_R2_X, col sep=comma] {data/data_X_rhombic2_L=9_ccz=0.csv};
            \addplot[
                color=cb-orange!50!white,
                mark=square,
                mark size=1,
                error bars/.cd,
                y dir=both,
                y explicit
            ] table [x=p, y=pfail_R2_X, y error=err_R2_X, col sep=comma] {data/data_X_rhombic2_L=11_ccz=0.csv};
            \addplot[
                color=cb-teal!50!white,
                mark=square,
                mark size=1,
                error bars/.cd,
                y dir=both,
                y explicit
            ] table [x=p, y=pfail_R2_X, y error=err_R2_X, col sep=comma] {data/data_X_rhombic2_L=13_ccz=0.csv};
            \addplot[
                color=cb-magenta!50!white,
                mark=square,
                mark size=1,
                error bars/.cd,
                y dir=both,
                y explicit
            ] table [x=p, y=pfail_R2_X, y error=err_R2_X, col sep=comma] {data/data_X_rhombic2_L=15_ccz=0.csv};
            \end{axis}
        \end{tikzpicture}
        \subcaption{Cuboctahedral 2 $\overline X$ errors}
        \label{subfig:rhombic2_pth_x_nb}
    \end{subfigure}
    \caption{(Colour)
        For each of the three codes, we show a plot of the logical error rate $p_{\mathrm{fail}}$ (separately for $\overline X$ and $\overline Z$ errors) as a function of the depolarising/measurement error rate $p$ for various lattice sizes $L$ (with and without the $CCZ$ gate).
        For these simulations, the depolarising noise channel was applied \emph{before} the $CCZ$ gate.
        The error bars show the Agresti-Coull 95\% confidence intervals~\cite{agresti1998,dasgupta2001}.
    }
    \label{fig:threshold_plots_noise_before}
\end{figure*}
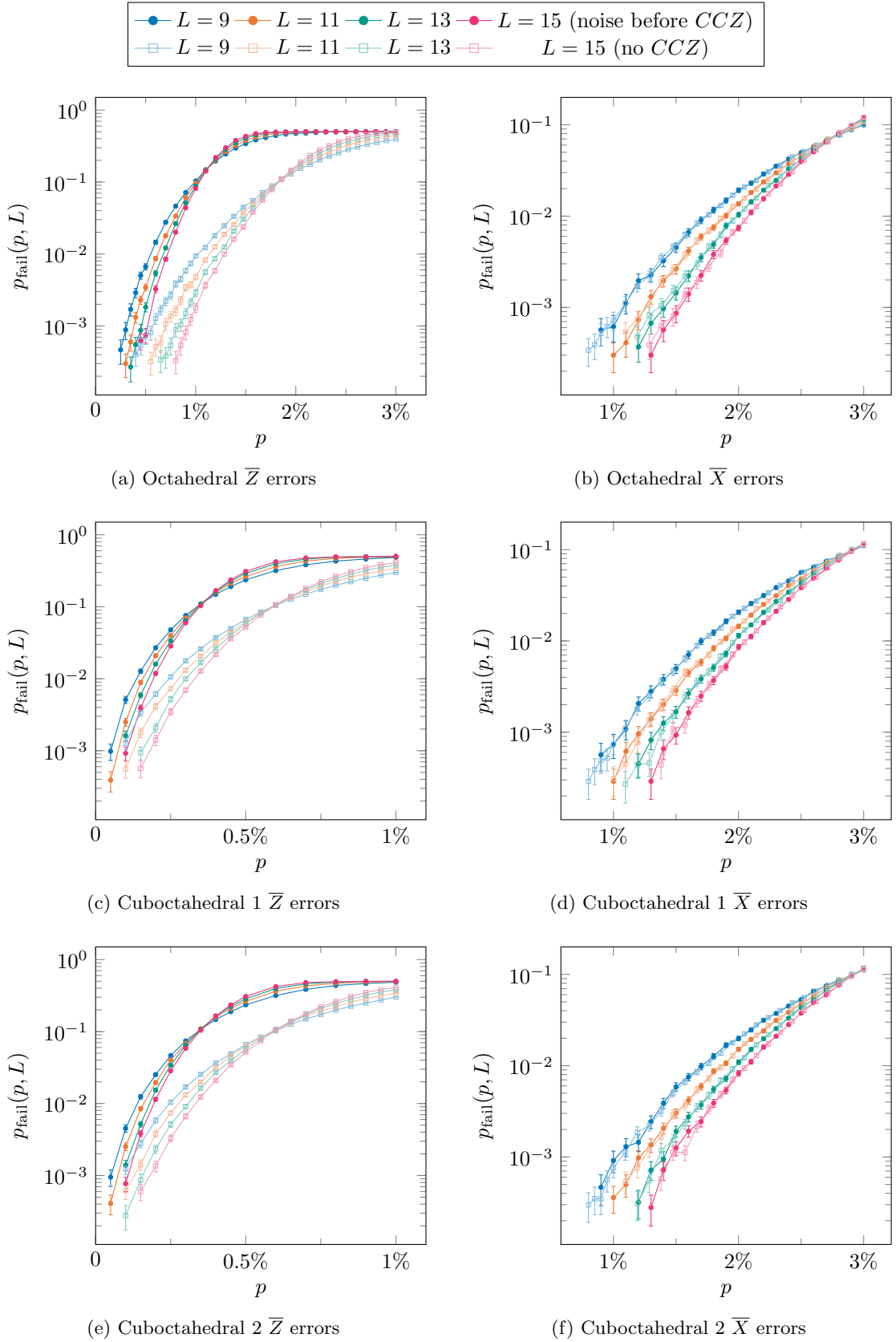

We apply measurement errors during the initial $Z$ stabiliser measurements and also apply a depolarising error either just before or just after the $CCZ$ gate, all with the same probability $p$. After all the steps are complete we use perfect stabiliser measurements to calculate a final correction for the 2D codes, check for logical errors and declare success or failure depending on the outcome. The results can be seen in \cref{fig:threshold_plots_noise_after} and \cref{fig:threshold_plots_noise_before}, with observed thresholds as shown in \cref{tab:thresholds}. To estimate the error thresholds we use standard finite-size scaling analysis~\cite{wang2003}. 
Specifically, in the vicinity of the threshold we fit the data to the following ansatz
\begin{equation}
    p_{\mathrm{fail}}(x) = a_0 + a_1 x + a_2 x^2,
    \label{eq:fit_ansatz}
\end{equation}
where $x=(p-p_{\mathrm{th}})L^{1/\nu}$ and $a_0$, $a_1$, $a_2$, $p_{\mathrm{th}}$ and $\nu$ are the parameters of the fit.

We observe that when the depolarising channel is applied after the $CCZ$ gate the inclusion of $CZ$ errors (in this case arising only due to measurement errors) results in only a marginally lower $\overline Z$ error threshold estimate in the octahedral surface code, and actually leads to a slight increase in the estimated $\overline Z$ error threshold value in the cuboctahedral surface codes. On the other hand, when the depolarising channel is applied just before the $CCZ$ gate the $CZ$ errors (now due to both measurement and qubit errors) have a significant impact on the threshold. In \cref{subsection:noise_after} and \cref{subsection:noise_before} we consider each of these cases in detail, and following this analysis we provide a more detailed description of the above procedure, with explanations of state preparation, $CCZ$ application and $3D \rightarrow 2D$ dimension jumping given in \cref{subsection:state_prep}, \cref{subsection:CCZ} and \cref{subsection:jump} respectively. 

\subsection{Noise After \texorpdfstring{$CCZ$}{CCZ}}
\label{subsection:noise_after}

To understand the minimal effect seen when the depolarising channel comes after the $CCZ$ we first need to observe that the $CZ$ error probability in this case is actually a function of lattice size $L$ (because the $CZ$ errors occur due to uncorrected $X$ errors). When we are above the $X$ threshold the likelihood of $CZ$ errors increases with increasing $L$, whereas below the $X$ threshold it decreases. We can then define an effective $Z$ error probability

\begin{equation}
    p^{\mathrm{eff}}_Z = p_Z + q(L)
\end{equation}

\noindent where $p_Z$ is the per-qubit $Z$ error probability due to the depolarising channel (i.e. $2p/3$ where $p$ is the error probability used in our simulations) and $q(L)$ is the probability of additional $Z$ errors produced by $CZ$ errors + stabiliser measurement. If we consider only values of $p_Z$ close to the $Z$ error threshold then we are below the $X$ error threshold so $q(L)$ will be small and will decrease with increasing $L$. In this region we can approximate the curves $p_\mathrm{fail}^L$ as straight lines $p_\mathrm{fail}^L = m_L p_Z + c_L$, and if we choose a pair of lines $p_\mathrm{fail}^{L_1}$ and $p_\mathrm{fail}^{L_2}$ and replace $p_Z$ with $p^\mathrm{eff}_Z$ then the point where $p_\mathrm{fail}^{L_1} = p_\mathrm{fail}^{L_2}$ shifts by an amount 

\begin{equation}
    \Delta p(L_1, L_2) = \frac{m_{L_2} q(L_2) - m_{L_1} q(L_1)}{m_{L_1} - m_{L_2}}.
\end{equation}

If $L_2 > L_1$ then this shift is positive if $m_{L_2}/m_{L_1} < q(L_1)/q(L_2)$ and negative (or zero) otherwise. $m_{L_2}/m_{L_1}$ is large in codes that deal well with $Z$ errors, while $q(L_1)/q(L_2)$ is large in codes that deal well with $X$ errors. Hence we see a positive shift of the threshold  estimate in the cuboctahedral codes (where performance against $X$ errors is very good but performance against $Z$ errors is fairly poor) and a negative shift in the octahedral code (which deals better with $Z$ errors but worse with $X$ errors relative to the cuboctahedral code). 

Notice also that although we use the term ``threshold'' in this discussion these are not true thresholds as are they are not independent of lattice size (the shift $\Delta p$ depends on $L$). Estimating a threshold value using a set of relatively small lattices gives values like those in \cref{tab:thresholds}, but if we estimated a threshold using only very large lattices the impact of the $CZ$ errors would be negligible and the calculated value would be almost identical to the case of purely IID noise. If we consider the true threshold to be the point below which $p_\mathrm{fail}^{L_2} < p_\mathrm{fail}^{L_1}$ for all $L_2 > L_1$ then this is the value 

\begin{equation}
    p_\mathrm{th}^\mathrm{true} = \min_{L_2 > L_1} (p_\mathrm{th}^\mathrm{IID} + \Delta p(L_1,L_2))
\end{equation}

Our numerical results suggest that for the cuboctahedral lattices $p_\mathrm{th}^\mathrm{true} = p_\mathrm{th}^\mathrm{IID}$ (with the minimum value of $\Delta p(L_1, L_2)$ being $\lim_{L \to \infty} \Delta p(L, L+1) = 0$) while $p_\mathrm{th}^\mathrm{true}$ will be very close to the value shown in \cref{tab:thresholds} for the octahedral lattice (with the minimum (negative) value of $\Delta p(L_1, L_2)$ being when $L_1$ and $L_2$ are the two smallest members of the code family). 

\subsection{Noise Before \texorpdfstring{$CCZ$}{CCZ}}
\label{subsection:noise_before}

In the previous case the relative immunity of the threshold to $CZ$ errors was due to the fact that $X$ error probability in the code fell with increasing lattice size. This is no longer true if additional $X$ errors occur after the $Z$ stabiliser measurements but before the application of the $CCZ$ gate, as these errors will not be accounted for by the $X$ correction calculated using those measurements. More explicitly, the per-qubit $X$ error probability at the point in time when the $CCZ$ gate is applied will be 

\begin{equation}
    p^{\mathrm{eff}}_X = p_X + p_{\mathrm{res}}
\end{equation}

\noindent where $p_{\mathrm{res}}$ is the probability of a residual $X$ error from decoding using faulty syndrome measurements and $p_X = 2p/3$  $(= p_Z)$ is once again the probability of a specific Pauli error due to the depolarising channel. $p_{\mathrm{res}}$ is a function of $L$ (and decreases with increasing $L$ when we are below the $X$ threshold) while $p_X$ is independent of $L$ and so for large $L$ we will have $p^{\mathrm{eff}}_X \approx p_X$, and if we are far enough below the threshold then we expect $p_X$ to dominate even for small $L$. This means that if we define the same effective $Z$ error probability $p^{\mathrm{eff}}$ as in the previous section the quantity $q(L)$ will also be almost constant in $L$ for values of $p_Z$ close to the $Z$ threshold and the quantity $\Delta p(L_1,L_2)$ will then be an almost constant negative shift of this threshold. This is reflected in \cref{fig:threshold_plots_noise_before}.

This kind of problem is not unique to non-Clifford gates (and in general we expect any non-trivial logical operation to have a negative impact on the performance of the code~\cite{bombin_logical_2021}). Consider, for example, the simple case of stabiliser measurement $\rightarrow$ correction $\rightarrow$ transversal $CNOT$ in the 2D surface code. All $X$ errors in the control code which occurred between measurement and correction will now also be applied to the target code, while all such $Z$ errors in the target will be applied to the control, effectively doubling the $X$/$Z$ error probability in the target/control for this step of the computation. However, for the case of transversal Clifford gates we can instead consider a process of stabiliser measurement $\rightarrow$ transversal gate $\rightarrow$ correction, where the correction we apply to the code is the correction produced by the decoder commuted through the transversal gate. For transversal non-Cliffords things are not so simple because commuting the Pauli correction through the non-Clifford gives a Clifford correction. This will be a valid correction for the Clifford error which exists in the code, but if we wish to restrict ourselves to only applying Pauli corrections (e.g. because single-qubit gates typically have better fidelities than multi-qubit gates) then we must once again measure the stabilisers to project the Clifford error to a Pauli error and there is then no way to deterministically infer a suitable Pauli correction from our previously calculated Clifford correction as the projection is probabilistic.

This may seem like a serious drawback for transversal non-Clifford gates (and particularly multi-qubit ones), as the decoding calculation will take a non-negligible amount of time and so it is reasonable to expect that a non-negligible number of errors will occur in the period between measurement of stabilisers and application of a correction. Fortunately, this problem may not be as significant as it seems. Notice that if we consider $p^{\mathrm{eff}}_X \approx p_X$ then we can write $p^{\mathrm{eff}}_Z$ as 

\begin{equation}
    p^{\mathrm{eff}}_Z = p_Z + \alpha p_X
\end{equation}

\noindent for some $0 \leq \alpha \leq 1$ (in practise we expect $\alpha \approx 0.5$ since most $X$ errors from the depolarising channel will be isolated single-qubit errors that will be mapped to $Z$ errors with (per-code) probability 0.5). In our simulations we have $p_X = p_Z$ and so we observe a large increase in $p^{\mathrm{eff}}_Z$ and a large decrease in the threshold. However, in many architectures we have $p_Z \gg p_X$~\cite{Aliferis_2009,Shulman2012,Nigg2014,Lescanne2020,Grimm2020}, in which case the increase in $p^{\mathrm{eff}}_Z$ due to errors of this type would be much less significant. 

Another practical consideration is the trade-off between decoding speed and performance. Faster decoders will allow less time for errors to build up before the gate application, but it is often the case that faster decoders have weaker performance and so if error rates are low enough then using a slower but more accurate decoder may be preferable. 

\subsection{State Preparation}
\label{subsection:state_prep}

While the state preparation step is not of particular interest by itself, the details of this step will be relevant when we discuss the generation of the $Z$ error distributions that result from the $CZ$ errors. Throughout the simulation we track the locations of both $X$ and $Z$ errors. 

To simulate the preparation of the initial state we start with no errors on any qubit (i.e. we assume perfect single-qubit state preparation) and then apply an $X$ error to each qubit with probability $0.5$. This reproduces the effect of measuring the $Z$ stabilisers of the code when all qubits are in $\ket{+}$. We then calculate the $Z$ stabiliser syndrome and, for each stabiliser, flip the outcome of this calculation with probability $p$ to simulate measurement errors. We use a minimum-weight perfect matching decoder~\cite{dennis_topological_2002} to generate a correction for these measurement errors and we use BP-OSD~\cite{Panteleev2021degeneratequantum,roffe_decoding_2020,quintavalle2021a} to calculate a correction for the qubit errors. 

In the absence of measurement errors (or if we make no mistakes in calculating a correction for these errors) the pattern of single-qubit $X$s after correction will correspond to a random configuration of $X$ stabilisers and possibly an $X$ logical. This pattern of $X$s represents one randomly selected term of the superposition 

\begin{equation}
    \label{eq:codeword}
    \begin{split}
        \ket{\overline{+}} &= \frac{1}{\sqrt{2}}(\ket{\overline{0}} + \ket{\overline{1}}) \\
                           &= \frac{1}{\sqrt{2}}(\frac{1}{\sqrt{n}}\sum_i X_{\beta_i}\ket{\bm{0}} 
                           + \frac{1}{\sqrt{n}}\overline{X}_L \sum_i X_{\beta_i}\ket{\bm{0}})
    \end{split}
\end{equation}

In the presence of measurement errors (and if we make mistakes correcting them) this pattern of $X$s will differ from a term of this superposition by errors on a subset of qubits\footnote{Of course, it technically differs from all terms of the superposition by errors on a subset of qubits, but one (or perhaps a small number) of the terms will be the closest in terms of Hamming distance}. Notice that an ``$X$ error'' can now mean either an application of $X$ to a qubit where we should have applied identity or the reverse, so the locations of these errors are not explicitly known or tracked by the simulation. 

\subsection{Application of \texorpdfstring{$CCZ$}{CCZ}}
\label{subsection:CCZ}

The pattern of $X$'s we have prepared makes the creation of an appropriate $Z$ error configuration extremely simple: we simply apply a $Z$ error to any qubit of code $k$ where we have applied $X$ to the corresponding qubits in codes $i$ and $j$ (i.e. just the natural local action of $CCZ$). We know from \cref{eq:sc_link} that this will produce linking charge strings so we only need to check that it will produce appropriate error configurations on the boundaries of error membranes and also that it will not produce errors elsewhere. 

To check the former we can consider the intersection of a membrane of $X$ errors in code $i$ with an $X$ stabiliser of code $j$ (we assume that there are no $X$ errors in code $j$ so that the pattern of $X$ operators for this code matches a term of \cref{eq:codeword}). We know from the analysis in \cref{section:sc} that a $Z$ error supported on this intersection in code $k$ anticommutes with a pair of $X$ stabilisers of code $k$ that lie on the boundary of the code $i$ error membrane. We also know that each such stabiliser should be violated with probability $0.5$ while the total number of violated stabilisers should be even. Each $X$ stabiliser of code $j$ appears in the pattern of $X$s for this code with probability $0.5$, and so for each error-stabiliser intersection we create a $Z$ error string with this same probability. The resulting pattern of all these strings will anticommute with the $X$ stabilisers of code $k$ exactly as described above.  

\begin{figure*}
    \centering
    \begin{minipage}{.5\textwidth}
    \includegraphics[width=\textwidth]{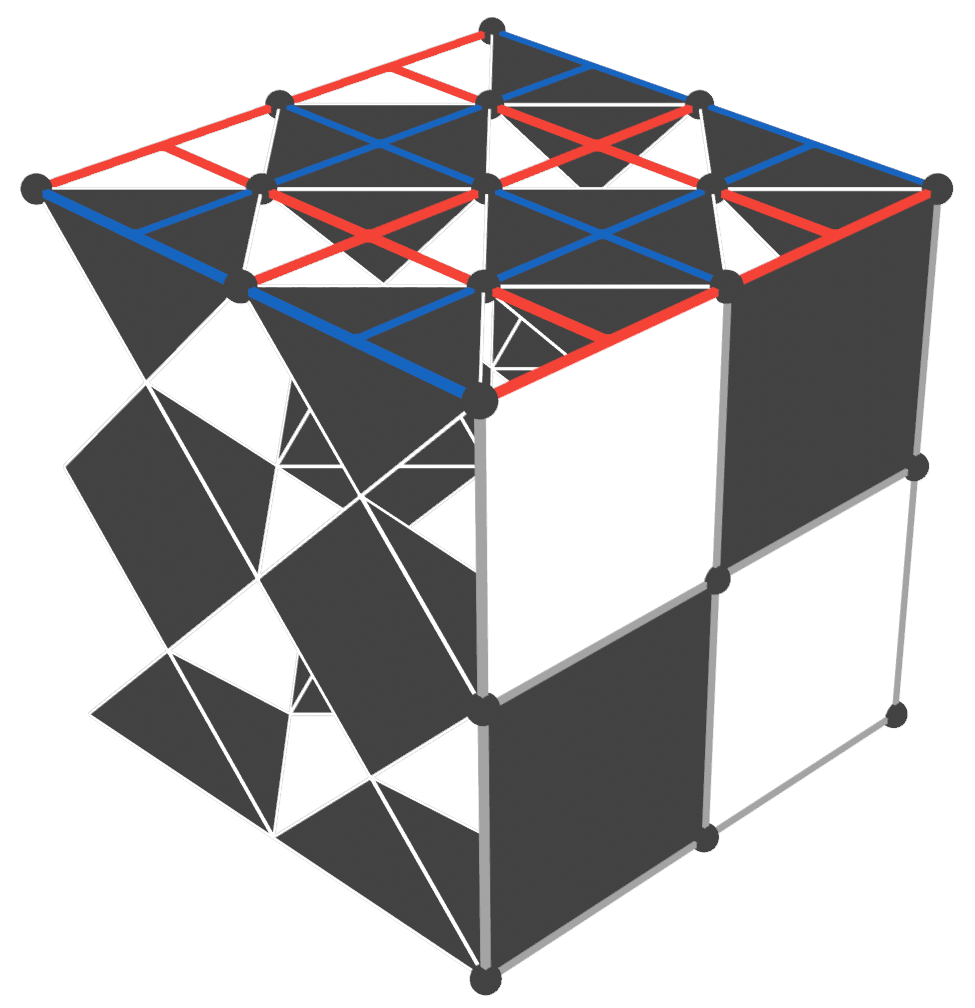}
    \end{minipage}
    ~~~~~~
    \begin{minipage}{.3\textwidth}
    \includegraphics[width=.96\textwidth]{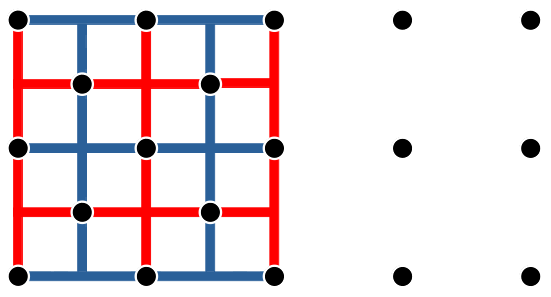}
    \\
    \includegraphics[width=\textwidth]{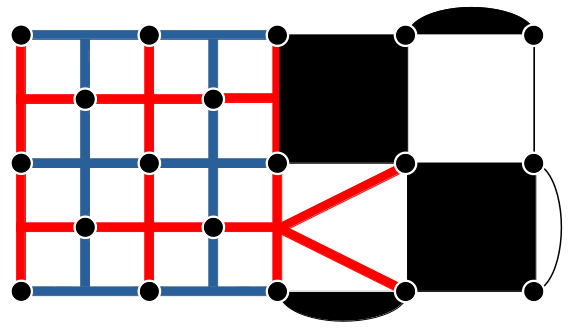}
    \\
    \includegraphics[width=.99\textwidth]{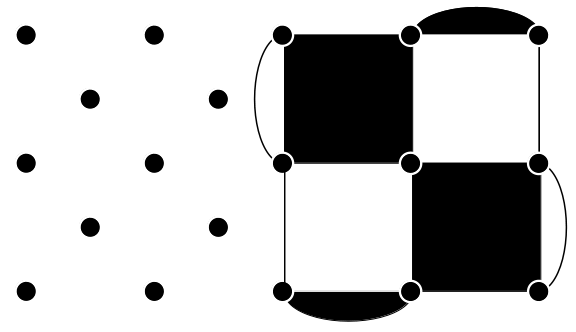}
    \end{minipage}
    \caption{(Colour) Code deformation operation for turning an unrotated surface code (supported on a half-octahedral boundary of the rectified lattice) into a rotated surface code (supported on a cuboctahedral boundary of the rectified lattice).}
    \label{fig:surgery}
\end{figure*}

To check that we do not create $Z$ errors in other cases we only need to observe that, in the absence of errors, $X$ will only be applied to a qubit if this forms part of a stabiliser or logical in the pattern of $X$'s for this code. The fact that intersections of stabilisers and logicals in codes $i$ and $j$ correspond to the supports of $Z$ stabilisers and logicals in code $k$ is what makes the $CCZ$ gate transversal so no errors can be created in this case. 

\subsection{Dimension Jump}
\label{subsection:jump}

Full descriptions of dimension jumping for 3D surface codes can be found in in~\cite{brown_fault-tolerant_2020,scruby_numerical_2021} but we review the collapse part here for completeness. It will be helpful to once again write our codewords in terms of stabilisers, but this time we use the $Z$ stabilisers rather than the $X$, so that, for example

\begin{equation}
    \ket{\overline{+}} = \sum_i Z_{\beta_i} \ket{\bm{+}}.
\end{equation}

Measuring out all qubits of the code in the $X$ basis will project to one specific term of this superposition. If we split the qubits into two sets $M$ and $N$ and measure out only qubits in $M$ then we project to a subset of terms

\begin{equation}
    \ket{\psi} = Z_{\beta_j} \sum_n Z_{\beta_n} \ket{\bm{+}},
\end{equation}

\noindent  where $\beta_j \cap M \neq \emptyset$ while $\beta_n \cap M = \emptyset$. If $\sum_n Z_{\beta_n} \ket{\bm{+}}^{\otimes \abs{N}}$ is a codeword $\ket{\overline{+}}$ of a smaller code then this code will be subject to a $Z$ error $Z_{\beta_j \cap N}$ and if we wish to transform from the larger code to the smaller one fault-tolerantly then this error must be correctable. A correction can be calculated by finding a stabiliser of the larger code that matches $Z_{\beta_j}$ on the qubits in $M$ and then applying this operator to qubits in $N$, but whether or not this reliably corrects the error will depend on the stabiliser structure of the code and on the choice of regions $M$ and $N$. Additionally, in order to correctly transfer the logical information we require that there is at least one representation $Z_L$ of logical $Z$ for the larger code for which $Z_{L \cap N}$ is a valid logical $Z$ of the smaller code. This means that $\ket{\overline{-}} = Z_L \sum_i Z_{\beta_i} \ket{+}$ in the larger code is mapped to 

\begin{equation}
    \ket{\psi} = Z_{L \cap M} Z_{\beta_j} (Z_{L \cap N} \sum_n Z_{\beta_n} \ket{\bm{+}}^{\otimes \abs{N}}) \ket{\bm{+}}^{\otimes \abs{M}},
\end{equation}

\noindent where $Z_{L \cap N} \sum_n Z_{\beta_n} \ket{\bm{+}}^{\otimes \abs{N}}$ will then be the codeword $\ket{\overline{-}}$ in the smaller code. 

An extra complication is added when the qubits of $M$ are subject to $Z$ errors prior to measurement. In this case the outcomes of our single-qubit $X$ measurements will not be consistent with any $Z$ stabiliser $Z_{\beta_j}$, and before we can calculate a correction for the qubits in $N$ we must first find one for the qubits in $M$. We can do this by using the single-qubit measurement outcomes along with the measurement outcomes of the $X$ stabilisers of the smaller code to reconstruct an $X$ stabiliser syndrome for the larger code, and then use this syndrome to find a correction. 

In our simulation the ``larger code'' is the full 3D surface code while the ``smaller code'' is a 2D surface code supported on one of the boundaries of the 3D code. The qubits in sets $M$ and $N$ are referred to as the ``inner'' and ``outer'' qubits respectively. Because the logical $Z$ operators of the three different surface codes are all perpendicular there is no way to collapse all three codes to the same boundary, since for any boundary $N$ of the rectified lattice there is always one code where $Z_{L \cap N} \neq \overline{Z}^{\textrm{2D}}$ for every representation $Z_L$. Instead we choose to collapse the two codes defined on cuboctohedral lattices to one of the half-cuboctahedral boundaries (giving two rotated 2D surface codes) while the code defined on the octahedral lattice collapses to a half-octahedral boundary (giving an unrotated 2D surface code) as shown in \cref{fig:surgery}. Code deformation can be used to transform the unrotated code into a rotated code that overlaps with the other two. 

In each of the three codes the error $Z_{\beta_j \cap N}$ comes from faces of the 3D lattice that meet the boundary at an edge, since these faces support $Z$ stabilisers of the 3D code and when restricted to the 2D code these edges form strings of errors. As mentioned above, a correction can be calculated by finding a $Z$ stabiliser of the 3D code that matches the measurement outcomes from the measured-out qubits, and because each edge of the 2D code corresponds to a single (non-boundary) face of the 3D lattice this correction will exactly match the error in the 2D code. An algorithm for calculating this correction in each of the three codes is given in \cref{appendix:jump}.

If there were $Z$ errors on the measured out qubits (or if the measurements themselves were faulty) we will need to find a correction for these by reconstructing a 3D $X$ stabiliser syndrome. These stabilisers are supported on cells of the 3D lattice so for cells supported entirely on the inner qubits we can simply take the product of all single-qubit measurement outcomes from that cell. If a cell is partially supported on the outer qubits then its restriction to these qubits gives an $X$ stabiliser of the 2D code, so to calculate a measurement outcome for this cell we can take the product of the measurement outcome of this 2D stabiliser and all the single-qubit measurement outcomes from the other qubits in the cell. However, obtaining a reliable set of stabiliser measurement outcomes for the 2D code requires repeated measurements of these stabilisers, and we have promised a single-shot magic state preparation process. Fortunately, there is a way to fault-tolerantly calculate a correction using only the single-qubit measurement outcomes although this comes at the cost of a reduction in code performance. This calculation is described in \cref{appendix:decoding}.

\section{Discussion}
\label{section:discussion}
We have generalised results regarding Clifford errors in the colour code to the case of the 3D surface code and found that not only do these results translate straightforwardly but they are much more easily understood in this setting. In the surface code the deterministic linking charge error string and the random error strings on the Clifford error boundary have separate mathematical origins, the former being caused by commuting the transversal non-Clifford gate through a pair of Pauli errors and the latter being caused by the action of a Clifford error on the codestate. In the colour code this distinction is less clear and these two kinds of $Z$ error have the same origin (the action of the Clifford error on the codestate). The transversal $T$ gate in the colour code and transversal $CCZ$ in the surface code are related by a local unitary mapping~\cite{kubica_abcs_2018,vasmer_morphing_2021} and so we propose that linking charge arising from the $T$ gate in the colour code is best understood as a surface code phenomenon whose origin is obscured by the mapping to the colour code. On the other hand, we showed in \cref{appendix:cc} that linking charge effects occurring when $CS$ and $CCZ$ are applied between three colour codes are identical to the surface code case (i.e. they have a local origin), so we see that the non-local linking charge previously observed in the colour code is specifically a property of the transversal $T$, rather than of the colour code itself. 

Clifford errors from the $CCZ$ gate are more approachable not only from an analytic perspective but also from a numerical one. Full simulation of Clifford errors due to $T$ requires a non-local method of detecting linked syndromes and as a result linking charge contributions have been left out of previous numerical works~\citep{beverland_cost_2021}. In contrast, our simulations reproduce the full effect of $CZ$ errors in the 3D surface code with only a simple local check at each lattice site. The results of these simulations show that the occurrence of a large number of $X$ errors before the $CCZ$ but after the most recent stabiliser measurement can result in a significant reduction of the $Z$ threshold for the code, but in practical implementations we expect that the number of such errors will be small when compared to the number of $Z$ errors which occur naturally during this period.  

Finally, we note that in addition to the 3D surface and colour codes, our proof techniques can be applied straightforwardly to any CSS code with a transversal non-Clifford that is diagonal in the $Z$ basis. For instance, the 4D surface code admits a transversal $CCCZ$ between four copies of the code~\citep{jochym-oconnor_four-dimensional_2021} and in the same way as \cref{eq:sc_link} we have

\begin{equation}
    \begin{split}
        \overline{CCCZ}&X_\alpha^1 X_\beta^2 X_\gamma^3 \ket*{\overline{\psi}} \\
        &= X_\alpha^1 CCZ_\alpha^{234} X_\beta^2 CCZ_\beta^{134} X_\gamma^3 CCZ_\gamma^{124} \ket*{\overline{\psi'}} \\
        &= X_\alpha^1 X_\beta^2 CZ_{\alpha \cap \beta}^{34} X_\gamma^3 CZ_{\alpha \cap \gamma}^{24} CZ_{\beta \cap \gamma}^{14}\ket*{\psi''} \\
        &= X_\alpha^1 X_\beta^2 X_\gamma^3 Z_{\alpha \cap \beta \cap \gamma}^4 \ket*{\psi'''}
    \end{split}
\end{equation}

\noindent and so we observe a $Z$ linking charge string on the intersection of these three $X$ errors in addition to whatever effect the Clifford $CZ$ and non-Clifford $CCZ$ errors have on the codestate.  This effect can be calculated in the same way as for the $CZ$ errors shown previously, i.e by writing the codestates as superpositions of stabilisers, commuting the non-Pauli errors through these stabilisers, then examining the effect of the phases produced on the possible stabiliser measurement outcomes.

\section*{Acknowledgements}
TRS acknowledges support from University College London and the Engineering and Physical Sciences Research Council [grant number EP/L015242/1] and also the JST Moonshot R\&D Grant [grant number JPMJMS2061]. Research at Perimeter Institute is supported in part by the Government of Canada through the Department of Innovation, Science and Economic Development Canada and by the Province of Ontario through the Ministry of Colleges and Universities. This research was enabled in part by support provided by \href{www.computeontario.ca}{Compute Ontario} and \href{www.computecanada.ca}{Compute Canada}. 

This work was partially completed while TRS was at University College London and some parts appeared previously in \cite{scruby_logical_2021}. 

The authors thank B. Brown, M. Kesselring, A. Kubica and M. Beverland for valuable discussions regarding linking charge in the colour code. 

\end{multicols}

\section*{Appendices}

\appendix

\section{Clifford Errors in the 3D Colour Code}
\label{appendix:cc}
In this appendix we apply the proof techniques used in the main text to the case of the 3D colour code. We recover the result of linking charge for the case of transversal $T$ in this setting, although it requires more effort and is considerably less intuitive than the surface code case. We also examine the cases where the non-Clifford gates $CS$ and $CCZ$ are applied between two and three copies of the code.

\subsection{Single Error Membranes in Cleanable Code Regions}

The colour code has a transversal $T$ gate corresponding to an application of $T$ and $T^\dag$ to white and black vertices in a bicolouring of the lattice. Consider a membranelike error $X_\alpha$ (Pauli $X$ on all qubits in set $\alpha$ and identity otherwise) supported on a subset of faces of colour $\kappa_1\kappa_2$ and detected by $Z$ stabilisers on faces of colour $\kappa_3\kappa_4$ at the boundary of the membrane. Using that $TX = e^{-i\pi/4}SXT$ and $T^\dag X = e^{i\pi/4}S^\dag X T^\dag$ we have that

\begin{equation}
    \overline{T}X_\alpha\ket*{\overline{\psi}} = e^{-i\pi N_w^\alpha/4}e^{i\pi N_b^\alpha/4}A_\alpha X_\alpha \ket*{\overline{\psi'}},
\end{equation}

\noindent where $N_w^\alpha$ and $N_b^\alpha$ are the numbers of white and black vertices in $\alpha$ and $A_\alpha$ is a tensor product of $S$ on all white vertices of $\alpha$ and $S^\dag$ on all black vertices of $\alpha$. $\ket*{\overline{\psi}}$ and $\ket*{\overline{\psi'}}$ are states in the codespace. Using $SX = YS$, $S^\dag X = -Y S^\dag$ and $Y = iXZ$ we have

\begin{equation}
    e^{-i\pi N_w^\alpha/4}e^{i\pi N_b^\alpha/4}A_\alpha X_\alpha \ket*{\overline{\psi'}} = e^{-i\pi N_w^\alpha/4}e^{i\pi N_b^\alpha/4}(-1)^{N_b^\alpha}i^{|\alpha|} X_\alpha Z_\alpha A_\alpha \ket*{\overline{\psi'}}.
\end{equation}

$\alpha$ is a product of faces of the code and faces are cycles in the lattice so must contain an equal number of $b$ and $w$ vertices so $e^{-i\pi N_w^\alpha/4}e^{i\pi N_b^\alpha/4} = 1$. If $|\alpha| = 0 \mod 4$ then $i^{|\alpha|} = (-1)^{N_b^\alpha} = 1$ and if $|\alpha| = 2 \mod 4$ then $i^{|\alpha|} = (-1)^{N_b^\alpha} = -1$. $Z_\alpha$ is a $Z$ stabiliser of the code (since it is a $Z$ operator supported on a set of faces) and commutes with $A_\alpha$ as they are both diagonal in the computational basis. In summary

\begin{equation}
    \label{eq:TXcommute}
    \overline{T}X_\alpha\ket*{\overline{\psi}} = X_\alpha A_\alpha \ket*{\overline{\psi'}}.
\end{equation}

We then want to know what effect $A_\alpha$ has on codestates. Since we are considering an error in a cleanable region of the code it is sufficient to consider only the effect on $\ket*{\overline{0}}$ as the analysis for $\ket*{\overline{1}}$ will be identical. We have that

\begin{equation}
    \ket*{\overline{0}} = \frac{1}{\sqrt{n}}\sum_{i=1}^n X_{\beta_i} \ket*{\bm{0}}
\end{equation}

\noindent where $X_{\beta_i}$ are stabilisers of the code (which are cells or products of cells of the lattice) and $\ket*{\bm{0}}$ is the all-zeros state. We can then use the same commutation relations as above to show

\begin{equation}
    A_\alpha \frac{1}{\sqrt{n}} \sum_{i=1}^n X_{\beta_i} \ket*{\bm{0}} = \frac{1}{\sqrt{n}}\sum_{i=1}^n A_\alpha X_{\beta_i} \ket*{\bm{0}} = \frac{1}{\sqrt{n}}\sum_{i=1}^n (-1)^{N_b^{\alpha \cap \beta_i}}i^{|\alpha \cap \beta_i|} X_{\beta_i} Z_{\alpha \cap \beta_i} \ket*{\bm{0}}
\end{equation}

\noindent $Z_{\alpha \cap \beta_i}$ acts trivially on the all-zeros state so we can rewrite this as

\begin{equation}
    A_\alpha \ket*{\overline{0}} = \frac{1}{\sqrt{n}}\sum_{i=1}^n i^{g(\alpha \cap \beta_i)} X_{\beta_i} \ket*{\bm{0}}
\end{equation}

\noindent where $g(\alpha \cap \beta_i) = |\alpha \cap \beta_i| + 2N_b^{\alpha \cap \beta_i}$. We can see that 

\begin{equation}
    \begin{split}
        (|\alpha \cap \beta_i| + 2N_b^{\alpha \cap \beta_i}) \mod 4 &= (N_w^{\alpha \cap \beta_i} + 3N_b^{\alpha \cap \beta_i}) \mod 4 \\
        &= (N_w^{\alpha \cap \beta_i} - N_b^{\alpha \cap \beta_i}) \mod 4
    \end{split}
\end{equation}

\noindent so $g(\alpha \cap \beta_i)$ can be understood as the difference (mod 4) between the number of $b$ and $w$ vertices in $\alpha \cap \beta_i$. $\alpha$ is a set of faces of the code and the intersection of any face (or product of faces) with a cell (or product of cells) of the 3D colour code is even so $i^{g(\alpha \cap \beta_i)} = \pm 1$ and the effect of $A_\alpha$ on $\ket*{\overline{0}}$ is to flip the sign of some of the terms in this superposition as in the surface code case. We can then once again write our state as 

\begin{equation}
    A_\alpha \ket*{\overline{0}} = a\ket*{\phi_i^+} + b\ket*{\phi_i^-}
\end{equation}

\noindent where $X_{\beta_i} \ket*{\phi_i^+} = \ket*{\phi_i^+}$ and $X_{\beta_i} \ket*{\phi_i^-} = - \ket*{\phi_i^-}$. The following lemma will be useful in finding values for $a$ and $b$. 

\begin{lemma}
    $i^{g(\alpha \cap (\beta_i + \beta_j))} \neq i^{g(\alpha \cap \beta_i)}i^{g(\alpha \cap \beta_j)}$ only if $|\alpha \cap \beta_i \cap \beta_j|$ is odd ($\beta_i + \beta_j$ is pointwise addition modulo 2)
\end{lemma}

If $\beta_i$ and $\beta_j$ are disjoint then $\beta_i + \beta_j = \beta_i \cup \beta_j$ and so $\alpha \cap (\beta_i \cup \beta_j) = (\alpha \cap \beta_i) \cup (\alpha \cap \beta_j) = (\alpha \cap \beta_i) + (\alpha \cap \beta_j)$. This means 

\begin{equation}
    \begin{split}
        g(\alpha \cap (\beta_i + \beta_j)) &= |\alpha \cap (\beta_i + \beta_j)| + 2N_b^{\alpha \cap (\beta_i + \beta_j)} \\
                                           &= |(\alpha \cap \beta_i) + (\alpha \cap \beta_j)| + 2N_b^{(\alpha \cap \beta_i) + (\alpha \cap \beta_j)} \\
                                           &= |\alpha \cap \beta_i| + |\alpha \cap \beta_j| + 2N_b^{\alpha \cap \beta_i} + 2N_b^{\alpha \cap \beta_j} \\
                                           &= g(\alpha \cap \beta_i) + g(\alpha \cap \beta_j)
    \end{split}
\end{equation}

\noindent where we have used that $|a + b| = |a| + |b|$ for disjoint $a$ and $b$. Therefore we only have $i^{g(\alpha \cap (\beta_i + \beta_j))} \neq i^{g(\alpha \cap \beta_i)}i^{g(\alpha \cap \beta_j)}$ if $\beta_i \cap \beta_j$ is nonempty. In this case we can write $\beta_i \cup \beta_j = \beta_i' + \beta_j' + \beta_i \cap \beta_j$ where $\beta_i'$ ($\beta_j'$) are the elements of $\beta_i$ ($\beta_j$) not in $\beta_i \cap \beta_j$. $\beta_i + \beta_j = \beta_i' + \beta_j'$ and $\beta_i'$, $\beta_j'$ and $\beta_i \cap \beta_j$ are all disjoint, so by the same method as above we can show

\begin{equation}
    g(\alpha \cap (\beta_i + \beta_j)) = g(\alpha \cap (\beta_i' + \beta_j')) = g(\alpha \cap \beta_i') + g(\alpha \cap \beta_j')
\end{equation}

\noindent and

\begin{equation}
    \begin{split}
        g(\alpha \cap \beta_i) + g(\alpha \cap \beta_j) &= g(\alpha \cap (\beta_i' + \beta_i \cap \beta_j)) + g(\alpha \cap (\beta_j' + \beta_i \cap \beta_j)) \\
                                                        &= g(\alpha \cap \beta_i') + g(\alpha \cap \beta_j') + 2g(\alpha \cap \beta_i \cap \beta_j)
    \end{split}
\end{equation}

Thus for general $\beta_i$ and $\beta_j$ we have

\begin{equation}
    \label{eq:odd}
    i^{g(\alpha \cap \beta_i)}i^{g(\alpha \cap \beta_j)} = (-1)^{g(\alpha \cap \beta_i \cap \beta_j)}i^{g(\alpha \cap (\beta_i + \beta_j))}
\end{equation}

\noindent and so $i^{g(\alpha \cap (\beta_i + \beta_j))} \neq i^{g(\alpha \cap \beta_i)}i^{g(\alpha \cap \beta_j)}$ only if $g(\alpha \cap \beta_i \cap \beta_j)$ is odd which means $|\alpha \cap \beta_i \cap \beta_j|$ is also odd. \hfill $\square$

\hfill

\textbf{Corollary 1:} If $Z_{\alpha \cap \beta_i}$ is a stabiliser then $i^{g(\alpha \cap (\beta_i + \beta_j))} = i^{g(\alpha \cap \beta_i)}i^{g(\alpha \cap \beta_j)} ~ \forall ~ \beta_j$. This is because the intersection of any $X$ and $Z$ stabiliser of the code must be even and so $\alpha \cap \beta_i \cap \beta_j$ must be even if $\alpha \cap \beta_i$ is the support of a $Z$ stabiliser. 

\hfill

Now let us reconsider the state 

\begin{equation}
    \ket*{\phi} = A_\alpha \ket*{\overline{0}} = \frac{1}{\sqrt{n}}\sum_{i=1}^n i^{g(\alpha \cap \beta_i)} X_i \ket*{\bm{0}} =  a\ket*{\phi_i^+} + b\ket*{\phi_i^-} 
\end{equation}

Note that once again $\ket*{\phi_i^+}$ must have the form 

\begin{equation}
    \ket*{\phi_i^+} = \sum_j (X_{\beta_j} + X_{\beta_i}X_{\beta_j})\ket*{\bm{0}}
\end{equation}

\noindent while 

\begin{equation}
    \ket*{\phi_i^-} = \sum_j (X_{\beta_j} - X_{\beta_i}X_{\beta_j})\ket*{\bm{0}}.
\end{equation}

The pairs in $\ket*{\phi_i^+}$ are those for which $i^{g(\alpha\cap\beta_j)} = i^{g(\alpha \cap (\beta_i + \beta_j))}$ while the pairs in $\ket*{\phi_i^-}$ are those where $i^{g(\alpha\cap\beta_j)} = -i^{g(\alpha \cap (\beta_i + \beta_j))}$. This means that if $i^{g(\alpha\cap\beta_i)} = 1$ we must have $i^{g(\alpha \cap (\beta_i + \beta_j))} = i^{g(\alpha \cap \beta_i)}i^{g(\alpha \cap \beta_j)}$ for terms of $\ket*{\phi_i^+}$ and $i^{g(\alpha \cap (\beta_i + \beta_j))} = -i^{g(\alpha \cap \beta_i)}i^{g(\alpha \cap \beta_j)}$ for terms of $\ket*{\phi_i^-}$, whereas if $i^{g(\alpha\cap\beta_i)} = -1$ then the reverse is true. 

Now consider the same options for $X_i$ as in the surface code (readers may find it helpful to return to \cref{subfig:s_error_3d_a}, which shows the kind of error we are considering here): 

\begin{itemize}
    \item Stabilisers not supported on the membrane boundary: The fact that $i^{g(\alpha \cap \beta_i)} = 1$ in this case is part of the requirement for transversal $T$ (see corollary 7 of \citep{rengaswamy_optimality_2020}). Additionally, the intersection of these $X$ stabilisers with the membrane is either nothing or the support of a $Z$ stabiliser and so we always have $i^{g(\alpha \cap (\beta_i + \beta_j))} = i^{g(\alpha \cap \beta_i)}i^{g(\alpha \cap \beta_j)}$ for this choice of $X_i$ by corollary 1. This means that $\ket*{\phi} = \ket*{\phi_i^+}$ in this case, and so we are in a $+1$ eigenstate of these stabilisers.
    \item Stabiliser generators on the membrane boundary: Our error membrane is supported on faces of colour $\kappa_1\kappa_2$ and detected by faces of colour $\kappa_3\kappa_4$, which are the interfaces of $\kappa_1$ and $\kappa_2$ cells on the membrane boundary. A $\kappa_1$ cell meets the membrane at $\kappa_2$ coloured edges which each contain one $w$ and one $b$ vertex and so the total numbers of each in $\alpha \cap \beta_i$ are equal and $i^{g(\alpha \cap \beta_i)} = 1$ for these stabilisers. The intersection of a $\kappa_1\kappa_2$ face, a $\kappa_1$ cell and a $\kappa_2$ cell is a single vertex (since the $\kappa_1$ cell meets the face at a $\kappa_2$ edge, the $\kappa_2$ cell meets the face at a $\kappa_1$ edge and these edges meet at a vertex) and therefore $|\alpha \cap \beta_i \cap \beta_j|$ is odd for a pair of neighbouring cells on the membrane boundary. Thus $\ket*{\phi_i^+}$ is formed from pairs $(X_{\beta_j} + X_{\beta_i}X_{\beta_j})\ket*{\bm{0}}$ where $X_{\beta_j}$ contains either 0 or 2 such neighbours of $X_{\beta_i}$ while $\ket*{\phi_i^-}$ contains all pairs $(X_{\beta_j} - X_{\beta_i}X_{\beta_j})\ket*{\bm{0}}$ where $X_{\beta_j}$ contains only one of the neighbours of $X_{\beta_i}$. There are equal numbers of each type of pair so $\ket*{\phi} = \frac{1}{\sqrt{2}} (\ket*{\phi_i^+} + \ket*{\phi_i^-})$ and we expect a random $\pm1$ outcome from a measurement of $X_{\beta_i}$. 
    \item All stabiliser generators of one colour on the membrane boundary: For any $\kappa_1$ cell $X_{\beta_i}$ and neighbouring $\kappa_2$ cell $X_{\beta_j}$ that are both on the membrane boundary we have that $|\alpha \cap \beta_i \cap \beta_j|$ is odd, and so $Z_{\alpha \cap \beta_i}$ is an error string that anticommutes with the two $\kappa_2$ coloured neighbours of $X_{\beta_i}$. If $X_{\beta_i}$ is instead the product of all $\kappa_1$ coloured cells on the membrane boundary then each $\kappa_2$ cell anticommutes individually with the string from each of its two $\kappa_1$ coloured neighbours and so commutes with their product. Thus $Z_{\alpha \cap \beta_i}$ is a stabiliser for this choice of $X_{\beta_i}$. Additionally, $i^{g(\alpha \cap \beta_i)} = 1$ (since $i^{g(\alpha\cap\beta_i)} = 1$ when $X_{\beta_i}$ is an individual cell and cells of the same colour are disjoint). Therefore we have $\ket*{\phi} = \ket*{\phi_i^+}$ for this $X_{\beta_i}$ as well.  
\end{itemize}

We have recoved the expected result for an isolated membrane: If we measure all stabiliser generators of the code then stabilisers not on the membrane boundary will return $+1$. Stabilisers on the membrane boundary return random $\pm1$ outcomes, but the total parity of $-1$ stabilisers of any colour will always be even. 

\subsection{Linked Error Membranes in Cleanable Code Regions}

Now we consider a pair of membranes of $X$ errors with one defined on $\kappa_1\kappa_2$ faces and detected by $\kappa_3\kappa_4$ faces and one defined on $\kappa_3\kappa_4$ faces and detected by $\kappa_1\kappa_2$ faces as in \cref{subfig:s_error_3d_b}. We will refer to these as $\alpha_{\kappa_1\kappa_2}$ and $\alpha_{\kappa_3\kappa_4}$ respectively. Following the application of $\overline{T}$ we have a state

\begin{equation}
    \begin{split}
    \overline{T}X_{\alpha_{\kappa_1\kappa_2}}X_{\alpha_{\kappa_3\kappa_4}}\ket*{\overline{\psi}}
        &= \overline{T}X_{\alpha_{\kappa_1\kappa_2} + \alpha_{\kappa_3\kappa_4}}\ket*{\overline{\psi}} \\
        &= X_{\alpha_{\kappa_1\kappa_2} + \alpha_{\kappa_3\kappa_4}} A_{\alpha_{\kappa_1\kappa_2} + \alpha_{\kappa_3\kappa_4}} \ket*{\overline{\psi'}}.
    \end{split}
\end{equation}

\noindent by the same reasoning as \cref{eq:TXcommute} and using the fact that the membranes are individually defined on the supports of $Z$ stabilisers (sets of faces) and so their product is also the support of a $Z$ stabiliser. Notice that, unlike in the surface code, we do not observe the emergence of a linking charge string at this point and in fact we observe no errors on the intersection at all since $\alpha_{\kappa_1\kappa_2} + \alpha_{\kappa_3\kappa_4} = \alpha_{\kappa_1\kappa_2} \cup \alpha_{\kappa_3\kappa_4} - \alpha_{\kappa_1\kappa_2} \cap \alpha_{\kappa_3\kappa_4}$. The linking charge string in this case will come from the action of the Clifford error on the codestate rather than directly from the commutation of the transversal non-Clifford through the original $X$ error.  

Much of the analysis from above carries over to this case, and the only difference will be for $X_{\beta_i}$ partially supported on the intersection of the two membranes. Note that the same method used to prove (\ref{eq:odd}) can equivalently be used to show

\begin{equation}
    \label{eq:odd2}
    i^{g((\alpha_{\kappa_1\kappa_2} + \alpha_{\kappa_3\kappa_4}) \cap \beta_i)} = (-1)^{g(\alpha_{\kappa_1\kappa_2} \cap \alpha_{\kappa_3\kappa_4} \cap \beta_i)}i^{g(\alpha_{\kappa_1\kappa_2} \cap \beta_i)}i^{g(\alpha_{\kappa_3\kappa_4} \cap \beta_i)}
\end{equation}

Then we have that

\begin{itemize}
    \item $X$ stabilisers at intersection endpoints: These stabilisers are on the boundary of one membrane and in the interior of the other and contain a single qubit in $\alpha_{\kappa_1\kappa_2} \cap \alpha_{\kappa_3\kappa_4}$. If this cell has colour $\kappa_1$ then it must meet the $\kappa_1\kappa_2$ membrane at a set of $\kappa_2$ coloured edges and the $\kappa_3\kappa_4$ membrane at a $\kappa_3\kappa_4$ coloured face. Sets of disjoint edges and individual faces both contain equal numbers of $b$ and $w$ vertices so $i^{g(\alpha_{\kappa_1\kappa_2} \cap \beta_i)} = i^{g(\alpha_{\kappa_3\kappa_4} \cap \beta_i)} = 1$. $\alpha_{\kappa_1\kappa_2} \cap \alpha_{\kappa_3\kappa_4} \cap \beta_i$ is a single qubit so $i^{g(\alpha \cap \beta_i)} = -1$ by (\ref{eq:odd2}). 
    \item $X$ stabilisers on the intersection (not endpoints). $Z_{\alpha_{\kappa_1\kappa_2} \cap \beta_i}$ and $Z_{\alpha_{\kappa_3\kappa_4} \cap \beta_i}$ are both $Z$ stabilisers so $i^{g(\alpha_{\kappa_1\kappa_2} \cap \beta_i)} = i^{g(\alpha_{\kappa_3\kappa_4} \cap \beta_i)} = 1$. The product of two $Z$ stabilisers is another $Z$ stabiliser, and all $Z$ stabilisers have even weight so $|(\alpha_{\kappa_1\kappa_2} \cap \beta_i) + (\alpha_{\kappa_3\kappa_4} \cap \beta_i)|$ is even and so $|\alpha_{\kappa_1\kappa_2} \cap \alpha_{\kappa_3\kappa_4} \cap \beta_i|$ is also even and $i^{g(\alpha \cap \beta_i)} = 1$. 
\end{itemize}

For non-endpoint stabilisers everything is as before. For endpoint stabiliser generators we once again have $\ket*{\phi} = \frac{1}{\sqrt{2}} (\ket*{\phi_i^+} + \ket*{\phi_i^-})$ but we have swapped which pairs of states are in $\ket*{\phi_i^+}$ and $\ket*{\phi_i^-}$ since, e.g. $\ket*{\phi_i^-}$ now contains the pair $\ket*{\bm{0}} - X_{\beta_i}\ket*{\bm{0}}$ whereas previously we had $\ket*{\bm{0}} + X_{\beta_i}\ket*{\bm{0}}$ in $\ket*{\phi_i^+}$. If $X_{\beta_i}$ is all cells of one colour on one of the membrane boundaries and $X_{\beta_j}$ is a single cell on this boundary then as before $Z_{\alpha\cap\beta_i}$ is a stabiliser and as before $i^{g(\alpha \cap \beta_i)}$ is the product of $i^{g(\alpha\cap\beta_j)}$ for all individual cells. One of these cells sits at an intersection endpoint and so has $i^{g(\alpha\cap\beta_j)} = -1$ whereas the rest have $i^{g(\alpha\cap\beta_j)} = 1$ and so $i^{g(\alpha \cap \beta_i)} = -1$. Thus we now have $\ket*{\phi} = \ket*{\phi_i^-}$ whereas before we had $\ket*{\phi} = \ket*{\phi_i^+}$. This implies that when we measure all the stabiliser generators of the code we will once again get random outcomes from the membrane boundary stabilisers, but instead of having an even parity of each colour on each boundary we have an odd parity, and this is consistent with a linking charge string running between the two membrane boundaries. 

\subsection{Error Membranes in Non-Cleanable Regions}

Consider an $X$ error membrane in the colour code supported on a subset of faces of colour $\kappa_1\kappa_2$ and also on the support of a logical $Z$ operator. As before we have

\begin{equation}
    \overline{T} X_\alpha \ket*{\overline{\psi}} = e^{-i\pi N_w^\alpha/4}e^{-i\pi N_b^\alpha/4}i^{g(\alpha)} X_\alpha Z_\alpha A_\alpha \ket*{\overline{\psi'}}. 
\end{equation}

$\alpha$ is a product of the supports of $Z$ stabilisers and a $Z$ logical. For the former $N_w = N_b$ and for the latter $N_w = N_b + 1$ so $e^{-i\pi N_w^\alpha/4}e^{-i\pi N_b^\alpha/4} = e^{-i\pi/4}$. Also $g(\alpha) = |\alpha| + 2N_b^\alpha = N_w^\alpha + N_b^\alpha + 2N_b^\alpha = 4N_b^\alpha + 1$ so $i^{g(\alpha)} = i$. This gives

\begin{equation}
    \label{eq:zerror}
    \overline{T} X_\alpha \ket*{\overline{\psi}} = e^{i\pi/4}X_\alpha Z_\alpha A_\alpha \ket*{\overline{\psi'}}
\end{equation}

Notice that $Z_\alpha$ is a logical $Z$ operator, whereas for cleanable $\alpha$ it was a stabiliser. Now we want to consider the effect of $A_\alpha$ on codestates. For $\ket*{\overline{0}}$ the analysis is the same as before, but for $\ket*{\overline{1}}$ we must now consider the interaction of logical $X$ operators with $A_\alpha$. $\ket*{\overline{1}}$ can be written 

\begin{equation}
    \ket*{\overline{1}} = \frac{1}{\sqrt{n}} \sum_{i=1}^n \overline{X} X_{\beta_i} \ket*{\bm{0}}
\end{equation}

\noindent where $\overline{X}$ is a logical $X$ operator. It does not matter which implementation of $\overline{X}$ we choose, so we choose it to be $X$ on all qubits. We then have that

\begin{equation}
    A_\alpha \ket*{\overline{1}} = \frac{1}{\sqrt{n}} \sum_{i=1}^n i^{g(\alpha)} \overline{X} Z_\alpha i^{g(\alpha \cap \beta_i)} X_{\beta_i} Z_{\alpha \cap \beta_i} \ket*{\bm{0}}
                                = \frac{i}{\sqrt{n}} \sum_{i=1}^n i^{g(\alpha \cap \beta_i)} \overline{X} X_{\beta_i} \ket*{\bm{0}}
\end{equation}

\noindent where we have used that $Z_\alpha$ is a logical $Z$ operator and so commutes with $X_{\beta_i}$ which is a stabiliser. Thus we see that the action of $A_\alpha$ on $\ket*{\overline{1}}$ is the same as in the cleanable case except for a global factor of $i$. This is consistent with a logical $S$ error and so we conclude that in addition to creating distributions of $Z$ errors $A_\alpha$ also applies a logical $S$ to the codestate. Notice that in addition to this, commuting $\overline{T}$ through $X_\alpha$ created a logical $Z$ error $Z_\alpha$ in (\ref{eq:zerror}).

\subsection{$CCZ$ and $CS$ Between Multiple Colour Codes}

We can also apply $CCZ$ between three copies of the 3D colour code and $CS$ between two copies (this follows from the fact that $CNOT$ and $T$ are transversal in this code and $CCZ$/$CS$ can be synthesised exactly using these gates \cite{beverland_lower_2020,nielsen_quantum_2011}). For $CCZ$ the effect of the resulting $CZ$ errors on terms in the colour code codestates will be identical to \cref{eq:cz_codestate} as this equation is not code-specific (provided the code is CSS and so has codestates which can be written in this form). Then to find $\ket{\phi_q^+}$ and $\ket{\phi_q^-}$ we can note that the only code properties we assumed in this calculation for the surface code were a.) that an $X$ error membrane is detected by $Z$ stabilisers at its boundary and b). that the intersection of $X$ stabilisers of codes $i$ and $j$ is a $Z$ stabiliser of code $k$. Both of these properties are true for the colour code, so the result is exactly the same as for the surface code case. 

The case of $CS$ (which can be applied transversally using the same bicolouring as the $T$ gate) between two colour codes is more interesting. $CS$ has the commutation relations

\begin{equation}
    \begin{split}
        (CS)(X \otimes I) &= (X \otimes S)(CZ)(CS) \\
        (CS^\dag)(X \otimes I) &= (X \otimes S^\dag)(CS)
    \end{split}
\end{equation}

If we then consider applying $CS$ to a pair of colour codes, one of which contains an $X$ error $X_\alpha^1$, then we have

\begin{equation}
    \overline{CS}X_\alpha^1\ket{\overline{\psi}} = X_\alpha^1A_\alpha^2CZ_\alpha^{12}\ket{\overline{\psi '}}
\end{equation}

\noindent and so we obtain both an $S$ error and a $CZ$ error. We have already examined both of these errors individually so we can use those results to see that the effect on the terms of the codestate will be 

\begin{equation}
    A_\alpha^2CZ_\alpha^{12}\ket{\overline{00}} = 
    \frac{1}{\sqrt{n}}\sum_{ij}i^{g(\alpha \cap \beta_j)}(-1)^{\abs{\alpha \cap \beta_i \cap \beta_j}}X_{\beta_i}^1X_{\beta_j}^2\ket{\bm{0}}
\end{equation}

From our previous analysis we know that $i^{g(\alpha \cap \beta_j)} = -1$ when $\beta_j$ contains the support of two neighbouring cells on the membrane boundary while $(-1)^{\abs{\alpha \cap \beta_i \cap \beta_j}} = -1$ when $\beta_i$ contains the support of a cell on the membrane boundary and $\beta_j$ contains the support of a single neighbour of that cell (also on the membrane boundary). For a given $X$ stabiliser generator in code $1$ only the latter contribution is relevant (because the qubits of this code are acted on only by $CZ$ and not by $S$) so the argument is identical to the case of errors due to $CCZ$. For a given $X$ stabiliser generator in code $2$ the former contribution depends only on $\beta_j$ while the latter depends only on $\beta_i$ and so the two are independent.

Thus we conclude that application of $CS$ to a pair of codes containing error $X_\alpha^1$ results in random $Z$ error distributions on the boundary of $\alpha$ in both codes $1$ and $2$, with the former being due just to a $CZ$ error and the latter being a product of errors produced by $CZ$ and $S$ (but still being identical to a distribution that could be produced by either of these errors individually, i.e. a $-1$ outcome from any given stabiliser with probability $p=0.5$ but an even number of $-1$s overall). 

We can also examine linking charge for the case of $CS$. The calculation is simply

\begin{equation}
    \begin{split}
        \overline{CS}X_\alpha^1X_\gamma^2\ket{\overline{\psi}} 
        &= X_\alpha^1A_\alpha^2CZ_\alpha^{12}X_\gamma^2A_\gamma^1CZ_\gamma^{12}\ket{\overline{\psi'}}\\
        &= X_\alpha^1X_\gamma^2((-1)^{N_b^\gamma}i^{\abs{\alpha \cap \gamma}}Z_{\alpha \cap \gamma}^2)(Z_{\alpha \cap \gamma}^1)\ket{\psi''}
    \end{split}
\end{equation}

\noindent where the first bracketed linking charge term (in code 2) comes from the $S$ part of the error and the second one (in code 1) comes from the $CZ$ part. The phases incurred are global and so not a problem. $\ket{\psi''}$ is the state $\ket{\overline{\psi'}}$ multiplied by the $CZ$ and $A$ Clifford errors. 

\section{Algorithms for Calculating Dimension Jump Corrections}
\label{appendix:jump}
In this appendix we describe the algorithms used in our simulation for calculating the corrections applied to the 2D code after the dimension jump. At this point in the procedure we have measured out all inner qubits of the 3D code in $X$ and calculated a correction for these qubits. We therefore have a (corrected) pattern of $\pm1$ outcomes from these measurements which should match some $Z$ stabiliser of the 3D code. The aim of the algorithms in this appendix is to use these outcomes to find a correction for any error which may have arisen in the 2D code due to these projective measurements.  

\begin{figure}
    \centering
    \begin{subfigure}{.48\textwidth}
        \includegraphics[width=\textwidth]{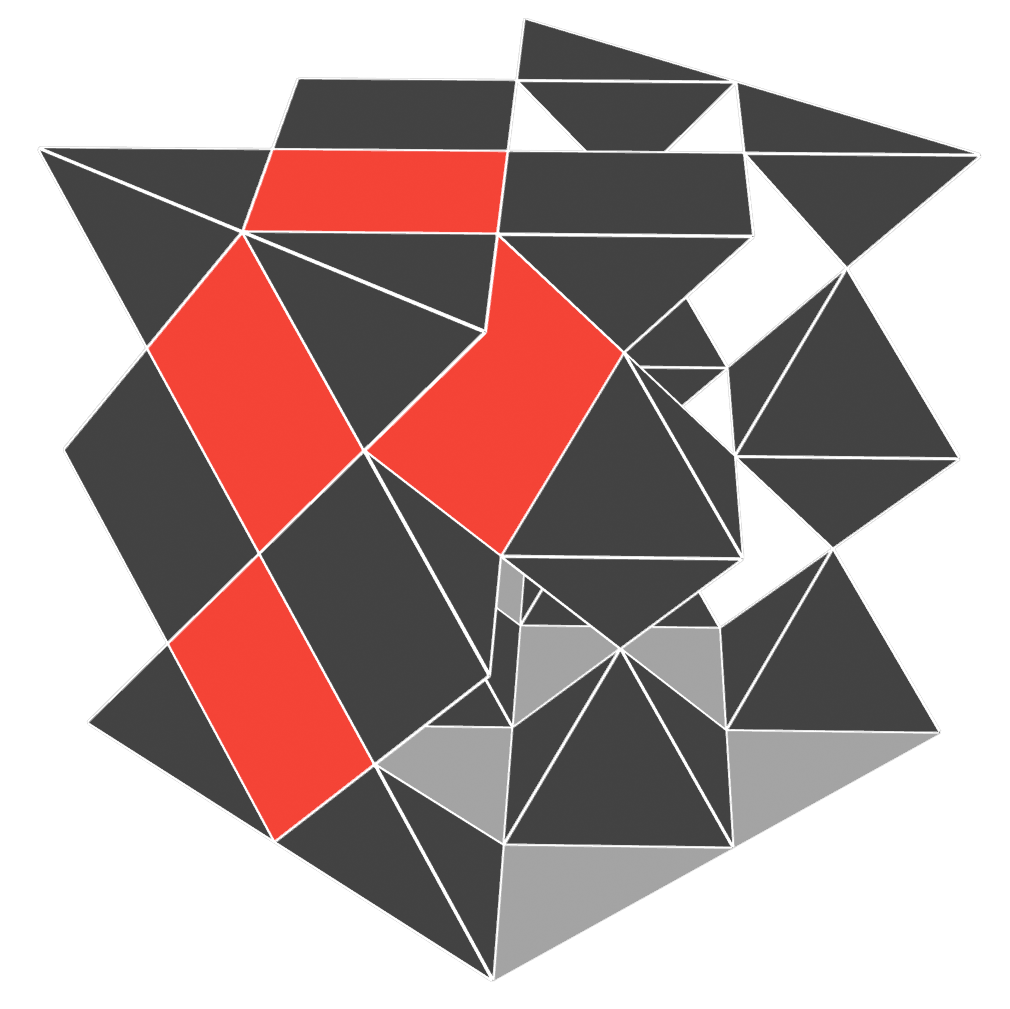}
        \subcaption{}
        \label{subfig:jump_corrA}
    \end{subfigure}
    ~
    \begin{subfigure}{.48\textwidth}
        \includegraphics[width=\textwidth]{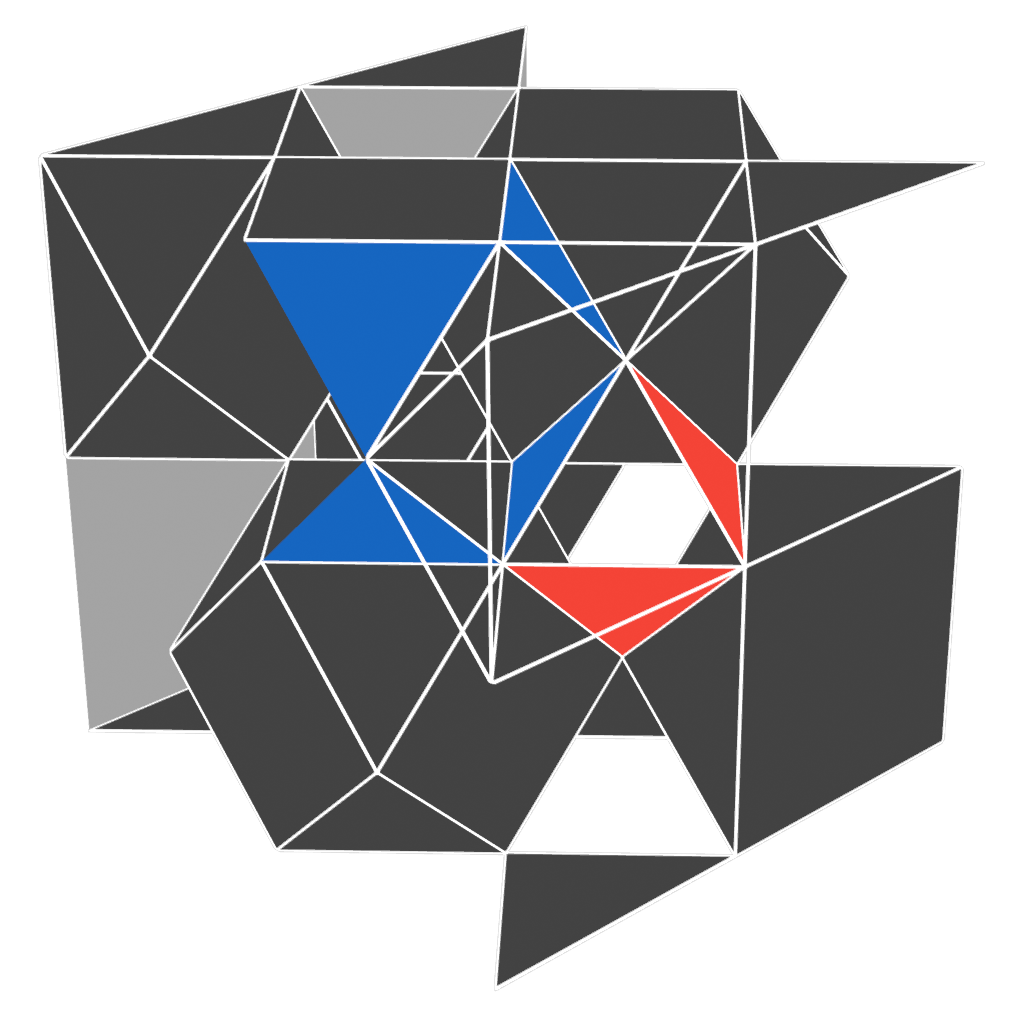}
        \subcaption{}
        \label{subfig:jump_corrB}
    \end{subfigure}
    \caption{(Colour) (a) Z stabiliser structure in the octahedral lattice. (b) Z stabiliser structure in the cuboctahedral lattice}
    \label{fig:jump_corr}
\end{figure}

We first discuss the simpler case of the octahedral surface code. An example is shown in \cref{subfig:jump_corrA}. In this code the $Z$ stabiliser generators are weight-4 (some shown in red) and we wish to collapse to a half-octahedral boundary (shown in grey). To calculate a correction we start at the top boundary (opposite to the boundary we wish to collapse to) and examine the single-qubit measurement outcomes from qubits on this boundary. From \cref{subfig:jump_corrA} we can see that the $Z$ stabiliser generators that touch this boundary are either supported entirely on the boundary or only on a single qubit of it. We assume that any $-1$ outcomes on this boundary are due to this second type of stabiliser, so that even if we obtain four $-1$ outcomes in the support of a single $Z$ stabiliser generator we assume that this is due instead to the product of four separate generators. In this way we obtain one generator for each $-1$ outcome on this boundary. We then flip the recorded signs of the measurement outcomes from qubits within the support of the product of all these generators. This guarantees that the top boundary now contains only $+1$s, and also that the plane of qubits immediately below this will also contain only $+1$s since (up to composition with generators supported only on the top boundary) the pattern of $-1$s on the top boundary uniquely specifies a pattern of generators that matches the $-1$s on this plane. The only mistake we can have made is to apply four generators partially supported on the boundary instead of a single generator fully supported on the boundary, as mentioned above. The effect of this is to remove these four $-1$s from the top plane of qubits but also to flip the four corresponding outcomes in the plane two layers below. This plane is identical to the top boundary and so we can repeat the process, terminating when we arrive at the bottom boundary. The final result will be the removal of all $Z$ errors in the 2D code which arise due to 3D code stabilisers partially supported on the outer qubits. Any extra $Z$s applied due to ``mistakes'' of the kind described above will actually be $Z$ stabilisers of the 2D code. 

The cuboctahedral lattices are more complicated. We see one such lattice and some example $Z$ stabilisers in \cref{subfig:jump_corrB}, where one cell has been removed so that we can see the interior. The boundary we wish to collapse to is once again shown in grey, and this time it is a half-cuboctahedral boundary. Once again we start at the boundary opposite the one we wish to collapse to, which this time is the front-right boundary. There are no $Z$ stabiliser generators fully supported on this boundary, and partially supported generators always meet it at a single qubit (examples shown in red). These generators come in pairs, so unlike in the octahedral lattice we cannot uniquely specify a stabiliser generator based on a $-1$ outcome of a qubit on this boundary. Instead we choose one of these two generators at random. Flipping the signs of the associated measurement outcomes as before means this boundary now contains only $+1$s and although we may have mistakenly chosen the ``wrong'' generator this will turn out not to matter.

\begin{figure}
    \centering
    \includegraphics[width=.6\textwidth]{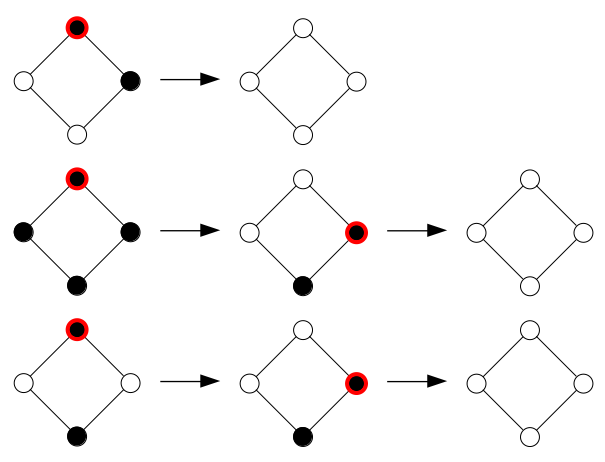}
    \caption{(Colour) Examples of corrections calculated for $-1$ measurement outcomes from quadruples of qubits in the cuboctahedral lattice. $+1$ ($-1$) outcomes are shown by empty (filled) circles. At each step we examine the neighbours of the qubit marked with a red circle and modify the recorded measurement outcomes according to the algorithm described in the text.}
    \label{fig:quad_corrections}
\end{figure}

In the next plane the $Z$ stabiliser generators are supported on two qubits of the plane rather than one. Notice also that if we consider only the generators that ``lead towards'' the boundary we wish to collapse to (shown in blue) and ignore the ones that ``lead away'' (red) then they come in sets of four. Each blue generator shares qubits (in this plane) only with other blue generators in this set of four and so each set can be considered individually. The component of the correction based on this layer is calculated by proceeding clockwise around each set of four qubits. If the measurement outcome from a given qubit is recorded as $-1$ then we check the outcomes of the clockwise and anticlockwise neighbours of this qubit. If exactly one of these two outcomes is $-1$ then we identify the generator supported on this pair of qubits and flip all recorded signs for measurement outcomes in its support. If both or neither of the neighbours of this qubit have outcome $-1$ then we randomly choose one of the two generators with this qubit in its support and flip the associated measurement outcomes. Examples are shown in \cref{fig:quad_corrections}. Once this process has been completed for all quadruples of qubits in the plane all recorded outcomes from qubits within this plane will be $+1$. The plane below this matches the front boundary plane and so the process starts again and repeats until we reach the final boundary where the 2D code is supported.

Now we can consider the effects of making the wrong selection during the random choices of generator involved in this algorithm. Firstly, consider the case were we choose the wrong generator while finding a correction for the front boundary (or any other matching plane). These are the red faces in \cref{subfig:jump_corrB}, so we can see that the product of the two generators that share a qubit of this boundary is a four-qubit operator supported on two different qubit quadruples on the next plane. This will lead us to flip the outcomes from qubits associated with two generators (one supported on each quadruple) that meet at a single qubit in the subsequent plane. This means that the outcome from this qubit will be flipped twice, and so not flipped at all and thus the original ``incorrect'' choice of generator causes no problems. For the random choices involved in finding corrections from the quadruples, the different choices of generator lead to corrections for the final 2D code that differ only by stabilisers of the code, and so once again cause no issues. 

\section{Constant-Time Decoding at the 2D/3D Code Boundary}
\label{appendix:decoding}
\begin{figure}
    \centering
    \begin{subfigure}{.3\textwidth}
        \includegraphics[width=\textwidth]{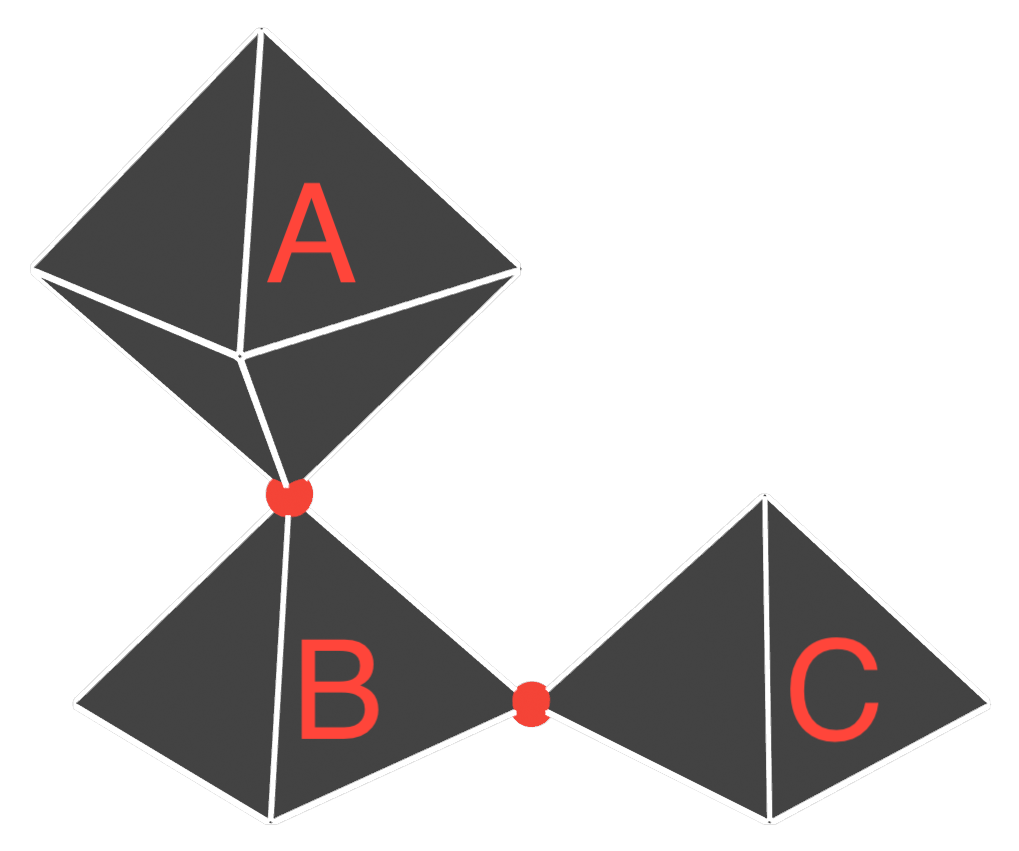}
        \subcaption{}
        \label{subfig:boundaryA}
    \end{subfigure}
    ~~~
    \begin{subfigure}{.3\textwidth}
        \includegraphics[width=\textwidth]{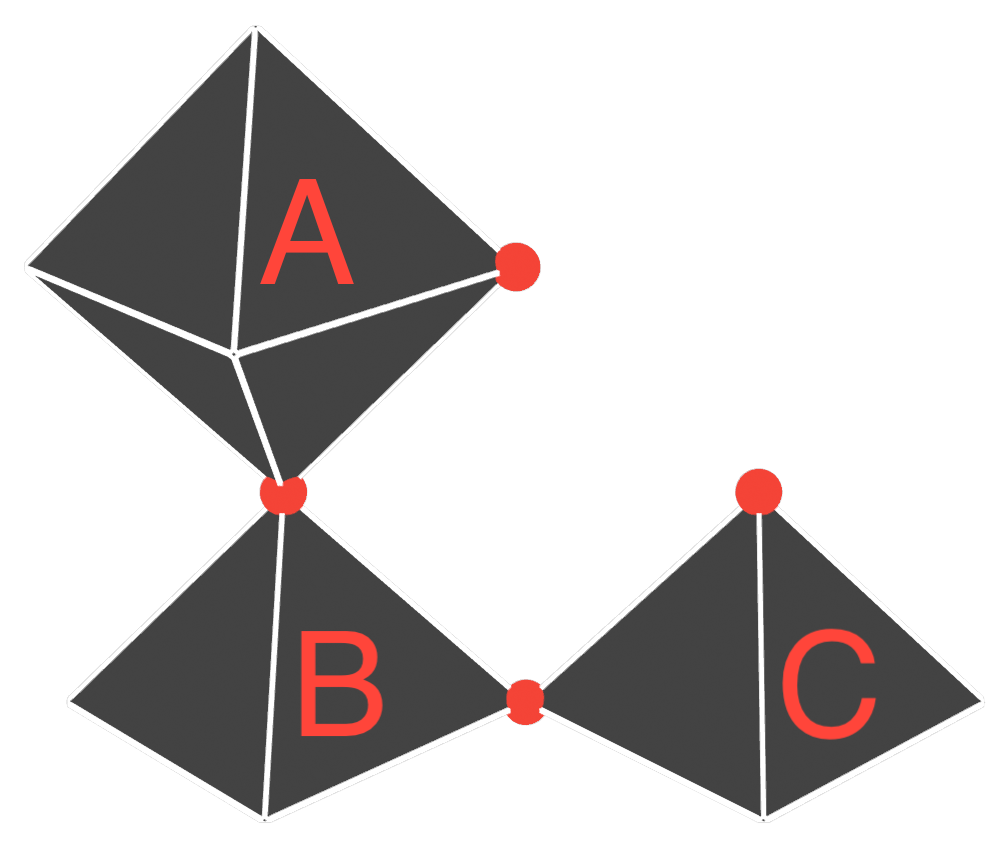}
        \subcaption{}
        \label{subfig:boundaryB}
    \end{subfigure}
    \caption{(Colour) (a) A two-qubit $Z$ error which anticommutes with the $X$ stabilisers on cells labelled A and C. (b) A four-qubit $Z$ stabiliser}
    \label{fig:boundary}
\end{figure}

To calculate this correction we assume that all 2D stabiliser measurement outcomes will return a value of $+1$. This means that the measurement outcome we calculate for cells that are partially supported on the outer qubits is just the product of all the outcomes of the measured-out qubits of this cell. To understand why this works we can consider the example in \cref{fig:boundary}. In \cref{subfig:boundaryA} we have a two-qubit $Z$ error on the marked qubits, for which the correct 3D syndrome would be a $-1$ outcome from the cells labelled $A$ and $C$. We imagine that the bottom boundary of the figure is the 2D code we wish to collapse to, so a syndrome calculation that uses only the inner (non-boundary) qubits will give $-1$ outcomes for cells $A$ and $B$, for which a valid correction would be the qubit in the intersection of these two cells. In other words, calculating stabiliser outcomes using only inner qubit measurements will only give corrections for inner qubit errors. This may seem like a rather trivial statement, but the fact that it can be made consistent with the rest of the dimension jump such that the entire procedure remains fault-tolerant actually relies on a rather specific feature of the codes we have chosen. To see this, we can consider \cref{subfig:boundaryB} where the 4-qubit $Z$ error is actually a $Z$ stabiliser of the 3D code. In this case, calculating a 3D $X$ syndrome using only the inner qubits will give a $-1$ outcome from cells $B$ and $C$ even though in reality we should have no syndrome at all. It is important that we do not try to correct this ``error'' because it will be dealt with separately in the next step of the dimension jump when we find a 3D $Z$ stabiliser that matches the single-qubit measurement outcomes and apply its restriction to the 2D code (as described in the previous appendix). For example, a decoder given this syndrome would return a correction supported on the outer qubit that is in the intersection of cells $B$ and $C$, and this same correction will be returned in the next step of the jump when we identify the other three single-qubit $Z$s as the restriction of this $Z$ stabiliser to the inner qubits. As a result, the error in the outer code will be corrected twice and so will not be corrected at all.

The way around this is to recall that we began by assuming that all 2D stabiliser measurements would return an outcome of $+1$, meaning that there are no $Z$ errors on the outer qubits. For the code and boundary depicted above all 3D $Z$ stabilisers are either also 2D $Z$ stabilisers or intersect the 2D code at a single qubit. As a result, a MWPM decoder given a syndrome calculated from only the inner qubits will always return a correction supported on the outer code if the ``error'' that caused the syndrome was actually a 3D $Z$ stabiliser (since e.g. for the case of \cref{subfig:boundaryB} a correction supported on the outer qubits is weight-1 while a correction supported on the inner qubits is weight-3). We can therefore add an extra rule to our 3D $X$ decoder which says that after calculating a correction we only apply the parts of the correction that are supported on the inner qubits of the 3D code, and this ensures that ``errors'' due to 3D $Z$ stabilisers are left untouched. Notice that for the other two 3D surface codes (where the $Z$ stabilisers are supported on the triangular faces of cells $A$, $B$ and $C$) the 3D $Z$ stabilisers at the boundary shown in \cref{fig:boundary} are supported on two outer qubits and one inner qubit, so this modification of the decoder would not work if we tried to collapse these codes to this boundary. Fortunately however, we do not try to do this. Instead we collapse them to the half-cuboctahedral boundaries where the $Z$ stabilisers are supported on one outer qubit and two inner qubits. 

This decoding strategy is not without cost. We have created a large number of new $-1$ $X$ stabiliser outcomes at one boundary of the 3D code, and while these can be corrected properly in isolation 3D $Z$ errors which occur close to this boundary can interfere with this and the performance of the MWPM decoder will suffer as a result. Despite this, we still observe thresholds for the dimension jump in all three codes.

\bibliographystyle{unsrtnat}
\bibliography{references,extra_refs}

\begin{thebibliography}{52}
\providecommand{\natexlab}[1]{#1}
\providecommand{\url}[1]{\texttt{#1}}
\expandafter\ifx\csname urlstyle\endcsname\relax
  \providecommand{\doi}[1]{doi: #1}\else
  \providecommand{\doi}{doi: \begingroup \urlstyle{rm}\Url}\fi

\bibitem[Shor(1995)]{shor1995}
Peter~W. Shor.
\newblock Scheme for reducing decoherence in quantum computer memory.
\newblock \emph{Physical Review A}, 52\penalty0 (4):\penalty0 R2493--R2496,
  1995.
\newblock \doi{10.1103/PhysRevA.52.R2493}.

\bibitem[Ekert and Macchiavello(1996)]{ekert1996error}
Artur Ekert and Chiara Macchiavello.
\newblock Error correction in quantum communication.
\newblock \emph{arXiv:quant-ph/9602022}, 1996.
\newblock URL \url{https://arxiv.org/abs/quant-ph/9602022}.

\bibitem[Nielsen and Chuang(2011)]{nielsen_quantum_2011}
Michael~A. Nielsen and Isaac~L. Chuang.
\newblock \emph{Quantum {Computation} and {Quantum} {Information}: 10th
  {Anniversary} {Edition}}.
\newblock Cambridge University Press, New York, NY, USA, 10th edition, 2011.

\bibitem[Gottesman(1998)]{gottesman1998a}
Daniel Gottesman.
\newblock The {{Heisenberg Representation}} of {{Quantum Computers}}.
\newblock \emph{arXiv:quant-ph/9807006}, 1998.
\newblock URL \url{https://arxiv.org/abs/quant-ph/9807006}.

\bibitem[Wang et~al.(2003)Wang, Harrington, and Preskill]{wang2003}
Chenyang Wang, Jim Harrington, and John Preskill.
\newblock Confinement-{{Higgs}} transition in a disordered gauge theory and the
  accuracy threshold for quantum memory.
\newblock \emph{Annals of Physics}, 303\penalty0 (1):\penalty0 31--58, 2003.
\newblock \doi{10.1016/S0003-4916(02)00019-2}.

\bibitem[Tuckett(2020)]{qecsim}
David~Kingsley Tuckett.
\newblock \emph{Tailoring surface codes: Improvements in quantum error
  correction with biased noise}.
\newblock PhD thesis, University of Sydney, 2020.
\newblock (qecsim: \url{https://github.com/qecsim/qecsim}).

\bibitem[Gidney(2021)]{gidney2021stim}
Craig Gidney.
\newblock Stim: a fast stabilizer circuit simulator.
\newblock \emph{{Quantum}}, 5:\penalty0 497, 2021.
\newblock ISSN 2521-327X.
\newblock \doi{10.22331/q-2021-07-06-497}.

\bibitem[Wallman et~al.(2015)Wallman, Granade, Harper, and
  Flammia]{wallman_estimating_2015}
Joel Wallman, Chris Granade, Robin Harper, and Steven~T. Flammia.
\newblock Estimating the coherence of noise.
\newblock \emph{New Journal of Physics}, 17\penalty0 (11):\penalty0 113020,
  2015.
\newblock \doi{10.1088/1367-2630/17/11/113020}.

\bibitem[Kueng et~al.(2016)Kueng, Long, Doherty, and
  Flammia]{kueng_comparing_2016}
Richard Kueng, David~M. Long, Andrew~C. Doherty, and Steven~T. Flammia.
\newblock Comparing {Experiments} to the {Fault}-{Tolerance} {Threshold}.
\newblock \emph{Physical Review Letters}, 117\penalty0 (17):\penalty0 170502,
  2016.
\newblock \doi{10.1103/PhysRevLett.117.170502}.

\bibitem[Greenbaum and Dutton(2017)]{greenbaum_modeling_2017}
Daniel Greenbaum and Zachary Dutton.
\newblock Modeling coherent errors in quantum error correction.
\newblock \emph{Quantum Science and Technology}, 3\penalty0 (1):\penalty0
  015007, 2017.
\newblock \doi{10.1088/2058-9565/aa9a06}.

\bibitem[Bravyi et~al.(2018)Bravyi, Englbrecht, König, and
  Peard]{bravyi_correcting_2018}
Sergey Bravyi, Matthias Englbrecht, Robert König, and Nolan Peard.
\newblock Correcting coherent errors with surface codes.
\newblock \emph{npj Quantum Information}, 4\penalty0 (1):\penalty0 1--6, 2018.
\newblock \doi{10.1038/s41534-018-0106-y}.

\bibitem[Beale et~al.(2018)Beale, Wallman, Guti{\'e}rrez, Brown, and
  Laflamme]{beale2018a}
Stefanie~J. Beale, Joel~J. Wallman, Mauricio Guti{\'e}rrez, Kenneth~R. Brown,
  and Raymond Laflamme.
\newblock Quantum {{Error Correction Decoheres Noise}}.
\newblock \emph{Physical Review Letters}, 121\penalty0 (19):\penalty0 190501,
  2018.
\newblock \doi{10.1103/PhysRevLett.121.190501}.

\bibitem[Iverson and Preskill(2020)]{iverson2020}
Joseph~K. Iverson and John Preskill.
\newblock Coherence in logical quantum channels.
\newblock \emph{New Journal of Physics}, 22\penalty0 (7):\penalty0 073066,
  2020.
\newblock \doi{10.1088/1367-2630/ab8e5c}.

\bibitem[Gottesman and Chuang(1999)]{gottesman1999a}
Daniel Gottesman and Isaac~L. Chuang.
\newblock Demonstrating the viability of universal quantum computation using
  teleportation and single-qubit operations.
\newblock \emph{Nature}, 402\penalty0 (6760):\penalty0 390--393, 1999.
\newblock \doi{10.1038/46503}.

\bibitem[Yoshida(2015)]{yoshida_topological_2015}
Beni Yoshida.
\newblock Topological color code and symmetry-protected topological phases.
\newblock \emph{Physical Review B}, 91\penalty0 (24):\penalty0 245131, 2015.
\newblock \doi{10.1103/PhysRevB.91.245131}.

\bibitem[Bomb{\'i}n(2018)]{bombin_transversal_2018}
H{\'e}ctor Bomb{\'i}n.
\newblock Transversal gates and error propagation in {3D} topological codes.
\newblock \emph{arXiv:1810.09575 [quant-ph]}, 2018.
\newblock URL \url{https://arxiv.org/abs/1810.09575}.

\bibitem[Beverland et~al.(2021)Beverland, Kubica, and
  Svore]{beverland_cost_2021}
Michael~E. Beverland, Aleksander Kubica, and Krysta~M. Svore.
\newblock Cost of {Universality}: {A} {Comparative} {Study} of the {Overhead}
  of {State} {Distillation} and {Code} {Switching} with {Color} {Codes}.
\newblock \emph{PRX Quantum}, 2\penalty0 (2):\penalty0 020341, 2021.
\newblock \doi{10.1103/PRXQuantum.2.020341}.

\bibitem[Chamberland et~al.(2018)Chamberland, Iyer, and
  Poulin]{chamberland_fault-tolerant_2018}
Christopher Chamberland, Pavithran Iyer, and David Poulin.
\newblock Fault-{Tolerant} {Quantum} {Computing} in the {Pauli} or {Clifford}
  {Frame} with {Slow} {Error} {Diagnostics}.
\newblock \emph{Quantum}, 2, 2018.
\newblock \doi{10.22331/q-2018-01-04-43}.

\bibitem[Chamberland et~al.(2017)Chamberland, Wallman, Beale, and
  Laflamme]{chamberland_hard_2017}
Christopher Chamberland, Joel~J. Wallman, Stefanie Beale, and Raymond Laflamme.
\newblock Hard decoding algorithm for optimizing thresholds under general
  {Markovian} noise.
\newblock \emph{Physical Review A}, 95\penalty0 (4), 2017.
\newblock \doi{10.1103/PhysRevA.95.042332}.

\bibitem[Zhu et~al.(2021)Zhu, Jochym-O'Connor, and Dua]{zhu_topological_2021}
Guanyu Zhu, Tomas Jochym-O'Connor, and Arpit Dua.
\newblock Topological {Order}, {Quantum} {Codes} and {Quantum} {Computation} on
  {Fractal} {Geometries}.
\newblock \emph{arXiv:2108.00018 [cond-mat, physics:hep-th, physics:math-ph,
  physics:quant-ph]}, 2021.

\bibitem[Bomb{\'i}n and Martin-Delgado(2007)]{bombin_topological_2007}
H.~Bomb{\'i}n and M.~A. Martin-Delgado.
\newblock Topological {Computation} without {Braiding}.
\newblock \emph{Physical Review Letters}, 98\penalty0 (16):\penalty0 160502,
  2007.
\newblock \doi{10.1103/PhysRevLett.98.160502}.

\bibitem[Bomb{\'i}n and {Martin-Delgado}(2006)]{bombin2006}
H.~Bomb{\'i}n and M.~A. {Martin-Delgado}.
\newblock Topological {{Quantum Distillation}}.
\newblock \emph{Physical Review Letters}, 97\penalty0 (18):\penalty0 180501,
  October 2006.
\newblock \doi{10.1103/PhysRevLett.97.180501}.

\bibitem[Kubica and Beverland(2015)]{kubica2015}
Aleksander Kubica and Michael~E. Beverland.
\newblock Universal transversal gates with color codes - a simplified approach.
\newblock \emph{Physical Review A}, 91\penalty0 (3):\penalty0 032330, 2015.
\newblock \doi{10.1103/PhysRevA.91.032330}.

\bibitem[Vasmer and Browne(2019)]{vasmer_three-dimensional_2019}
Michael Vasmer and Dan~E. Browne.
\newblock Three-dimensional surface codes: {Transversal} gates and
  fault-tolerant architectures.
\newblock \emph{Physical Review A}, 100\penalty0 (1):\penalty0 012312, 2019.
\newblock \doi{10.1103/PhysRevA.100.012312}.

\bibitem[Bomb{\'i}n(2016)]{bombin_dimensional_2016}
H{\'e}ctor Bomb{\'i}n.
\newblock Dimensional {Jump} in {Quantum} {Error} {Correction}.
\newblock \emph{arXiv:1412.5079 [quant-ph]}, 2016.
\newblock URL \url{https://arxiv.org/abs/1412.5079}.

\bibitem[Kitaev(2003)]{kitaev_fault-tolerant_2003}
A.~Yu. Kitaev.
\newblock Fault-tolerant quantum computation by anyons.
\newblock \emph{Annals of Physics}, 303\penalty0 (1):\penalty0 2--30, 2003.
\newblock ISSN 0003-4916.
\newblock \doi{10.1016/S0003-4916(02)00018-0}.

\bibitem[Dennis et~al.(2002)Dennis, Kitaev, Landahl, and
  Preskill]{dennis_topological_2002}
Eric Dennis, Alexei Kitaev, Andrew Landahl, and John Preskill.
\newblock Topological quantum memory.
\newblock \emph{Journal of Mathematical Physics}, 43\penalty0 (9):\penalty0
  4452--4505, 2002.
\newblock \doi{10.1063/1.1499754}.

\bibitem[Horsman et~al.(2012)Horsman, Fowler, Devitt, and
  Meter]{horsman_surface_2012}
Clare Horsman, Austin~G. Fowler, Simon Devitt, and Rodney~Van Meter.
\newblock Surface code quantum computing by lattice surgery.
\newblock \emph{New Journal of Physics}, 14\penalty0 (12):\penalty0 123011,
  2012.
\newblock \doi{10.1088/1367-2630/14/12/123011}.

\bibitem[Kubica et~al.(2015)Kubica, Yoshida, and Pastawski]{kubica2015a}
Aleksander Kubica, Beni Yoshida, and Fernando Pastawski.
\newblock Unfolding the color code.
\newblock \emph{New Journal of Physics}, 17\penalty0 (8):\penalty0 083026,
  August 2015.
\newblock \doi{10.1088/1367-2630/17/8/083026}.

\bibitem[Bomb{\'i}n(2015)]{bombin_gauge_2015}
H{\'e}ctor Bomb{\'i}n.
\newblock Gauge color codes: optimal transversal gates and gauge fixing in
  topological stabilizer codes.
\newblock \emph{New Journal of Physics}, 17\penalty0 (8):\penalty0 083002,
  2015.
\newblock \doi{10.1088/1367-2630/17/8/083002}.

\bibitem[Brown and Kesselring()]{brown_kesselring_unreleased}
B.~Brown and M.~Kesselring.
\newblock In preparation.

\bibitem[Bravyi and Terhal(2009)]{bravyi2009}
Sergey Bravyi and Barbara Terhal.
\newblock A no-go theorem for a two-dimensional self-correcting quantum memory
  based on stabilizer codes.
\newblock \emph{New Journal of Physics}, 11\penalty0 (4):\penalty0 043029,
  2009.
\newblock \doi{10.1088/1367-2630/11/4/043029}.

\bibitem[sou()]{source_code}
URL \url{https://github.com/tRowans/clifford-errors}.

\bibitem[Agresti and Coull(1998)]{agresti1998}
Alan Agresti and Brent~A. Coull.
\newblock Approximate {{Is Better}} than ``{{Exact}}" for {{Interval
  Estimation}} of {{Binomial Proportions}}.
\newblock \emph{The American Statistician}, 52\penalty0 (2):\penalty0 119--126,
  1998.
\newblock \doi{10.2307/2685469}.

\bibitem[DasGupta et~al.(2001)DasGupta, Cai, and Brown]{dasgupta2001}
Anirban DasGupta, T.~Tony Cai, and Lawrence~D. Brown.
\newblock Interval {{Estimation}} for a {{Binomial Proportion}}.
\newblock \emph{Statistical Science}, 16\penalty0 (2):\penalty0 101--133, 2001.
\newblock \doi{10.1214/ss/1009213286}.

\bibitem[Bombin et~al.(2021)Bombin, Dawson, Mishmash, Nickerson, Pastawski, and
  Roberts]{bombin_logical_2021}
Hector Bombin, Chris Dawson, Ryan~V. Mishmash, Naomi Nickerson, Fernando
  Pastawski, and Sam Roberts.
\newblock Logical blocks for fault-tolerant topological quantum computation.
\newblock \emph{arXiv:2112.12160 [quant-ph]}, 2021.

\bibitem[Aliferis et~al.(2009)Aliferis, Brito, DiVincenzo, Preskill, Steffen,
  and Terhal]{Aliferis_2009}
P~Aliferis, F~Brito, D~P DiVincenzo, J~Preskill, M~Steffen, and B~M Terhal.
\newblock Fault-tolerant computing with biased-noise superconducting qubits: a
  case study.
\newblock \emph{New Journal of Physics}, 11\penalty0 (1):\penalty0 013061, jan
  2009.
\newblock \doi{10.1088/1367-2630/11/1/013061}.

\bibitem[Shulman et~al.(2012)Shulman, Dial, Harvey, Bluhm, Umansky, and
  Yacoby]{Shulman2012}
M.~D. Shulman, O.~E. Dial, S.~P. Harvey, H.~Bluhm, V.~Umansky, and A.~Yacoby.
\newblock Demonstration of entanglement of electrostatically coupled
  singlet-triplet qubits.
\newblock \emph{Science}, 336\penalty0 (6078):\penalty0 202--205, 2012.
\newblock \doi{10.1126/science.1217692}.

\bibitem[Nigg et~al.(2014)Nigg, M{\"u}ller, Martinez, Schindler, Hennrich,
  Monz, Martin-Delgado, and Blatt]{Nigg2014}
D.~Nigg, M.~M{\"u}ller, E.~A. Martinez, P.~Schindler, M.~Hennrich, T.~Monz,
  M.~A. Martin-Delgado, and R.~Blatt.
\newblock Quantum computations on a topologically encoded qubit.
\newblock \emph{Science}, 345\penalty0 (6194):\penalty0 302--305, 2014.
\newblock \doi{10.1126/science.1253742}.

\bibitem[Lescanne et~al.(2020)Lescanne, Villiers, Peronnin, Sarlette, Delbecq,
  Huard, Kontos, Mirrahimi, and Leghtas]{Lescanne2020}
Rapha{\"e}l Lescanne, Marius Villiers, Th{\'e}au Peronnin, Alain Sarlette,
  Matthieu Delbecq, Benjamin Huard, Takis Kontos, Mazyar Mirrahimi, and Zaki
  Leghtas.
\newblock Exponential suppression of bit-flips in a qubit encoded in an
  oscillator.
\newblock \emph{Nature Physics}, 16\penalty0 (5):\penalty0 509--513, May 2020.
\newblock \doi{10.1038/s41567-020-0824-x}.

\bibitem[Grimm et~al.(2020)Grimm, Frattini, Puri, Mundhada, Touzard, Mirrahimi,
  Girvin, Shankar, and Devoret]{Grimm2020}
A.~Grimm, N.~E. Frattini, S.~Puri, S.~O. Mundhada, S.~Touzard, M.~Mirrahimi,
  S.~M. Girvin, S.~Shankar, and M.~H. Devoret.
\newblock Stabilization and operation of a kerr-cat qubit.
\newblock \emph{Nature}, 584\penalty0 (7820):\penalty0 205--209, Aug 2020.
\newblock \doi{10.1038/s41586-020-2587-z}.

\bibitem[Panteleev and Kalachev(2021)]{Panteleev2021degeneratequantum}
Pavel Panteleev and Gleb Kalachev.
\newblock Degenerate {Q}uantum {LDPC} {C}odes {W}ith {G}ood {F}inite {L}ength
  {P}erformance.
\newblock \emph{{Quantum}}, 5:\penalty0 585, 2021.
\newblock \doi{10.22331/q-2021-11-22-585}.

\bibitem[Roffe et~al.(2020)Roffe, White, Burton, and
  Campbell]{roffe_decoding_2020}
Joschka Roffe, David~R. White, Simon Burton, and Earl Campbell.
\newblock Decoding across the quantum low-density parity-check code landscape.
\newblock \emph{Physical Review Research}, 2\penalty0 (4):\penalty0 043423,
  2020.
\newblock \doi{10.1103/PhysRevResearch.2.043423}.

\bibitem[Quintavalle et~al.(2021)Quintavalle, Vasmer, Roffe, and
  Campbell]{quintavalle2021a}
Armanda~O. Quintavalle, Michael Vasmer, Joschka Roffe, and Earl~T. Campbell.
\newblock Single-{{Shot Error Correction}} of {{Three-Dimensional Homological
  Product Codes}}.
\newblock \emph{PRX Quantum}, 2\penalty0 (2):\penalty0 020340, 2021.
\newblock \doi{10.1103/PRXQuantum.2.020340}.

\bibitem[Brown(2020)]{brown_fault-tolerant_2020}
Benjamin~J. Brown.
\newblock A fault-tolerant non-{Clifford} gate for the surface code in two
  dimensions.
\newblock \emph{Science Advances}, 2020.
\newblock \doi{10.1126/sciadv.aay4929}.

\bibitem[Scruby et~al.(2021)Scruby, Browne, Webster, and
  Vasmer]{scruby_numerical_2021}
T.~R. Scruby, D.~E. Browne, P.~Webster, and M.~Vasmer.
\newblock Numerical {Implementation} of {Just}-{In}-{Time} {Decoding} in
  {Novel} {Lattice} {Slices} {Through} the {Three}-{Dimensional} {Surface}
  {Code}.
\newblock \emph{arXiv:2012.08536 [quant-ph]}, 2021.
\newblock URL \url{https://arxiv.org/abs/2012.08536}.

\bibitem[Kubica(2018)]{kubica_abcs_2018}
Aleksander~Marek Kubica.
\newblock \emph{The {ABCs} of the {Color} {Code}: {A} {Study} of {Topological}
  {Quantum} {Codes} as {Toy} {Models} for {Fault}-{Tolerant} {Quantum}
  {Computation} and {Quantum} {Phases} {Of} {Matter}}.
\newblock {PhD} {Thesis}, California Institute of Technology, 2018.
\newblock URL \url{http://dx.doi.org/10.7907/059V-MG69}.

\bibitem[Vasmer and Kubica(2021)]{vasmer_morphing_2021}
Michael Vasmer and Aleksander Kubica.
\newblock Morphing quantum codes.
\newblock \emph{arXiv:2112.01446 [quant-ph]}, 2021.
\newblock URL \url{https://arxiv.org/abs/2112.01446}.

\bibitem[Jochym-O'Connor and Yoder(2021)]{jochym-oconnor_four-dimensional_2021}
Tomas Jochym-O'Connor and Theodore~J. Yoder.
\newblock Four-dimensional toric code with non-{Clifford} transversal gates.
\newblock \emph{Physical Review Research}, 3\penalty0 (1):\penalty0 013118,
  2021.
\newblock \doi{10.1103/PhysRevResearch.3.013118}.

\bibitem[Scruby(2021)]{scruby_logical_2021}
Thomas~Rowan Scruby.
\newblock \emph{Logical gates by code deformation in topological quantum
  codes}.
\newblock {Phd} {Thesis}, UCL (University College London), 2021.
\newblock URL \url{https://discovery.ucl.ac.uk/id/eprint/10135040}.

\bibitem[Rengaswamy et~al.(2020)Rengaswamy, Calderbank, Newman, and
  Pfister]{rengaswamy_optimality_2020}
Narayanan Rengaswamy, Robert Calderbank, Michael Newman, and Henry~D. Pfister.
\newblock On optimality of {CSS} codes for transversal {$T$}.
\newblock \emph{IEEE Journal on Selected Areas in Information Theory},
  1\penalty0 (2):\penalty0 499--514, 2020.
\newblock \doi{10.1109/JSAIT.2020.3012914}.

\bibitem[Beverland et~al.(2020)Beverland, Campbell, Howard, and
  Kliuchnikov]{beverland_lower_2020}
Michael Beverland, Earl Campbell, Mark Howard, and Vadym Kliuchnikov.
\newblock Lower bounds on the non-{Clifford} resources for quantum
  computations.
\newblock \emph{Quantum Science and Technology}, 5\penalty0 (3):\penalty0
  035009, 2020.
\newblock \doi{10.1088/2058-9565/ab8963}.

\end{thebibliography}

\end{document}